\documentclass[aip,jcp]{revtex4-1}
\usepackage{amsmath}
\usepackage{amsfonts}
\usepackage{braket}
\usepackage{color}
\usepackage{graphicx}

\begin{document}
\title{Matrix Product Operators, Matrix Product States, and ab initio Density Matrix
Renormalization Group algorithms}
 \author{Garnet Kin-Lic Chan}
\affiliation{Department of Chemistry, Princeton University, New Jersey, NJ-08544, USA }
\author{Anna Keselman}
\affiliation{Department of Condensed Matter Physics, Weizmann Institute of Science, Rehovot, Israel 76100}
 \affiliation{Department of Physics and Astronomy, University of California, Irvine, CA 92697-4575, USA}
 \author{Naoki Nakatani}
\affiliation{Institute for Catalysis, Hokkaido University, Kita 21 Nishi 10, Sapporo, Hokkaido 001-0021, Japan}
 \author{Zhendong Li}
\affiliation{Department of Chemistry, Princeton University, New Jersey, NJ-08544, USA }
 \author{Steven R. White}
\affiliation{Department of Physics and Astronomy, University of California, Irvine, CA 92697-4575, USA}

\begin{abstract}
Current descriptions of the ab initio DMRG algorithm use two superficially different languages: an older
language of the renormalization group and renormalized operators, and a more recent language of
matrix product states and matrix product operators.
The same algorithm can appear dramatically different when written in the two different vocabularies. In this work,
we carefully describe the translation between the two languages in several contexts. First, we describe
how to efficiently implement the ab-initio DMRG sweep using a matrix product operator based code, and
the equivalence to the original renormalized operator implementation. Next we describe how to implement
the general matrix product operator/matrix product state algebra within a pure renormalized operator-based DMRG code. Finally,
we discuss two improvements of the ab initio DMRG sweep algorithm motivated by matrix product operator language:
Hamiltonian compression, and a sum over operators representation that allows for perfect computational parallelism.
The connections and correspondences described here serve to link the future developments with the past, and are important
in the efficient implementation of continuing advances in ab initio DMRG and related algorithms.
\end{abstract}

\maketitle
\section{Introduction}

The density matrix renormalization group (DMRG), introduced by White,\cite{white1992density,white1993density} is now more than two decades old.
Although originally presented as a computational framework for
one-dimensional lattice systems, in the last decade many of its most interesting applications
have been to a much broader class of problems.
In the context of quantum chemistry, it was recognized early on that, as
a non-perturbative method, the DMRG could
be a useful tool to replace configuration interaction. With the advent of
efficient ab initio algorithms in the early 2000's,\cite{xiang1996density,white1999ab,daul2000full,chan2002highly,chan2004algorithm,
legeza2003controlling,legeza2003qc} the DMRG has since established itself as an indispensable part of
the toolkit of quantum chemistry, especially in problems requiring the accurate treatment
of strongly correlated electrons
\cite{white1999ab,daul2000full,mitrushenkov2001quantum,chan2002highly,chan2003exact,
legeza2003controlling,legeza2003qc,legeza2003optimizing,legeza2004quantum,
chan2004algorithm,mitrushenkov2003quantum,chan2004state,chan2005density,
moritz2006construction,hachmann2006multireference,marti2008density,ghosh2008orbital,chan2008density,
zgid2008obtaining,marti2010density,luo2010optimizing,marti2011new,kurashige2011second,sharma2012spin,chan2012low,wouters2012longitudinal,mizukami2012more,
kurashige2013entangled,sharma2014low,wouters2014density,wouters2014chemps2,
fertitta2014investigation,knecht2014communication,szalay2015tensor,
yanai2015density,olivares2015ab,li2016hilbert,guo2016nelectron}.

The conceptual framework of the DMRG has further greatly expanded and deepened
in the last decade. In the early 2000's, it became clear that the
power of DMRG-like algorithms originates
from  the ``matrix product'' structure of the ansatz\cite{ostlund1995thermodynamic,rommer1997class}
which expresses the low entanglement nature of one-dimensional low-energy quantum eigenstates, such
as the ground-state.
This entanglement perspective made it possible to expand the ideas of the DMRG into new domains:
matrix product operator representations\cite{verstraete2004matrix,mcculloch2007density,verstraete2008matrix,pirvu2010matrix}, time-evolution\cite{vidal2004efficient,daley2004time,white2004real}, infinite systems\cite{vidal2007classical,orus2008infinite,mcculloch2008infinite}, finite temperatures\cite{verstraete2004matrix,feiguin2005finite}, and higher-dimensions\cite{nishino2001two,verstraete2004renormalization,verstraete2008matrix,orus2014practical,
murg2010simulating,nakatani2013efficient,murg2015tree}, to name a few. Beyond computation, the language
of matrix product and tensor network states is now widely used
to  reason about the structure of many-particle quantum states.\cite{mcculloch2007density,verstraete2008matrix,schollwock2011density,orus2014practical}
Within this greatly expanded setting, DMRG is often taken
to be synonymous with the sweep-like algorithms
commonly used with matrix product states (MPS) and matrix product operators (MPO), e.g. ``finite-temperature DMRG'', ``time-dependent DMRG'', and
``infinite DMRG'', while the term ``tensor network'' embodies the wider class of representations and
algorithms associated with higher dimensions.

%% Further, a new graphical language in which to conveniently represents states and operators (in t
%% work with DMRG algorithms and its generalizations
%% has been an
%% increasingly deeper understanding of the formal structure of DMRG and how to extend it.
%% The matrix product state (MPS) structure of the DMRG ansatz was recognized already
%% at the beginning, but it was not until the early 2000's
%% that the vocabulary of the matrix product state, came into widespread use.
%% In part, this was due to the adoption of a particularly efficient graphical notation,
%% which greatly simplified reasoning about MPS and DMRG. No less important was the explosion
%% of new algorithms which expanded the original concepts of the DMRG to new domains: to  An immediate extension was
%% to generalize matrix product states to matrix product operators, which allowed the same convenient language
%% and formalism to be used for operators.

Early ab initio DMRG work focused on how to efficiently implement
energy optimization and compute expectation values,\cite{white1999ab,daul2000full,
mitrushenkov2001quantum,chan2002highly,mitrushenkov2003quantum,chan2004algorithm,
chan2005density,legeza2003controlling}
such as reduced density matrices.\cite{ghosh2008orbital,zgid2008obtaining,kurashige2011second,guo2016nelectron}
These expectation value computations are performed via
a sweep algorithm that proceeds through orbitals one-by-one. In the ab initio context, the key
step is to identify and construct efficiently the appropriate renormalized operators
as one proceeds through the sweep.
This concept of renormalized operators arises naturally within the renormalization group
framework within which the DMRG was originally proposed.
In modern day MPO/MPS parlance, however, renormalized operators are simply the computational intermediates
corresponding to partial traces of the operator (MPO) with the bra and ket states (MPS) as one
includes successive orbitals in a sweep\cite{schollwock2011density}. In an expectation value computation (or optimization), only these
partial traces of the MPO are required, and the explicit MPO itself
never needs to appear. Thus the original implementations of the ab initio DMRG, which focus on
renormalized operator-based computation, are not structured around explicit MPO's but
rather the renormalized operators and matrix product states. We refer to these
original implementations as ``pure renormalized operator-based'' DMRG implementations.

%% and how to build these intermediates in parallel.
%% We refer to these descriptions of the DMRG implementation as ``sweep-centric''.
%% Since the renormalized intermediates are key to efficiency, extensions of the ground-state energy
%% algorithm, e.g. to compute $n$-particle
%% reduced density matrix elements efficiently, have also focused on the
%% intermediates to build along a sweep.
%% Within the language of MPS and MPO, however, it is possible to
%% However, the graphical notation and language of MPS and MPO is also very useful
%% in thinking about DMRG algorithms.
Within the MPO/MPS setting,  it is of course natural to implement codes where MPO's appear explicitly.
It is important to emphasize that using MPO's in a code does not in itself change the ab initio DMRG algorithm.
The most efficient serial formulation of expectation value computation (without further approximation) remains
to use the MPO and the bra and ket MPS to build the renormalized operators, in precisely the same manner
as in the original ab initio DMRG.
However, having explicit MPO's in the code is useful in connecting
to the modern notation and graphical language of MPO's and MPS. %% , which allow the
%% simple depiction of algorithmic operations at  high level.
We will refer to DMRG programs organized around explicit MPO representations as ``MPO-based'' DMRG implementations.
These MPO-based ab initio DMRG implementations have been carried out by several groups,
including Murg {\it et al},\cite{murg2010simulating} the authors,\cite{Naoki} and Keller {\it et al}.\cite{keller2014determining, keller2015efficient, keller2016spin}. The implementations typically rely on general MPO/MPS libraries, such as \textsc{ITensor},\cite{iTensor} \textsc{MPSXX},\cite{Naoki} and \textsc{Alps}.\cite{dolfi2014matrix}
The implementations have been used in publications and some are freely available. However, with the exception of
that of Keller {\it et al}\cite{keller2015efficient, keller2016spin} they have not previously been described in detail in
the literature.

The computational steps of an MPO-based implementation are essentially the same as in the traditional ``renormalized operator-based'' DMRG implementation.
However, while the mapping between renormalized operators, and explicit MPO/MPS representations,
 %% description of the quantities computed along a sweep in terms of the MPO/MPS language
is well-known in general terms to DMRG practitioners, the lack of an explicit translation between quantities
appears as a source of confusion in the wider quantum chemistry community.
This is because the language involved in MPO-based implementations and the pure renormalized operator-based implementations
can appear very different.
For example, in the description of the DMRG algorithm by Keller {\it et al} in Ref.\cite{keller2015efficient, keller2016spin}, the connection
to the identical quantities and operations in the original ab initio DMRG algorithm is not described. To the uninitiated,
the discussed algorithm may appear fundamentally different.
The problem is exacerbated by the fact that the optimized intermediates in an ab initio DMRG program
enter in a complicated way in both pure renormalized operator-based and MPO-based implementation.
A first goal of this  paper is to provide a pedagogical description of how to efficiently implement
the ab initio DMRG in an MPO-based setting, highlighting the translation to and from
the original language of renormalized operators used in renormalized operator-based implementations. This constitutes Section \ref{sec:overview}
of this paper.

%% We also describe how to set up a ``sweep-centric'' code implementation to support an MPO/MPS interface,
%% as used in codes such as \textsc{Block}. These various connections are the subjects of Sections \ref{sec:mps} to \ref{sec:impl}.

If the MPO/MPS language only supplied a re-writing of the existing DMRG algorithm, it would be of limited use.
However, the power of this language is that certain new perspectives become more natural, and this leads
to new algorithms. For example, the MPO/MPS algebra introduces many new operations
beyond the scalar computation of a bra-operator-ket expectation value. These operations provide
the basis for new algorithms such as DMRG time-evolution\cite{vidal2004efficient,daley2004time,white2004real}. One way to implement these algorithms
in a pure renormalized operator-based DMRG code would be to augment such codes with explicit MPO's.
However,  in almost all cases of interest,
 i.e. if the output of the algorithm is a scalar quantity or an MPS, the operations of the MPO/MPS algebra
can be carried out efficiently using only the renormalized operators. Thus, one can
build a light-weight layer on top of an existing renormalized operator-based DMRG code to support the relevant MPO/MPS algebra.
This  strategy is used, for example, in the \textsc{Block} code of some of the authors,
to support the MPO/MPS operations needed for
perturbation\cite{sharma2014communication,sharma2015multireference,sharma2016quasi} and response-based\cite{dorando2009analytic,nakatani2014linear} calculations on top of DMRG wavefunctions.
We here describe this further connection between sweep computations and  MPO/MPS algebra in detail in Section \ref{sec:impl} of this work.

%% the MPO/MPS language makes certain perspectives more natural.
%% For example, when MPO's appeared, it was immediately recognized that the DMRG could be trivially extended to finite temperatures, by regarding MPO's as DMRG ``states'' in Liouville space.
MPO/MPS concepts also suggest new formulations of the ab initio DMRG sweep algorithm itself. We describe two such formulations here, which can be implemented with equal facility in either a pure renormalized operator-based or MPO-based code.
The first is a particular version of Hamiltonian compression (in a particular MPO gauge), that can be directly applied to the
ab initio Hamiltonian integrals. This allows for a reduction in the number of renormalized
operators built in a DMRG sweep, which can lead to substantial speedups. This algorithm
is described in Section~\ref{sec:compress}, using the linear hydrogen chain as toy computational example.
The second is a new way to express the Hamiltonian as a sum of operators, which leads to perfect parallelization of the DMRG algorithm.
These ideas are described in Sections~\ref{sec:sum_over_ops} and ~\ref{sec:parallel}. Finally, our conclusions are provided in Section~\ref{sec:conclusions}.

\section{The DMRG algorithm in the MPO and MPS language}
\label{sec:overview}

\subsection{The DMRG in renormalization group language}

\label{sec:sweep-centric}
%% The purpose of these initial sections is to describe the efficient implementation
%% of the DMRG algorithm in the MPO-centric language, and in this context explicitly map between the original sweep-centric
%% description and the same algorithm in the MPO-centric language.
To make the connections clear, we provide a quick refresher of the main concepts
of the  ab initio DMRG sweep algorithm using renormalization group language, as described,
for example, in Refs.\cite{white1999ab,chan2002highly}.

The goal of the DMRG sweep is to compute and/or minimize the energy of the DMRG variational wavefunction.
There are variational parameters (renormalized wavefunctions) associated with each of the $K$ orbitals in the problem, thus
the sweep consists of iteratively solving a set of ground-state problems, one at a time, associated with each of the $K$ orbitals.
To start the procedure, we choose a sequence in which to traverse the orbitals by mapping the orbitals
onto the $K$ sites of a one-dimensional lattice. The sweep going from left to right
then consists of $K$ steps. At a given step $k$,
we can think of the orbitals as partitioned into two sets, the left block
of orbitals $L_k$ and a right block of orbitals $R_k$, which sets up a tensor product structure
of the Hilbert space and operators on the Hilbert space.
Associated with the left and right blocks are
 a set of left and right renormalized states (bases), and left and right renormalized operators. The latter
are used as an operator basis to reconstruct the Hamiltonian on all $K$ orbitals, and a proper left-right
decomposition of the Hamiltonian (into the so-called normal and complementary operators) is a key step
in implementing the ab initio DMRG algorithm efficiently.

For each of the $K$ steps of the sweep, three operations are carried out:
\begin{enumerate}
\item  blocking, which updates the set of left and right renormalized bases
and operators, from the renormalized representations at site $k-1$, to the ``blocked'' representation at site $k$.
% $\{ \ket{l_{\alpha_{k}}} \}$, $\{ \ket{r_{\alpha_k}} \}$, and $\{ \mathbf{O}^{L_k}_{\alpha_k} \}, \{ \mathbf{O}^{R_k}_{\alpha_k}\}$
\item  solving, which computes the (ground-state) renormalized wavefunction at site $k$ in the product
of the left- and right-renormalized bases,
\item  and decimation, which transforms the ``blocked'' bases and operators to the renormalized representation at site $k$.
\end{enumerate}
A complete sweep from left to right and back updates all  renormalized bases $\{ \ket{l_{\alpha_{k}}} \}$, $\{ \ket{r_{\alpha_k}} \}$, and renormalized operators $\{ \mathbf{O}^{L_k}_{\beta_k} \}, \{ \mathbf{O}^{R_k}_{\beta_k}\}$ for every
partition of the orbitals $k$.

The above operations, and the associated renormalized quantities, are the central objects
in the original pure renormalized operator-based ab initio DMRG algorithm.
{\it All the same steps and quantities will also appear in an efficient ``MPO-based'' DMRG implementation.}
To make the translation, we thus will be looking to highlight (i) the connection between
the renormalized left- and right-bases and renormalized wavefunctions,
 and the tensors in the MPS, (ii) the correspondence between the
left- and right- renormalized operators and the tensors in the Hamiltonian MPO,
(iii) the relation between the efficient implementation of DMRG energy minimization and expectation value evaluation with MPS and MPO, and the computational organization into a DMRG sweep algorithm using
normal and complementary operators, with the individual steps of blocking, solving, and decimation.

\subsection{Matrix product states}\label{sec:mps}

We now recall the basic concepts of MPS. This will also establish
some notation useful in discussing MPO's. The relationship between
the MPS and the renormalized bases and wavefunctions along a sweep has been discussed before in the chemistry
literature, for example in Refs.~\cite{chan2008introduction,chan2011density}. A particularly
detailed account is given in Ref.~\cite{schollwock2011density}.

Matrix product states (MPS) are the wavefunction representations that define the
variational space of the DMRG. Within the Fock space of an orthonormal basis of $K$ orbitals, the electronic wavefunction
is written in occupation representation as
\begin{align}
  \ket{\Psi} = \sum_{n_1 \cdots n_K} \Psi^{n_1 n_2 \cdots n_K} \ket{ n_1 n_2 \cdots n_K}
  \end{align}
where $\ket{n_1 n_2 \cdots n_K}$ is an occupancy basis state in the Fock space, and spin-labels have been suppressed. For a given particle number $N$, we have the condition
\begin{eqnarray}
\Psi^{n_1n_2\cdots n_K}=
\left\{\begin{array}{ll}
\Psi^{n_1n_2\cdots n_K},& \sum_{k=1}^K n_k=N,\\
0,& \mathrm{otherwise}.
\end{array}\right.\label{FStensor}
\end{eqnarray}
In a matrix product state,
the wavefunction amplitude for a given basis state is written as a product of matrices:
\begin{eqnarray}
\Psi^{n_1n_2\cdots n_K}
=\sum_{\{\alpha_k\}} A^{n_1}_{\alpha_1}[1]A^{n_2}_{\alpha_1\alpha_2}[2]\cdots A^{n_K}_{\alpha_{K-1}}[K],\label{FSMPS}
\end{eqnarray}
where the dimension of $A^{n_k}[k]$ is an $M \times M$ matrix
%% (or a $4 \times M \times M$ tensor if we include the $n_k$
%% index in the tensor $A[k]$
 (or a $2 \times M \times M$ tensor if we include the $n_k\in\{0,1\}$ index for spin-orbitals), and the leftmost and rightmost matrices are $1 \times M$ and $M \times 1$ vectors
to ensure that the matrix product results in the scalar amplitude $\Psi^{n_1n_2\cdots n_K}$. As the dimension $M$,
known variously as the bond-dimension or the number of renormalized states, increases, the
representation Eq.~(\ref{FSMPS}) becomes increasingly flexible. We will here assume for simplicity that all $A^{n_k}[k]$ are real.

\begin{figure}
\centering
\begin{tabular}{c}
{\resizebox{0.6\textwidth}{!}{\includegraphics{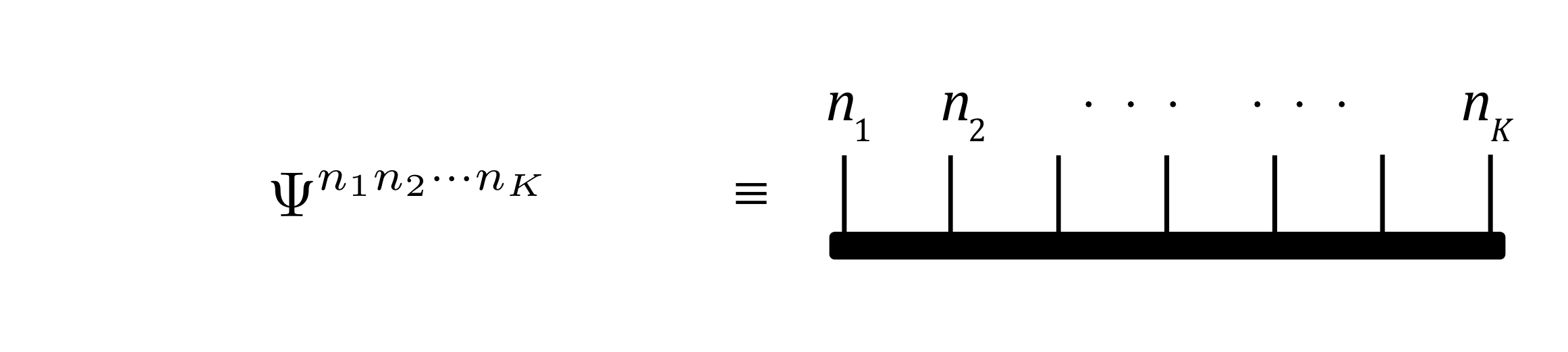}}}\\
(i) \\
{\resizebox{0.6\textwidth}{!}{\includegraphics{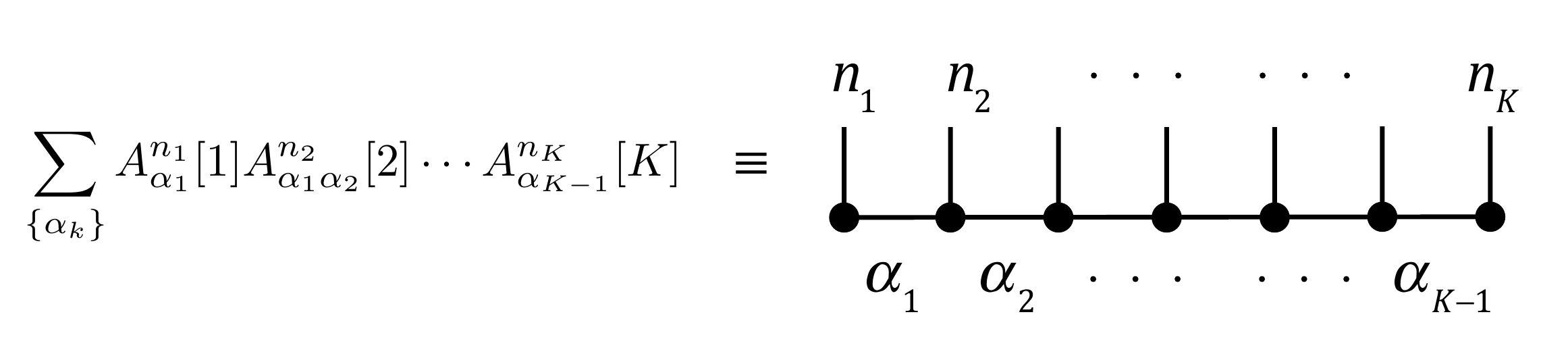}}}\\
(ii) \\
\end{tabular}
\caption{\label{fig:mps1} (i) Wavefunction coefficients in graphical notation. (ii) Representation
of wavefunction as a matrix product state in graphical notation (Eq.~(\ref{FSMPS})). }
\end{figure}

It is very useful to employ a graphical notation for the MPS. In this notation, the general wavefunction amplitude is represented
by a tensor with $K$ legs, while the MPS representation is a connected set of 2-index and 3-index tensors, each associated with a site. Contraction between the tensors represents summation, as shown in Fig.~\ref{fig:mps1}.

Note that there is a non-uniqueness in the representation
since we can always redefine two adjacent matrices
\begin{align}
  A^{n_k}[k] &\to A^{n_k}[k] G \\
  A^{n_{k+1}}[k+1] &\to G^{-1} A^{n_{k+1}}[k+1]
\end{align}
where $G$ is an invertible $M\times M$ ``gauge'' matrix, while leaving the matrix {\it product} invariant. This
redundancy is partially eliminated by placing additional constraints on the matrices, such as
the left orthonormality condition $\sum_{n_k} {A^{n_k}}^T A^{n_k} =1$ and right orthonormality condition
$\sum_{n_k} {A^{n_k}} {A^{n_k}}^T =1$. Applied to all the tensors,
this leads to  the left- and right- canonical forms of the MPS respectively. The DMRG sweep
algorithm employs a mixed-canonical form. In this case, at step $k$ of
the sweep, all tensors to the left of site $k$ are in left-canonical form, and all tensors
to the right of site $k$ are in right-canonical form. The MPS is then expressed as,
\begin{align}
\Psi^{n_1n_2\cdots n_K}
&=\sum_{\{\alpha_k\}} L^{n_1}_{\alpha_1}[1]L^{n_2}_{\alpha_1\alpha_2}[2]\cdots C^{n_k}_{\alpha_{k-1}\alpha_k}[k] \cdots
R^{n_K}_{\alpha_{K-1}}[K] \label{eq:mixed}
\end{align}
where we have emphasized the choice of gauge by using symbols $L$, $C$, $R$ for the different tensors. $C^{n_k}[k]$
is called the DMRG renormalized wavefunction.

\begin{figure}
\centering
\begin{tabular}{c}
{\resizebox{0.4\textwidth}{!}{\includegraphics{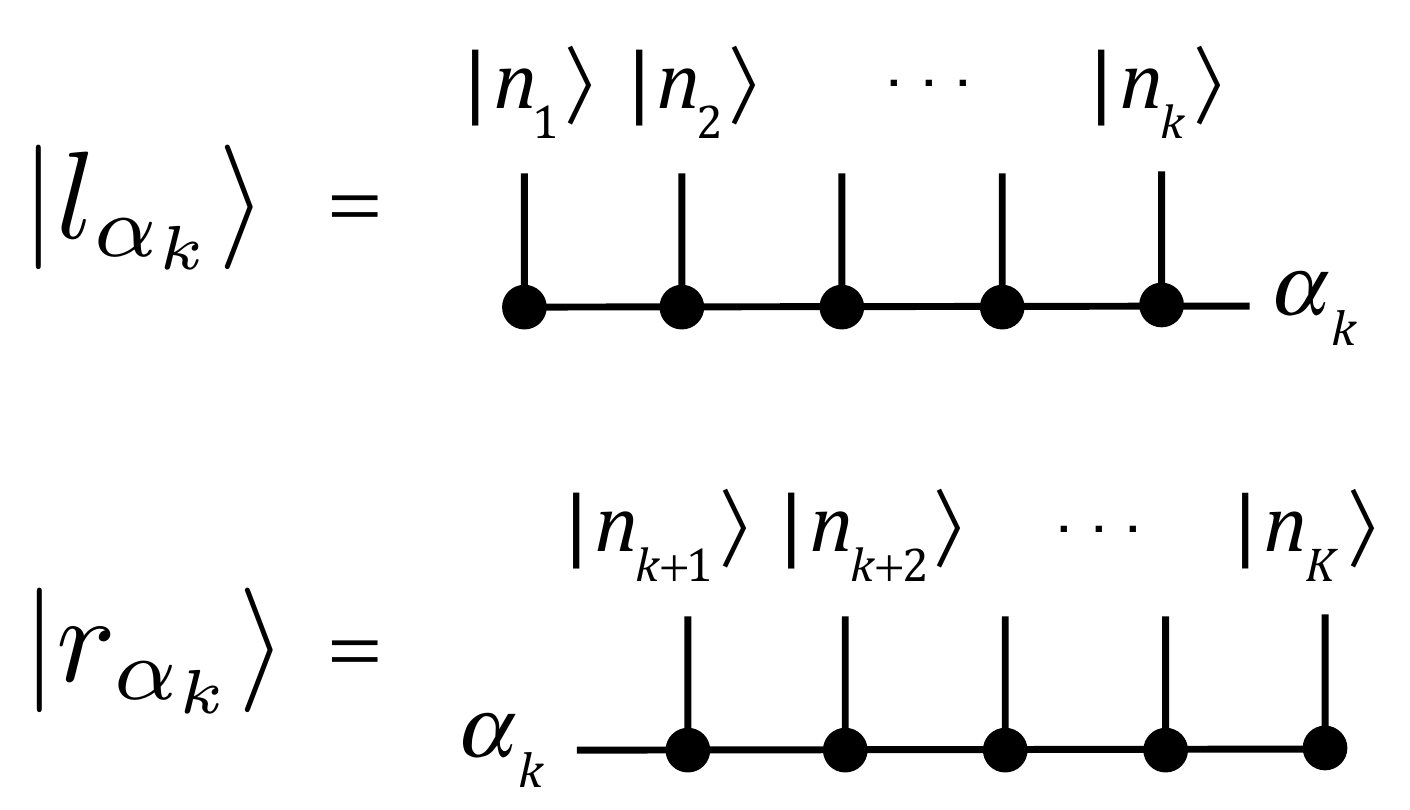}}}\\
\end{tabular}
\caption{\label{fig:mpslr} Left and right renormalized bases at site $k$ (Eq. (\ref{eq:lbasis}) and (\ref{eq:rbasis})) in
graphical notation.}
\end{figure}

The matrices in the MPS define a recursively constructed set of many-body renormalized basis states. These
are precisely the left- and right-renormalized bases that are constructed in the DMRG sweep. In this
context, the matrices are sometimes called  renormalization matrices.
For example, if we consider a (bi-)partitioning of the sites at site $k$, and consider
the left block of sites $1 \cdots k$, we obtain the left renormalized basis
\begin{align}
  \ket{l_{\alpha_k}} =  \sum_{  n_1 \cdots n_k  } (A^{n_1}[1] A^{n_2}[2] \cdots A^{n_k}[k])_{\alpha_k} |n_1 \cdots n_k\rangle
\label{eq:lbasis}
  \end{align}
and from the right block of sites $k+1 \to K$, we obtain the right renormalized basis
\begin{align}
  \ket{r_{\alpha_k}} =  \sum_{  n_{k+1} \cdots n_K } (A^{n_{k+1}}[k+1] A^{n_{k+2}}[k+2] \cdots A^{n_K}[K])_{\alpha_k} |n_{k+1} \cdots n_K\rangle \label{eq:rbasis}
\end{align}
The graphical representation of the left and right renormalized basis is shown in Fig.~\ref{fig:mpslr}.
Note that the renormalized states are defined for partitionings at any site $k$. Iterating
through the partitions from $1 \cdots K$ builds up the renormalized states
in the same recursive fashion as they are built up during a DMRG sweep.
In particular, the renormalized states at site $k+1$ are
explicitly defined from the renormalized states at site $k$ by the renormalization matrix $A^{n_{k+1}}[k]$, e.g. for
the left basis,
\begin{align}
  \ket{l_{\alpha_{k+1}}} = \sum_{\alpha_k n_{k+1}} A^{n_{k+1}}_{\alpha_k \alpha_{k+1}} \ket{l_{\alpha_k} n_{k+1}} \label{eq:blockdecimation}
\end{align}
and similarly for the right basis. The above transformation Eq.~(\ref{eq:blockdecimation}) is exactly
that of blocking and decimating the
states at step $k+1$ of the DMRG sweep:  blocking consists of
expanding the renormalized basis space, $\{ \ket{l_{\alpha_k}} \} \to \{ \ket{l_{\alpha_k} n_{k+1}} \}$, while
decimation consists of projecting $\{ \ket{l_{\alpha_k} n_{k+1}} \}\to \{ \ket{l_{\alpha_{k+1}}} \}$.

In determining the tensors $A^{n_k}[k]$ successively in the DMRG sweep, the tensor to be optimized at site $k$ is
expressed in the mixed-canonical gauge in Eq.~(\ref{eq:mixed}) ($C^{n_k}[k]$). In this gauge,
the MPS is written in terms of the renormalized states as
\begin{align}
|\Psi\rangle = \sum_{\alpha_{k-1} n_k \alpha_k} C^{n_k}_{\alpha_{k-1} \alpha_k} |l_{\alpha_{k-1}} n_k r_{\alpha_k}\rangle
\label{MPSCform}
\end{align}
and thus the coefficients $C^{n_k}_{\alpha_{k-1} \alpha_k}$ are the coefficients of the wavefunction in the DMRG
renormalized space.
We can also write the MPS more compactly in terms of the left renormalized states at site $k$, $\{\ket{l_{\alpha_k}}\}$
(rather than the blocked basis $\{ \ket{ l_{\alpha_{k-1}} n_k} \}$), giving the simpler form
\begin{align}
  |\Psi\rangle = \sum_{\alpha_k} \ket{l_{\alpha_k} r_{\alpha_k}}s_{\alpha_k}
\end{align}
This shows that the MPS corresponds to a wavefunction whose Schmidt decomposition, for the bi-partitioning at any site $k$,
contains at most $M$ singular values $s_{\alpha_k}$.

From the above, it is clear that there is no computational distinction to be made between
working with the renormalized representations (left, right renormalized bases and renormalized wavefunctions) in
a DMRG sweep and the underlying matrix product tensors: since one set is defined in terms of the other, both quantities
are always present, in any DMRG implementation, simultaneously.

%%In terms of these two sets of left and right renormalized states, the MPS takes the simple form
%% \begin{align}
%%   |\Psi\rangle = \sum_{\alpha} \ket{l_\alpha r_\alpha}
%% \end{align}

\subsection{Matrix product operators}\label{sec:mpo}

We now review the formalism of matrix product operators, emphasizing the similarity with the above analysis for
matrix product states.
A matrix product operator (MPO) is a generalization of a matrix product representation to the operator space\cite{verstraete2004matrix,mcculloch2007density,verstraete2008matrix,pirvu2010matrix}. Let us first define an operator basis that spans the operators associated with a given spin-orbital site, such as $\{\hat{z}\} = \{ 1, a, a^\dag, a^\dag a\}$. A general operator can be written as the expansion
\begin{align}
\hat{O} = \sum_{ \{ \hat{z} \}} O^{z_1 z_2 \cdots z_K} \ \hat{z}_1 \hat{z}_2 \cdots \hat{z}_K \label{eq:opexpand}
\end{align}

We introduce a matrix product operator representation as a representation
for the element $O^{z_1 z_2 \cdots z_K}$
%% \begin{align}
%% {O}^{z_1 z_2 \cdots z_K} = \sum_{\{ \beta_k \}} W^{{z}_1}_{\beta_1 }[1] W^{{z}_2}_{\beta_1 \beta_2}[2] \cdots W^{{z}_K}_{\beta_K} %\hat{z}_1 \hat{z}_2 \cdots \hat{z}_K
%% \label{eq:mpoopbasis}
%% \end{align}
%{\color{blue}
\begin{align}
{O}^{z_1 z_2 \cdots z_K} = \sum_{\{ \beta_k \}} W^{{z}_1}_{\beta_1 }[1] W^{{z}_2}_{\beta_1 \beta_2}[2] \cdots W^{{z}_K}_{\beta_K}[K] %\hat{z}_1 \hat{z}_2 \cdots \hat{z}_K
\label{eq:mpoopbasis}
\end{align}
%}
Note that the entries of $W^{{z_k}}[k]$ are simply scalars; the operators (which, for example,
describe the non-commuting nature of the fermions)
are contained within the operator string $\hat{z}_1 \hat{z}_2 \cdots \hat{z}_K$ in Eq.~(\ref{eq:opexpand}). Also,
the decomposition in Eq. (\ref{eq:mpoopbasis}) is not unique, and contains the same ``gauge'' redundancy
as in the case of the MPS.

\begin{figure}
\centering
\begin{tabular}{c}
{\resizebox{0.6\textwidth}{!}{\includegraphics{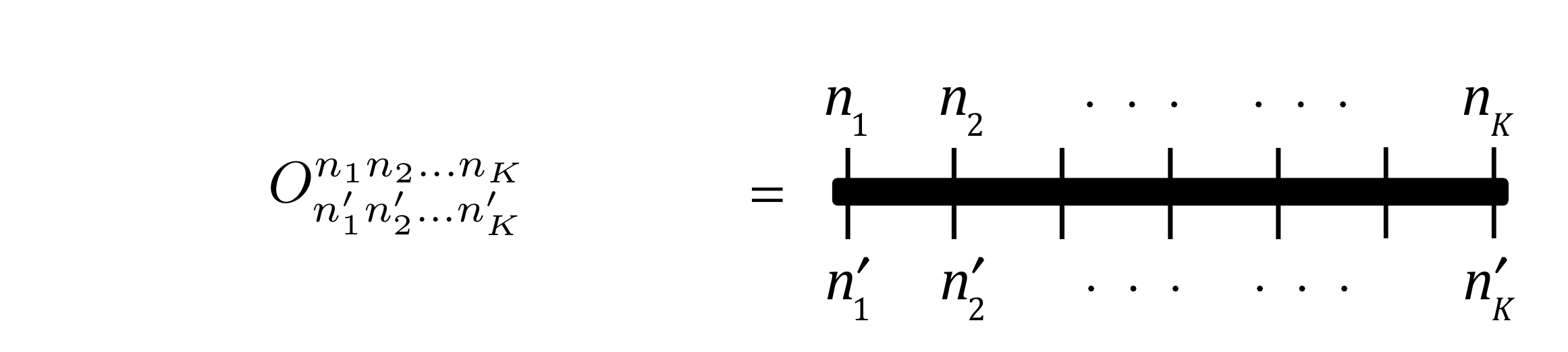}}}\\
(i) \\
{\resizebox{0.6\textwidth}{!}{\includegraphics{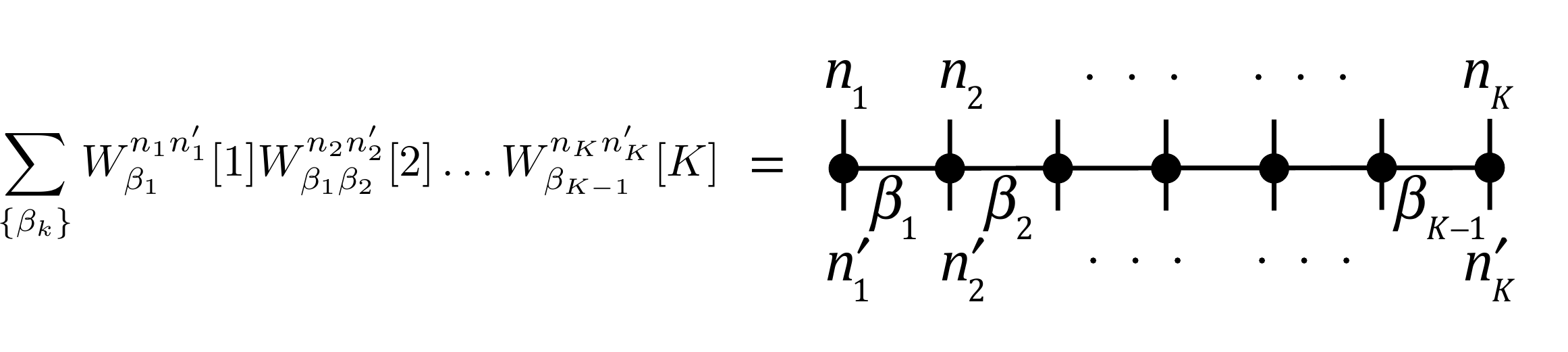}}}\\
(ii) \\
\end{tabular}
\caption{\label{fig:mpo1} Matrix product operator (Eq.~(\ref{eq:mponform})) in graphical notation.}
\end{figure}

It is convenient to define a matrix product operator form where the matrices appearing
are operator valued (i.e. the matrix elements are operators). This is done by grouping the operator $\hat{z}_k$ with the corresponding tensor ${W}^{z_k}[k]$,
to define the operator valued matrix $\hat{W}[k]$,
\begin{align}
\hat{W}_{\beta_{k-1} \beta_k}[k] = \sum_{z_k} W^{z_k}_{\beta_{k-1} \beta_k}[k]  \hat{z}_k
\end{align}
The full operator $\hat{O}$ is then a product over the operator valued matrices,
\begin{align}
\hat{O} = \hat{W}[1] \hat{W}[2] \cdots \hat{W}[K] \label{eq:wproduct}
\end{align}

An MPO can be expressed in graphical form. Here, it is more conventional to write the operator basis on each
site as $\{ \hat{z}\} = \{ \ket{n}\bra{n'} \} $, such that
\begin{align}
\hat{O} = \sum_{\{n_k n_k'\} } O^{n_1 n_2 \cdots n_K}_{n_1' n_2' \cdots n_K'} |n_1 n_2 \cdots n_K\rangle \langle n_1' n_2' \cdots n_K'|  \label{eq:mponform} \end{align}
The MPO representation of the operator matrix element is
%% \begin{align}
%% O^{n_1 n_2 \cdots n_K}_{n_1' n_2' \cdots n_K'}  = \sum_{\{ \beta_k\}} W^{n_1 n_1'}[1]_{\beta_1 } W^{n_2n_2'}[2]_{\beta_1 \beta_2} \cdots W^{n_Kn_K'}[K]_{\beta_K}
%% %|n_1 n_2 \cdots n_K\rangle \langle n_1' n_2' \cdots n_K'|
%% \label{eq:mpoopbasis2}
%% \end{align}
%{\color{blue}
\begin{align}
O^{n_1 n_2 \cdots n_K}_{n_1' n_2' \cdots n_K'}  = \sum_{\{ \beta_k\}} W^{n_1 n_1'}_{\beta_1 }[1] W^{n_2n_2'}_{\beta_1 \beta_2}[2] \cdots W^{n_Kn_K'}_{\beta_{K-1}}[K]
%|n_1 n_2 \cdots n_K\rangle \langle n_1' n_2' \cdots n_K'|
\label{eq:mpoopbasis2}
\end{align}
%}
A general operator is represented by a tensor with $K$ ``up'' legs and
$K$ ``down'' legs. The MPO is drawn as a connected set of 3-index and 4-index tensors, each associated with a site, as illustrated in Fig.~\ref{fig:mpo1}. Note that in this formulation, the non-commuting nature of fermion operators is implicit
in the values of the elements of the site-tensors $W^{n_kn_k'}_{\beta_{k-1}\beta_k}[k]$ in Eq. \eqref{eq:mpoopbasis2}. 

Similarly to as in the case of MPS, the MPO tensors define sets of many-body  operators over a partitioning of the sites.
For example, for a partitioning into a left block of sites $1 \cdots k$, we define the corresponding left  operators $\hat{O}^L_{\beta_k}$,
\begin{align}
  \hat{O}^L_{\beta_k} =  (\hat{W}[1] \hat{W}[2] \cdots \hat{W}[k])_{\beta_k}
%\sum_{  \hat{z}_1 \cdots \hat{z}_k  } (A^{\hat{z}_1}[1] A^{\hat{z}_2}[2] \cdots A^{\hat{z}_k}[k])_{\beta}
%\ \hat{z}_1 \cdots \hat{z}_k
  \end{align}
and from the right block of sites $k+1 \cdots K$, we define a set of right operators, $\hat{O}^R_{\beta_k}$,
\begin{align}
  \hat{O}^R_{\beta_k} =  (\hat{W}[k+1] \hat{W}[k+2] \cdots \hat{W}[K])_{\beta_k}
%\sum_{  \hat{z}_{k+1} \cdots \hat{z}_K } (A^{\hat{z}_{k+1}}[k+1] A^{\hat{z}_{k+2}}[k+2] \cdots A^{\hat{z}_K}[K])_{\beta} \ \hat{z}_{k+1} \cdots \hat{z}_K
\end{align}
Using the sets of left and right operators, the full operator at any partition can be expressed as
\begin{align}
\hat{O} = \sum_{\beta_k} \hat{O}^L_{\beta_k} \hat{O}^R_{\beta_k} \label{eq:Hpartition}
\end{align}
Note that the bond-dimension of the MPO at partition $k$ is equal to the number of terms in the summation over
$\beta$ in Eq.~(\ref{eq:Hpartition}).

\begin{figure}
\centering
\begin{tabular}{c}
{\resizebox{0.6\textwidth}{!}{\includegraphics{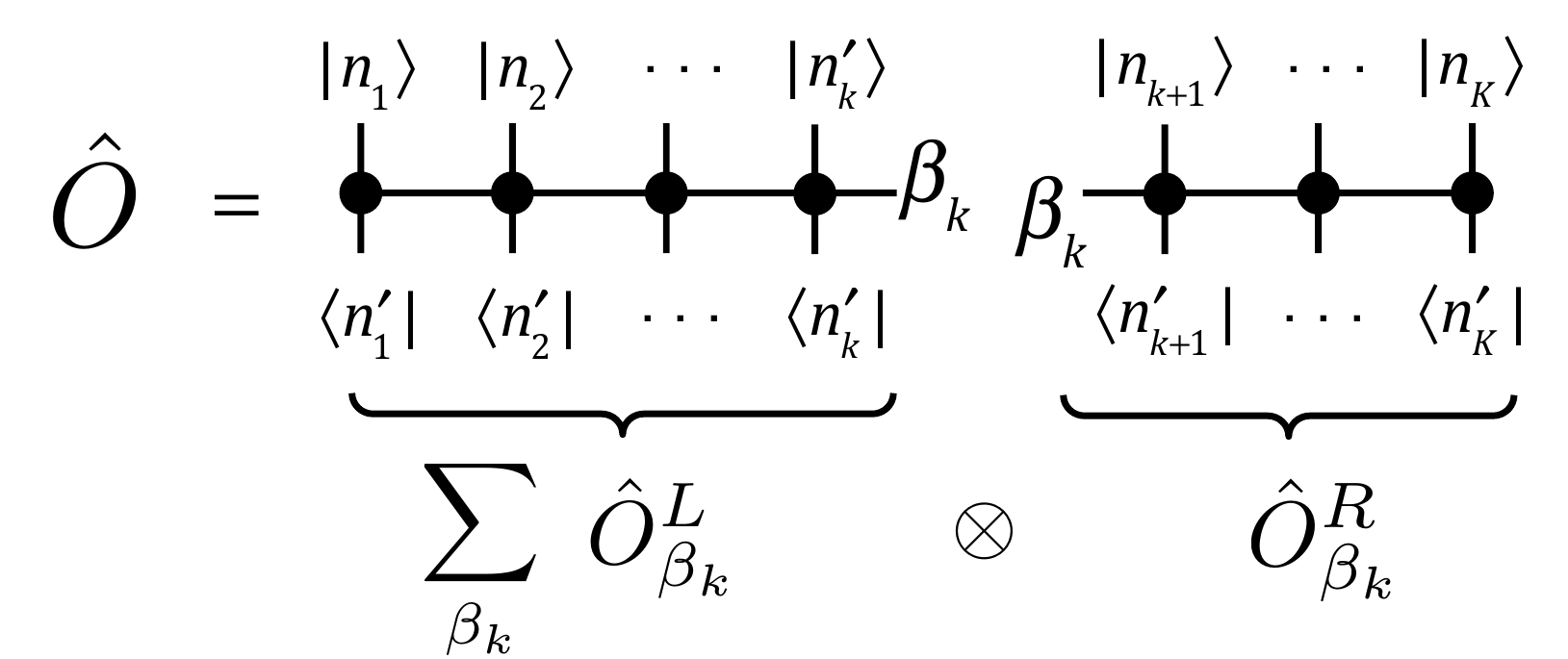}}}\\
(i) \\
{\resizebox{0.6\textwidth}{!}{\includegraphics{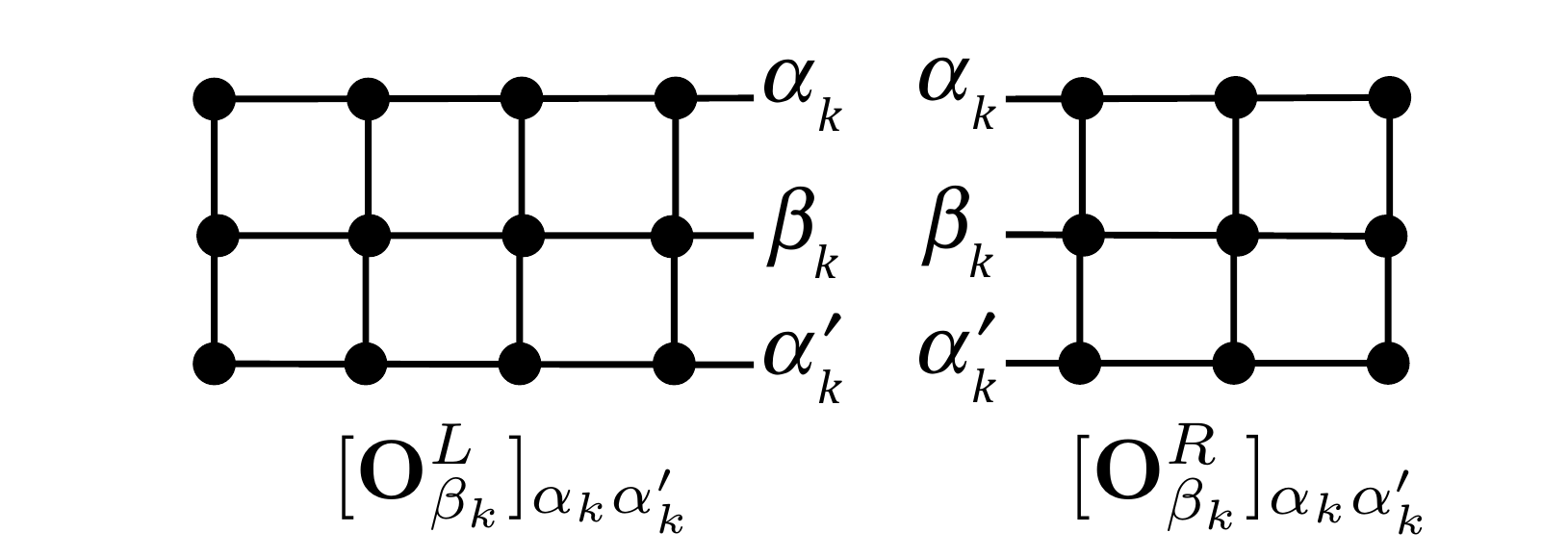}}}\\
(ii) \\
\end{tabular}
\caption{\label{fig:mpolr} (i) Left-right decomposition of the MPO at site $k$. (ii) Left- and right-block renormalized operators
at site $k$.
}
\end{figure}

The left-right decomposition of an operator at site $k$ described above is isomorphic to the left-right decomposition of an operator
at  step $k$ in a DMRG sweep. In particular,
the renormalized left-block operators $\mathbf{O}^{L}_{\beta_k}$ and renormalized right-block operators $\mathbf{O}^{R}_{\beta_k}$ at step $k$
correspond to projections of $\hat{O}^{L}_{\beta_k}$, $\hat{O}^{R}_{\beta_k}$ into the left- and right- renormalized bases,
\begin{align}
[\mathbf{O}^{L}_{\beta_k}]_{\alpha_k \alpha_k'} &= \langle l_{\alpha_k} | \hat{O}^{L}_{\beta_k} | l_{\alpha_k'} \rangle  \notag \\
[\mathbf{O}^{R}_{\beta_k}]_{\alpha_k \alpha_k'} &= \langle r_{\alpha_k} | \hat{O}^{R}_{\beta_k} | r_{\alpha_k'} \rangle  \label{eq:renormop}
\end{align}
These renormalized left- and right-block operators are, of course, the main computational intermediates in a pure renormalized operator-based
DMRG implementation, and they play the same role in an MPO-based implementation.
The relationship between the left-right decomposition of an operator and the renormalized left-block and right-block operators
is shown in graphical form in Fig.~\ref{fig:mpolr}.
We return to their role in efficient computation in section \ref{sec:efficient}.

The left and right operators at a given partition are
explicitly related to the left and right operators at the neighbouring partition. For example,
for the left operators, we have
%% \begin{align}
%% \hat{O}^L_{\beta_{k+1}} = \sum_{\alpha_k z_{k+1}} A^{z_{k+1}}_{\beta_k \beta_{k+1}} \hat{O}^L_{\beta_k} \hat{z}_{k+1}
%% \end{align}
%% This relationship is written more conveniently in terms of the operator valued matrices $\hat{W}[k]$. In particular, we find
%% \begin{align}
%% \hat{O}^L_{\beta_{k}} = \sum_{\beta_{k-1}} \hat{O}^L_{\beta_{k-1}} \hat{W}[k]_{\beta_{k-1} \beta_{k}} \label{eq:leftrenorm}
%% \end{align}
%% {\color{blue}
\begin{align}
\hat{O}^L_{\beta_{k}} = \sum_{\beta_{k-1}} \hat{O}^L_{\beta_{k-1}} \hat{W}_{\beta_{k-1} \beta_{k}}[k] \label{eq:leftrenorm}
\end{align}
%}
where we can interpret the above as a vector matrix product of the operator valued row-vector
$\hat{O}^L$ with the operator valued matrix $\hat{W}[k]$.
%%  may be seen to be a matrix vector product of the form
%% \begin{align}
%% \left( \hat{O}^L_{\beta_{k+1}} \right) = \left(\hat{O}^L_{\beta_k}\right) \left( \hat{W}[k+1] \right) \label{eq:leftrenorm}
%% \end{align}
%% where $( \hat{O}^L_{\beta_{k+1}} )$ is a row vector.
Analogously for the right operators, we have
%% \begin{align}
%%  \hat{O}^R_{\beta_{k-1}} = \sum_{\beta_k}
%%  \hat{W}[k]_{\beta_{k-1} \beta_{k}}   \hat{O}^R_{\beta_{k}} \label{eq:rightrenorm}
%% \end{align}
%% {\color{blue}
\begin{align}
 \hat{O}^R_{\beta_{k-1}} = \sum_{\beta_k}
 \hat{W}_{\beta_{k-1} \beta_{k}}[k] \hat{O}^R_{\beta_{k}} \label{eq:rightrenorm}
\end{align}
%}
which can be seen as a matrix vector product.
Eqs.~(\ref{eq:leftrenorm}) and (\ref{eq:rightrenorm})
explicitly define the recursion rules that relate the operators for one block of sites, e.g. $1 \cdots k-1$, to a neighbouring block
of sites, e.g. $1 \cdots k$. This process of recursively constructing the left- and right-operators at successive blocks
is isomorphic to the process of blocking as one proceeds through the sites in a DMRG sweep,
the only distinction being that the operators $\hat{O}^R_{\beta_k}$ in Eq.~(\ref{eq:rightrenorm}) are replaced by their matrix
representations $\mathbf{O}^R_{\beta_k}$. We thus refer to  the rules as {\it blocking rules}. As we explain in
Section \ref{sec:efficient},  to efficiently compute expectation values we
should in fact use the renormalized operators (i.e. operator matrix representations) as in the DMRG sweep, rather than the bare operators themselves, during the blocking process.

It is often convenient for the purposes of interpretation to write the left-right decomposition of $\hat{O}$ in Eq.~(\ref{eq:Hpartition}) a slightly different form,
\begin{align}
\hat{O} = \hat{O}^{L_k} \otimes \hat{1}^{R_k} + \hat{1}^{L_k} \otimes \hat{O}^{R_k}  + \sum_{\beta_k} \hat{o}^{L_k}_{\beta_k} \hat{o}^{R_k}_{\beta_k}
\end{align}
We have introduced 3 kinds of left and right operator terms: the identity operator ($\hat{1}^{L_k}$ or $\hat{1}^{R_k}$),
the operator $\hat{O}$ restricted to act on the left or right block of sites ($\hat{O}^{L_k}$ or $\hat{O}^{R_k}$),
and terms which express interactions between the left and right sites at
partition $k$ ($\hat{o}^{L_k}_{\beta_k} $, $\hat{o}^{R_k}_{\beta_k}$ respectively. Since there are 3 kinds
of terms, then the matrices and vectors appearing in the blocking rules Eq.~(\ref{eq:leftrenorm}), ~(\ref{eq:rightrenorm})
now have a $(3 \times 3)$ and $(3 \times 1)$ (or $(1 \times 3)$) block structure, for example Eq.~(\ref{eq:rightrenorm}) becomes in expanded form
%% In specifying the renormalization rules, it is often conventional to write the left-right decomposition in a slightly expanded form.
%% We write the operator as
%% where $\hat{O}_L$ denotes the part of $\hat{O}$ that acts only within the left block of sites $1 \cdots k$, and $\hat{O}_R$
%% denotes the part of $\hat{O}$ that acts only within the right block of sites.
%% Since the set of right operators (and similarly the left operators) are now separated into 3 kinds of operators,
%% namely $\hat{O}_{R_k}, \hat{o}_{R, \beta_k}, \hat{1}$, the renormalization
%% rule Eq.~(\ref{eq:rightrenorm} becomes expanded to
%% \begin{align}
%% \left(\begin{array}{c} \hat{O}^{R_{k-1}} \\ o^{R_{k-1}}_{\beta_{k-1}} \\ \hat{{1}}^{R_{k-1}} \end{array}\right) =
%% \left(\begin{array}{ccc}
%% \hat{{1}} & \hat{C} & \hat{O}^k \\
%% 0 & \hat{A} & \hat{B} \\
%% 0 & 0 & \hat{1}
%% \end{array}
%% \right)
%% \left(\begin{array}{c}
%% \hat{O}^{R_k} \\
%% \hat{o}^{R_k}_{ \beta_{k}} \\
%% \hat{{1}}^{R_k}
%% \end{array}
%% \right)
%% \label{eq:renormrule}
%% \end{align}
%{\color{blue}
%% To be consistent with Eq. \eqref{eq:prenormrule}, the convention $R_k=C_kR_{k+1}$ results in the following form
\begin{align}
\left(\begin{array}{c} \hat{O}^{R_{k}} \\ o^{R_{k}}_{\beta_{k}} \\ \hat{{1}}^{R_{k}} \end{array}\right) =
\left(\begin{array}{ccc}
\hat{{1}}^k & \hat{C}^k & \hat{O}^k \\
0 & \hat{A}^k & \hat{B}^k \\
0 & 0 & \hat{1}^k
\end{array}
\right)
\left(\begin{array}{c}
\hat{O}^{R_{k+1}} \\
\hat{o}^{R_{k+1}}_{ \beta_{k+1}} \\
\hat{{1}}^{R_{k+1}}
\end{array}
\right)
\label{eq:renormrule}
\end{align}
%}
where the superscript on $\hat{O}^k$ denotes the operator acts on site $k$.

From the above, we see
that building the left-right operator decompositions through the blocking rules in a DMRG sweep
is isomorphic to the operations required to construct the explicit MPO; the only difference being that
explicit operators are replaced
by operator matrices, which is necessary in the efficient computation of expectation values.

%% exactly the same computation as the operator blocking rules (recursions) appearing in the DMRG sweep algorithm

\subsection{ MPO representation of quantum chemistry Hamiltonians}

\label{sec:mpoqc}

Based on the efficient left-right decomposition
and blocking rules for the ab initio Hamiltonian in the standard DMRG algorithm, and the isomorphism
to the elements of the MPO tensors $\hat{W}[k]$ established above, we can now easily
identify the efficient MPO representation of the quantum chemistry Hamiltonian.

The ab initio Hamiltonian is written as
\begin{align}
\hat{H} = \sum_{pq} t_{pq} a^\dag_p a_q + \frac{1}{2}\sum_{pqrs} v_{pqrs} a^\dag_p a^\dag_q a_r a_s \label{eq:longrangeh}
\end{align}
where spin labels have been suppressed and $v_{pqrs}=\langle pq|sr\rangle=v_{qpsr}$. The summation over the indices is not restricted, thus for a system with $K$ sites, the indices range from $1 \cdots K$.

To obtain the MPO representation, we first identify
the left-right decomposition of the Hamiltonian, namely
\begin{align}
\hat{H} = \hat{H}^{L_k} \otimes \hat{{1}}^{R_k} + \hat{{1}}^{L_k} \otimes \hat{H}^{R_k} +
\sum_{\alpha_k} \hat{h}^{L_k}_{\alpha_k} \hat{h}^{R_k}_{\alpha_k}
\label{eq:lrdecomp}
\end{align}
where the left and right Hamiltonians, are explicitly
\begin{align}
\hat{H}^{L_k} &= \sum_{pq \in L_k} t_{pq} a^\dag_p a_q + \frac{1}{2}\sum_{pqrs \in L_k} v_{pqrs} a^\dag_p a^\dag_q a_r a_s
\label{eq:lefth}\\
\hat{H}^{R_k} &= \sum_{pq \in R_k} t_{pq} a^\dag_p a_q + \frac{1}{2}\sum_{pqrs \in R_k} v_{pqrs} a^\dag_p a^\dag_q a_r a_s \label{eq:righth}
\end{align}
where $L_k$ indicates the domain of indices $1 \cdots k$ (the left block of sites), and $R_k$ the
domain of indices $k+1 \cdots K$.

The operators $\hat{h}^{L_k}_{\alpha_k}$ and $\hat{h}^{R_k}_{\alpha_k}$ describe
the interactions between the left and right blocks of sites. Although these operators are not uniquely defined (only
 $\sum_{\alpha_k} \hat{h}^{L_k}_{\alpha_k} \hat{h}^{R_k}_{\alpha_k}$ need remain invariant)
the standard ab initio DMRG left-right decomposition of the quantum chemistry Hamiltonian provides an
efficient and convenient set. In this choice,
certain of the operators are associated with electronic integrals (the complementary operators) while
other operators are not (the normal operators).
Using the notation of Ref.~\cite{chan2002highly,chan2004algorithm} (see the Appendix of the above references) we can write down a normal/complementary operator decomposition
of the Hamiltonian as
\begin{align}
\hat{H} &= \hat{H}^{L_k} \otimes \hat{{1}}^{R_k} +  \hat{{1}}^{L_k} \otimes \hat{H}^{R_k}  \notag\\
&+  \frac{1}{2}\left(
\sum_{p \in L_k} a^\dag_{p} \hat{S}^{R_k}_p + h.c. +
\sum_{p \in R_k} a^\dag_p \hat{S}^{L_k}_p  +  h.c.\right) \notag \\
&+ \frac{1}{2}\left(\sum_{pq \in L_k} \hat{A}^{L_k}_{pq} \hat{P}^{R_k }_{pq}+ h.c.\right) \notag\\
&  -\frac{1}{2}\left(\sum_{pq \in L_k} \hat{B}_{pq}^{L_k}\hat{Q}^{R_k}_{pq}+h.c.\right)\label{eq:Hdecomp}
\end{align}
where the various operators are defined as (see also Ref.~\cite{chan2002highly,chan2004algorithm})
\begin{align}
\hat{S}^{L_k/R_k}_{p} &=  \sum_{q \in L_k/R_k} t_{pq} a_q + \sum_{qrs \in L_k/R_k} w_{pqrs} a^\dag_q a_r a_s \label{eq:Sm}\\
\hat{A}_{pq} &= a_p^\dagger a_q^\dagger  \\
\hat{B}_{pq} &= a^\dag_p a_q \\
\hat{P}_{pq}^{R_k} &= \sum_{rs \in R_k} v_{pqrs} a_r a_s \\
\hat{Q}_{pq}^{R_k} &= \sum_{rs \in R_k} \frac{1}{2}x_{prqs} a^\dag_r a_s
=\sum_{rs \in R_k} w_{prqs} a^\dag_r a_s\label{eq:norm_complementary}
\end{align}
with $w_{pqrs} = v_{pqrs} - v_{qprs}=v_{pqrs}-v_{pqsr}$, $x_{pqrs} = v_{pqrs} - v_{qprs} - v_{pqsr} + v_{qpsr}=2w_{pqrs}$.
%}
In the above, the two index complementary operators are chosen to be defined on the right block of sites only.
For efficiency, it is possible to use other decompositions where sets of complementary operators are defined on both the left and right blocks.
For example, the number of terms in the summation over normal/complementary two index operators will increase during a DMRG sweep
as the partition site $k$ is moved from $1 \cdots K$, and the size of the $L_k$ block increases. Thus, for $k > K/2$, efficient DMRG sweep implementations switch to a representation where the two index complementary operators are chosen to be defined on the left block of sites.
In addition, fermionic symmetries (such as $B_{pq} = -B_{qp}^\dag$ for $p>q$) are used.

From the double summation over $pq$, it is clear that the number of terms appearing in Eq. \eqref{eq:Hdecomp}
is $O(K^2)$, thus the total bond-dimension of the MPO representation of the Hamiltonian is also $O(K^2)$. The prefactor in the $O(K^2)$
bond-dimension depends on the particular choice of splitting between normal and complementary operators,
and how the integrals are distributed. Several cases are worked
out explicitly in the Appendix. For example, Fig.~\ref{fig:distribution} shows explicitly that the
bond-dimension is minimized by using the switch between left and right
complementary operators at the middle site $k=K/2$, as discussed above.

As we have explained, the $\hat{W}[k]$ matrix of the MPO encodes the blocking rule that takes
the left/right operators at one partitioning to a neighbouring partitioning.
For the choice of normal/complementary operators in Eq.~(\ref{eq:norm_complementary}), the blocking
rules can be found in the original DMRG quantum chemistry algorithm descriptions, see e.g.
Eqs. (A1)-(A10) in the Appendix of Ref.~\cite{chan2004algorithm}, and from these rules
we can read off the $\hat{A}, \hat{B}, \hat{C}$ in Eq.~(\ref{eq:renormrule}).
%%  $k$ to partition $k+1$. These renormalization rules can be found in
%% the descriptions of how renormalized operators are built in the original DMRG quantum chemistry algorithms, see, for example
%% Corresponding to the notation in Eq.~(\ref{eq:renormrule}), we can write the operator renormalization process as
%% \begin{align}
%% \left(\begin{array}{c} \hat{H}_{R_{k-1}} \\ h_{R_{k-1}, \alpha_{k-1}} \\ \hat{\mathbf{1}}_{R_{k-1}} \end{array}\right) =
%% \left(\begin{array}{ccc}
%% \hat{\mathbf{1}} & \hat{C} & \hat{H}_k \\
%% 0 & \hat{A} & \hat{B} \\
%% 0 & 0 & \mathbf{1}
%% \end{array}
%% \right)
%% \left(\begin{array}{c}
%% \hat{H}_{R_k} \\
%% \hat{h}_{R_k, \alpha_{k}} \\
%% \hat{\mathbf{1}}_{R_k}
%% \end{array}
%% \right)
%% \label{eq:Hrenormrule}
%% \end{align}
%% where the set of interaction operators $h_{R_{k-1}}$ is given by the set $a_i, S_i, P_{ij}, Q_{ij}$ and
%% their complex conjugates. (Choosing to renormalize from the left, we would obtain a different set $a_i, S_i, A_{ij}, B_{ij}$ and
%% their complex conjugates). We can read off the matrix elements of the MPO matrix $\hat{W}_k$ from these renormalization rules.
For example, the rule to construct the operator $\hat{P}_{pq}^{R_{k}}$ from the operators in the
previous partition is given in Eq. (A7) in Ref.~\cite{chan2004algorithm},
%% \begin{align}
%% \hat{P}^{R_{k}}_{pq} = \hat{1}^{k} \otimes \hat{P}_{pq}^{R_{k+1}} + \hat{P}_{pq}^{k} \otimes \hat{1}^{R_{k+1}} + \sum_{s \in R_{k+1}} v_{pqks} \ a_k \otimes a_s \label{eq:prenormrule}
%% \end{align}
%% {\color{blue}
\begin{align}
\hat{P}^{R_{k}}_{pq} = \hat{1}^{k} \otimes \hat{P}_{pq}^{R_{k+1}} + \hat{P}_{pq}^{k} \otimes \hat{1}^{R_{k+1}} + \sum_{s \in R_{k+1}}w_{pqks}a_k \otimes a_s \label{eq:prenormrule}
\end{align}
%}
where we have used the fact that the additional site relating $R_k$ and $R_{k+1}$ has orbital index $k$, and $\hat{1}^k$, $\hat{P}_{pq}^k$, $a_k$ denote the corresponding operators defined on site $k$.
%% \hat{S}^{R_k}_i = \hat{1}^{k} \otimes \hat{S}_{i}^{R_{k-1}} + \hat{S}^k_{i} \otimes \hat{1}^{R_k}
%% + 2 a_k^\dag \otimes P^{R_k}_{ik} + a_k \otimes Q_{ik}^{R_k}
%% \end{align}
The blocking rule (\ref{eq:prenormrule}) corresponds to a matrix vector product in Eq.~(\ref{eq:renormrule}),
\begin{align}
\left(\begin{array}{c}
\vdots \\
\vdots \\
\hat{P}_{pq}^{R_k} \\
\vdots
\end{array}\right)
=
\left(\begin{array}{cccc}
\vdots & \vdots & \vdots & \vdots  \\
\vdots & \vdots & \vdots & \vdots  \\
\vdots & \hat{A}_0       & \hat{A}_1 &  \hat{B} \\
\vdots & \vdots & \vdots & \vdots
\end{array}
\right)
\left(\begin{array}{c}
\vdots \\
a_n \\
\hat{P}_{pq}^{R_{k+1}} \\
1
\end{array}\right) \label{eq:pwform}
\end{align}
with the correspondence $\hat{A} = ( \hat{A}_0 , \hat{A}_1 )$, $\hat{o}^{R_{k+1}}_{\alpha_{k+1}} = (a_n, \hat{P}_{pq}^{R_{k+1}})$.
$\hat{A}_0$ has elements $[\hat{A}_0]_{pq,s} = v_{pqks} a_k$,
$\hat{A}_1$ is an identity matrix, and $\hat{B}$ is a diagonal
matrix with diagonal entries $\hat{P}_{pq}^k$.
%% \textcolor{red}{This analysis is not quite right, because
%% there is a different number of $P_{pq}^{R_k}$ and $P_{pq}^{R_{k+1}}$. So it may be necessary to add
%% some indices to $A_0$ and other matrices e.g. $[A_0]_{pq,p'q'}$.
%% The same should then be done to
%% modify the general relations in Eq.~(\ref{eq:renormrule}).}
%% {\color{blue}
%% This equation is correct for $p,q\in L_{k}$.
%% Suppose $L_{k+1}=L_kC_{k+1}$, the additional term $P_{pq}^{R_{k+1}}$ for $p\in L_k$ and $q\in C_{k+1}$ will appear in the recursion for $S^{R_k}$., while the term $P_{pq}^{R_{k+1}}$ for $p,q\in C_{k+1}$ will appear in the recursion for $H^{R_k}$.
%% These three parts merge into the block $P_{pq}^{R_{k+1}}$ where now $p,q\in L_{k+1}$.
%% }
The blocking rule of the operators used in the original DMRG algorithm thus explicitly carries
out the matrix-vector multiplication of $\hat{W}[k]$ in the MPO in an element-wise notation.

%{\bf {\color{blue}
Finally, we note that the approach used by  Keller {\it et al} in Ref. \cite{keller2015efficient},
the so-called \emph{fork-fork-merge} or \emph{fork-merge-merge} operations for reusing
common intermediate operators, is completely
equivalent to using $\hat{P}$ and $\hat{Q}$ complementary operators in the \emph{right}
or \emph{left} block, respectively. Specifically, in Figures 2 and 3 of Ref. \cite{keller2015efficient},
two-electron integrals with two indices on the left,
one index on site $k$, and one index on the right, are all collected into the MPO matrix $\hat{W}[k]$, similarly to
Eq. \eqref{eq:pwform}.
%}}

\subsection{Efficient implementation of expectation values}

\label{sec:efficient}

We have so far established the correspondence between the language of MPO/MPS and
the renormalized states and operators used in a pure renormalized operator-based DMRG implementation. %% , including
%% the efficient representation of the
%% renormalized bases and wavefunctions,
%% and the efficient MPO Hamiltonian and the left-right representation
%% and blocking rules, as appear along a DMRG sweep in the ``sweep-centric'' language of Sec.~\ref{sec:sweep-centric}.
We now discuss how to efficiently compute expectation values (such as the energy) in an MPO-based DMRG implementation.
%% where the MPO matrices $\hat{W}$
%% explicitly appear.
Two questions arise:
how to use the structure and sparsity of the MPO tensors, and
the order in which to perform contractions between the MPO and MPS.
In fact,  both aspects are addressed by the original DMRG sweep algorithm,
by using element-wise blocking operations, separate blocking and decimation steps, and by building the renormalized operators
mentioned in Sec.~\ref{sec:mpo} as computational intermediates. We now discuss how these individual
components arise in an MPO-based expectation value computation.
%% We now discuss these various issues of efficency, and how the renormalized operators appear
%% in an MPO-centric implementation.

To see why MPO tensor sparsity is important, we first observe that the cost of the quantum chemistry DMRG sweep algorithm
to compute (or minimize) the energy is $O(K^4)$, which is what one would
expect given that the Hamiltonian contains $O(K^4)$ fermionic terms.
However, if we try to reconstruct the Hamiltonian operator $\hat{H}$ from its $\hat{W}[k]$ MPO product (\ref{eq:wproduct})
using dense matrix algebra, we formally require $O(K^5)$ operations,
as first noted in Ref.~\cite{Naoki}. This is because multiplying out Eq.~(\ref{eq:wproduct})
requires $O(K)$ matrix vector products between the $O(K^2) \times O(K^2)$ dimension $\hat{W}[k]$ matrices, and the $K^2 \times 1$ boundary
$\hat{W}$ vectors ($\hat{W}[1]$ or $\hat{W}[K]$). %% As noted in Ref.~\cite{naoki} the $O(K^5)$ scaling is clearly too high given
%% that $\hat{H}$ only contains $O(K^4)$ fermionic operators. Furthermore the original DMRG algorithms
%% are formulated with a scaling of $O(K^4)$.

The reason for the incorrect scaling of $O(K^5)$
%% when using an MPO in
%% conjunction with dense matrix operations,
is that the above argument neglects the fact that $\hat{W}$ contains many zero elements~\cite{Naoki}. To see this explicitly, we consider Eq.~(\ref{eq:pwform}) that
determines the update of the $\hat{P}_{pq}$ elements of $\hat{W}$. Here, the multiplication of the $\hat{A}_1$ and $\hat{B}$ matrices
into the column vector formally takes $O(K^4)$ cost, which repeated over the $O(K)$ $\hat{W}[k]$ matrices leads to the incorrect $O(K^5)$ scaling.
However, the $\hat{A}_1$ and $\hat{B}$ matrices are in fact {\it diagonal} matrices, and can be multiplied with $O(K^3)$ cost over all the $\hat{W}$ matrices. The main cost in Eq.~(\ref{eq:pwform})
arises then from the multiplication of the $\hat{A}_0$ matrix (of dimension $O(K^2 \times K)$) into the $O(K)$ $a_s$ operators. This
is of $O(K^3)$ cost for a single multiplication, and of $O(K^4)$ cost over all the $\hat{W}$ matrices, leading to the correct scaling.

Note also, that there are many symmetries between different elements of  $\hat{W}$.
For example, although both  $a_p^\dag$, $a_p$ appear
as elements of $\hat{W}$, they are related by Hermitian conjugation (and
similarly for elements such as $a_p^\dag a_q^\dag$, $a_p a_q$ and
the $p>q$ and $p<q$ components of  $a_p^\dag a_q$). These elements would be manipulated
and multiplied separately in a simple implementation of an MPO. However,
such symmetries and relationships can further be used to reduce the prefactor of the
reconstruction of $\hat{H}$ as well as the storage of the MPO's.

The  explicit expressions for blocking in the original DMRG algorithm
are {\it element-wise} expressions
of the multiplications of $\hat{W}$ which already incorporate both the sparsity and symmetry between the elements,
and thus lead to efficient operations with  $\hat{H}$.
To efficiently carry out blocking in an
MPO-based implementation, the  same element-wise strategy should be used.
This can be achieved in practice by storing additional meta-information on the non-zero matrix elements and how to multiply them,
as is done, for example in \textsc{MPSXX}\cite{Naoki} and \textsc{QC-Maquis}\cite{keller2015efficient}.
%% This is the approach taken in \textsc{MPSXX} and \textsc{QC-Maquis}.

\begin{figure}
\centering
\begin{tabular}{c}
{\resizebox{0.4\textwidth}{!}{\includegraphics{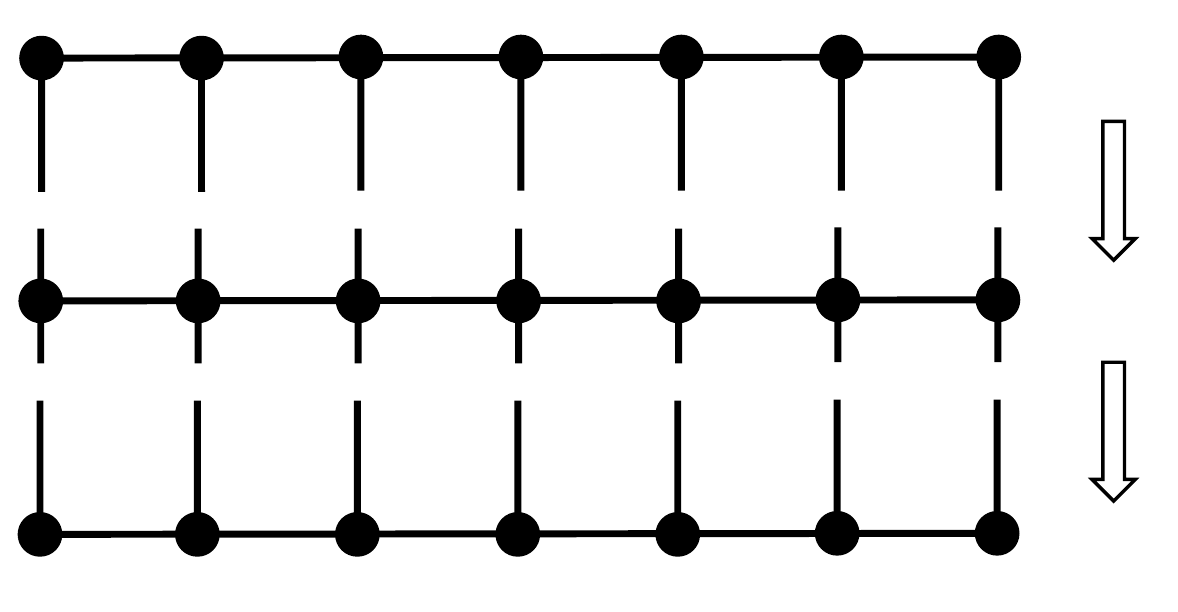}}}\\
(i) \\
{\resizebox{0.4\textwidth}{!}{\includegraphics{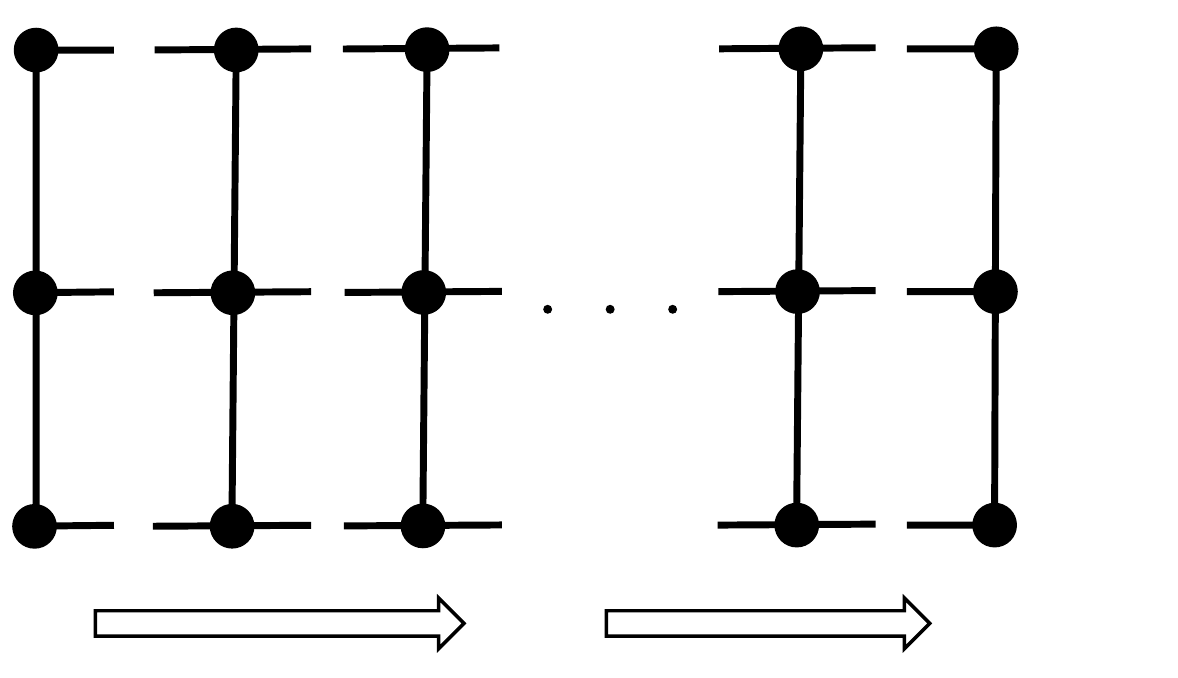}}}\\
(ii) \\
\end{tabular}
\caption{\label{fig:orderMPOMPS} (i) Incorrect contraction order for an expectation value. (ii) Contraction order,
leading to renormalized operators, for an expectation value. The individual tensors appearing correspond
to the transfer operators in Eq.~(\ref{eq:transferop}).}
\end{figure}

We now consider contracting the Hamiltonian MPO with the bra and ket MPS to
compute the energy, $E = \langle \Psi | \hat{H} | \Psi\rangle$. As $\ket{\Psi}$ is an MPS
and $\hat{H}$ is an MPO, we could imagine first computing $\hat{H}\ket{\Psi}$ (obtaining a new MPS) before contracting
with the bra. However, it is easy to see that this leads again to the wrong scaling, because the intermediate $\hat{H}\ket{\Psi}$
is now an MPS of very large bond dimension $O(M K^2)$, requiring a very large amount of storage. Instead,
one should contract the tensors of the  MPS bra and ket with the tensors
of the MPO, site by site from $1 \cdots K$. {\it This corresponds exactly to the recursive construction
of the renormalized operators through blocking and decimation along a sweep} (see Fig.~\ref{fig:orderMPOMPS}).

To illustrate how this recursive construction arises naturally, we first define
a partial expectation value over a site $k$ as the matrix $E[k]$ (sometimes called the transfer operator),
\begin{align}
E[k]_{\gamma_{k-1}, \gamma_k}
 = \sum_{n_k n_k'} \langle n_k | A^{n_k}_{\alpha_{k-1} \alpha_k} \hat{W}_{\beta_{k-1}, \beta_k}[k] A^{n_k'}_{\alpha'_{k-1} \alpha'_k} |n_k'\rangle \label{eq:transferop}
\end{align}
where the compound index
$(\gamma_{k-1}, \gamma_k) \equiv (\alpha_{k-1}\alpha'_{k-1}\beta_{k-1}, \alpha_k \alpha'_k \beta_{k})$.
The energy expectation value can be written as
\begin{align}
E = E[1] E[2] \cdots E[K]\label{eq:energy_expect}
\end{align}
where $E[k]$ is an $O(M^2 K^2) \times O(M^2 K^2)$ matrix, and $E[1]$ and $E[K]$ are $1 \times O(M^2 K^2)$ and
$O(M^2 K^2) \times 1$ vectors. Graphically, we  illustrate the energy expectation
computation by Fig.~\ref{fig:orderMPOMPS}.

When carrying out the energy computation, one naturally multiplies the matrices together
from the left or from the right. Multiplying up to site $k$ from the left or the right
respectively, defines
%% yields the
%% Each $E[k]$ represents the partial trace of a set of operators in the local site basis.
%% For a given partitioning into a left block $1 \cdots k$ and a right block $k+1 \cdots K$, we can
the left and
right operator matrix representations, namely
\begin{align}
[\mathbf{O}^{L}_{\beta_k}]_{\alpha_k\alpha_k'} &= (E[1]E[2] \cdots E[k])_{\gamma_k} \notag\\
[\mathbf{O}^{R}_{\beta_k}]_{\alpha_k\alpha_k'} &= (E[k+1]E[k+2] \cdots E[K])_{\gamma_k}
\end{align}
where $\alpha_k, \alpha_k'$ denote the matrix indices of the renormalized operator matrices, and the different renormalized
operators are indexed by $\beta_k$ (c.f. Eq.~(\ref{eq:renormop})).
$\mathbf{O}^{L}_{\beta_k}$ and
$\mathbf{O}^{R}_{\beta_k}$ are of course the same left- and right-renormalized operators
that appear in the left-right decomposition Hamiltonian at site $k$, and are the standard intermediates in a DMRG sweep.

%% The total energy can then be written as
%% \begin{align}
%% E = \sum_{\beta_k} \mathrm{tr} \left[\mathbf{E}^{L_k}_{\beta_k} \otimes \mathbf{E}^{R_k}_{\beta_k}\right] \label{eq:energyE}
%% \end{align}
%% The left and right renormalized operator matrices at one partitioning are obtained from those at the adjacent partitioning
%% by a blocking and decimation step
%% \begin{align}
%% \end{align}
%% In a slightly expanded form, similarly to as in Eq.~(\ref{eq:lrdecomp}),
%% this becomes
%% \begin{align}
%% E = \mathrm{tr} \left[\mathbf{H}^{L_k} \otimes \mathbf{1}^{R_k} + \mathbf{1}^{L_k} \otimes \mathbf{H}^{R_k} + \sum_{\beta_k} \mathbf{h}^{L_k}_{\beta_k} \mathbf{h}^{R_k}_{\beta_k}\right]
%% \end{align}
%% The above expression, the left-right decomposition of the Hamiltonian operator $\hat{H}$ in terms
%% of the renormalized operator matrices, and is exactly the left-right
%% decomposition of the Hamiltonian (in the renormalized basis space) at step $k$ of a DMRG sweep.
%% identical
%% the terms are now the renormalized operators matrices (i.e.
%% the operators
%% projected into the left and right renormalized basis).
%% These projected operators are the natural intermediates
%% to assemble within a DMRG
%% sweep algorithm to compute the energy, where they
%% are typically simply referred to as renormalized operators.
%% As noted previously, the renormalized operator matrices are the central intermediates to compute
%% in an efficient implementation.

\begin{figure}
\centering
\begin{tabular}{c}
{\resizebox{0.6\textwidth}{!}{\includegraphics{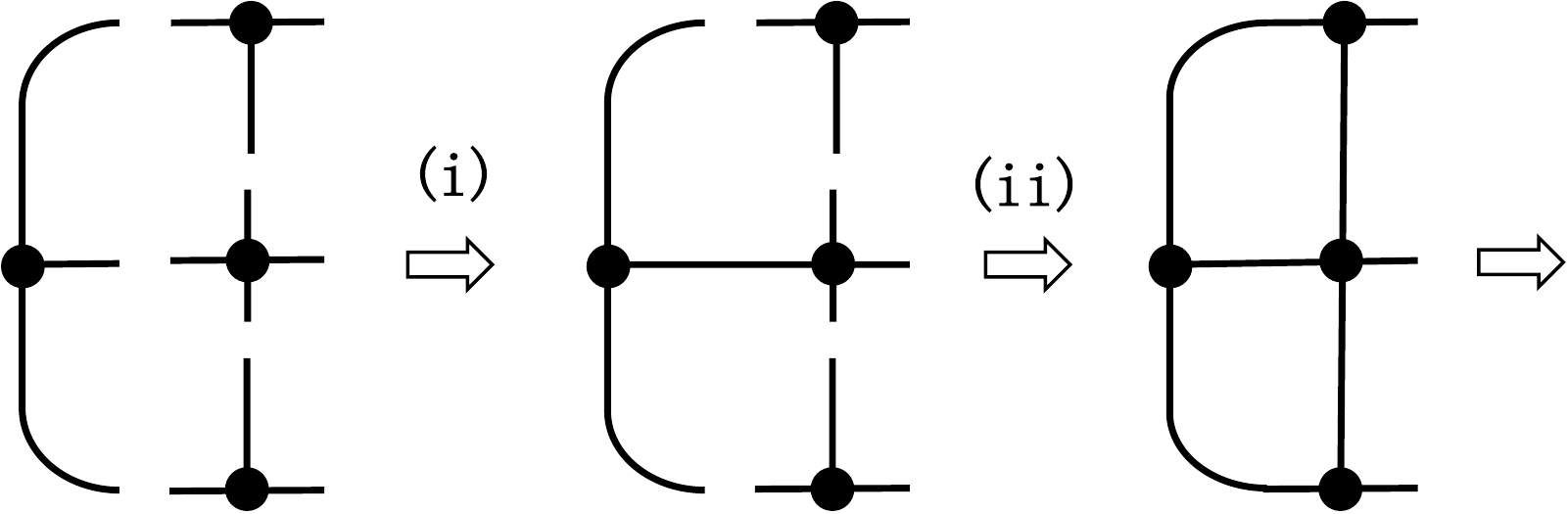}}}\\
\end{tabular}
\caption{\label{fig:econtraction} Individual steps in an expectation value contraction. (i) corresponds
to renormalized operator blocking, while (ii) corresponds to renormalized operator decimation.}
\end{figure}

What is the cost to build the renormalized operators?
%% If we recall that the operator matrices at partition $k+1$ are obtained from those
%% at the previous partition $k$.
A naive multiplication of
the $K$ $E[k]$ matrices is a multiplication
of $O(M^2 K^2) \times O(M^2 K^2)$ matrices into an $O(M^2 K^2)$ length vector. Carrying this out $O(K)$ times would appear
 to require $O(M^4 K^5)$ cost, which is higher than cost of the ab initio DMRG algorithm.
However, in a standard DMRG sweep implementation (c.f. Section \ref{sec:overview}), the renormalized operators are built in two steps: first blocking, then decimation.
This is equivalent to observing that
$E[k]$ is itself composed of a tensor contraction, and thus we can
perform multiplication of two $E[k]$ matrices in two
smaller steps (Fig.~\ref{fig:econtraction}). %% This corresponds precisely to
%% the blocking and decimation steps in the DMRG sweep,
%% as illustrated in
This reduces the cost of multiplying the $K$ $E[k]$ matrices (and building
the renormalized operators) to $O(M^3 K^3)$ + $O(M^2 K^5)$.
This is the lowest cost if we assume that the $E[k]$ matrices are dense, and
is the generic cost associated with evaluating the expectation value of an MPO with bond dimension $O(K^2)$ with an MPS of bond dimension $O(M^2)$. However, the high $O(K^5)$ scaling
is once again (as noted in Ref.~\cite{Naoki}) because we have not yet accounted for the
sparsity of the $E[k]$ matrices.
By using elementwise blocking rules (as in Eq.~(\ref{eq:prenormmatrix}), (\ref{eq:prenormdec})), we can explicitly carry out the elementwise multiplication of the $E[k]$ matrices
taking into account the appropriate sparsity, as well as the symmetries of the elements of $E[k]$.
For example, the blocking operation followed by decimation, for the $\mathbf{P}_{pq}$ element of the $\mathbf{O}^{R_k}$
corresponds to (blocking)
\begin{align}
\mathbf{P}_{pq}^{R_{k}} =
\mathbf{1}^{k} \otimes \mathbf{P}_{pq}^{R_{k+1}} + \mathbf{P}_{pq}^{k} \otimes \mathbf{1}^{R_{k+1}} + \sum_{n \in R_{k+1}} w_{pqks} \ \mathbf{a}_k \otimes \mathbf{a}_s \label{eq:prenormmatrix}
\end{align}
followed by (decimation)
\begin{align}
{[\mathbf{P}_{pq}^{R_{k}}]}_{\alpha_k, \alpha_k'} \leftarrow \sum_{n_k \alpha_k, n_k' \alpha_k'} A^{n_k}_{\alpha_k \alpha_{k+1}} {[\mathbf{P}_{pq}^{R_k}]}_{n_k \alpha_{k+1}, n_k' \alpha_{k+1}} A^{n'_k}_{\alpha_k' \alpha'_{k+1}} \label{eq:prenormdec}
\end{align}
%% Such operator matrices at each partition of the orbitals are built and stored in memory or on disk during a DMRG sweep over the orbitals in the evaluation
%% and optimization of the energy.

Incorporating elementwise blocking and decimation steps
then leads finally to the correct cost of $O(M^3 K^3)$ + $O(M^2 K^4)$ (the cost of the original DMRG quantum chemistry algorithm).
In summary, this allows an MPO-based implementation of DMRG to recover the same cost as a pure renormalized operator-based implementation,
through essentially an identical set of computations.
%% Keeping track of  sparsity and symmetry of the $E[k]$ matrices with meta-data in an
%% MPO-centric code allows the equivalent set of operations to be implemented.

%% However,  is clear that the mathematical operations to implement an {\it efficient} MPO algorithm are {\it exactly} the same as those
%% performed along a standard ab initio DMRG sweep, and {\it exactly} the same set of tensor contractions must be performed.

\section{MPO and MPS algebra in a renormalized operator-based implementation}
\label{sec:impl}

In the previous section, we focused on the relationship between the efficient computation of expectation values
 within an MPO-based DMRG implementation, and the same computation within a pure renormalized operator-based implementation.
We saw that a natural way to achieve the same scaling in an MPO-based implementation is
to map the computations  in the standard DMRG sweep to the MPO-based language.

Expectation values are the natural target of the DMRG sweep algorithm.
The algebra of matrix product operators and matrix product states extends beyond
expectation values, however, and many more general MPO-MPS operations appear in a variety
of algorithmic contexts. For example, to time-evolve an MPS with maximum bond-dimension $M$, involves repeating the following sequence of operations,
for each time-step,\cite{vidal2004efficient,daley2004time,white2004real}
\begin{enumerate}
\item $\ket{\Psi(t) [M]} \to e^{-i\epsilon \hat{H}} \ket{\Psi(t) [M]} \equiv \ket{\Psi(t+\epsilon)[M']}$ (evolution)
\item $\ket{\Psi(t+\epsilon)[M']} \to \ket{\Psi(t+\epsilon)[M]}$ (compression)
\end{enumerate}
%% where the operation $\ket{\Psi'(M')} \to \ket{\Psi(M)}$ is the operation of {\it compressing} a MPS
%% of bond-dimension $M'$ to a smaller bond dimension $M$.

An important question is whether or not this kind of
algorithm, involving a more general MPO/MPS algebra, can be supported within a pure renormalized operator-based DMRG implementation,
where only the renormalized operators appear.
The answer is that {\it any} MPO/MPS operation, whose final result is a scalar or an MPS,
can in fact be easily implemented
within a pure-sweep implementation without any major effort. Consider, for example, the time-evolution operation above. The
first step is an MPO $\times$ MPS product, which is not part of the standard DMRG sweep.
However, the combination of the two steps (including the compression) {\it is} in the form of a sweep computation,
since compression corresponds to maximizing the overlap (i.e. expectation value) $\langle \Phi[M] | e^{-i \epsilon \hat{H}} | \Psi [M]\rangle$ with respect to $\bra{\Phi}$. In fact, one can even obtain the full MPS $e^{-i\epsilon \hat{H}} \ket{\Psi [M]}$
with no compression,
by simply requiring, in the overlap maximization sweep, that the bond dimension of $\bra{\Phi}$ is kept as $M \times D$,
where $D$ is the bond dimension of the MPO $e^{-i \epsilon \hat{H}}$ (and thus no compression occurs).

To compute the action of a product of matrix product operators
on a matrix product state, one simply has to apply the above procedure multiple times. For example,
to obtain  $\braket{\Psi |\hat{O}\hat{O} |\Psi}$, we first maximize the overlap $\langle \Phi|\hat{O}|\Psi\rangle$
to determine $\bra{\Phi}$, and then compute the overlap $\braket{\Psi|\hat{O} |\Phi}$.

Only algorithms for whom the final output is an MPO itself (which is  rare in zero-temperature calculations)
require a full implementation of MPO functionality beyond renormalized operator computation. Implementing the
general MPO/MPS algebra as described above can be achieved by updating a renormalized operator-based DMRG code with a simple interface.
This is what is found, for example, in the MPO/MPS implementation
within the \textsc{Block} code, as is used in DMRG response\cite{dorando2009analytic,nakatani2014linear} and perturbation calculations\cite{sharma2014communication,sharma2015multireference,sharma2016quasi}.

\section{Improving DMRG through matrix product operators}

\subsection{Hamiltonian compression}
\label{sec:compress}

%% While the previous sections emphasized the equivalence and mapping between operations in a pure sweep implementation
%% of DMRG and efficient MPO and MPS implementations,
In this section, we focus on some of the new
ideas brought by matrix product operators to the implementation of DMRG-like algorithms.

The simplest observation is that, in the same way that it is possible to compress an MPS, it is
also possible to compress an MPO. Consequently, in all algorithms where, for example, an operator
appears, it is possible to carry out an approximate computation using a compressed version of the same operator.
In some cases, this can lead to very substantial savings. For example, for two-point interactions
that are a sum of $D$ exponentials, such as $\sum_{ij} V_{ij} n_i n_j$
where $V_{ij} = \sum_{\lambda} \exp (\lambda|i-j|)$ then the MPO can be compressed exactly to have bond dimension $D$.
This means that, for example, when carrying out a DMRG calculation in a one-dimension system using a short-ranged (e.g.
sum of exponentials) interaction, it is possible to carry out such a calculation with a cost that is linear
with the length of the system.

In general, a unique compression scheme in an MPO requires choosing a gauge convention.
A particularly simple way to arrange the compression is to start from a left-right decomposition of the Hamiltonian at each site $i$,
\begin{align}
\hat{H} = \hat{H}^{L_k} \otimes \hat{{1}}^{R_k} + \hat{{1}}^{L_k} \otimes \hat{H}^{R_k} +
\sum_{\alpha_k,\beta_k} h_{\alpha_k, \beta_k} \hat{h}^{L_k}_{\alpha_k} \hat{h}^{R_k}_{\beta_k }\label{eq:lrdecomp_compression}
\end{align}
In Eq.~(\ref{eq:lrdecomp_compression}) the operators $\hat{h}^{L_k}$ and $\hat{h}^{R_k}$ are purely fermionic operators (normal operators) and do not have any one- or two-particle integrals attached; the corresponding one- and two-particle integrals are stored in the matrix ${h}_{\alpha_k, \beta_k}$.
For example, considering only the one-particle part of the Hamiltonian, the interaction term in Eq.~(\ref{eq:lrdecomp_compression}) would become
\begin{align}
\sum_{\alpha_k,\beta_k} h_{\alpha_k, \beta_k} \hat{h}^{L_k}_{\alpha_k} \hat{h}^{R_k}_{\beta_k} \to \sum_{p \in L_k, q \in R_k} t_{pq} (a^\dag_p a_q + h.c.)
\end{align}
We can then compress the MPO by simply considering the singular value decomposition of the matrix $h_{\alpha_k, \beta_k}$, $h = U \lambda V^\dag$, defining the left and right operators as $\hat{h}^{L_k} U$ and $V^\dag \hat{h}^{R_k}$, and dropping small singular values. (Note that due to quantum number symmetries,
$h_{\alpha_k, \beta_k}$ is block diagonal, thus the singular value decomposition can be carried out on the separate blocks).

  The left-right decomposition of the Hamiltonian becomes
  \begin{align}
    \hat{H} = \hat{H}^{L_k} \otimes \hat{{1}}^{R_k} + \hat{{1}}^{L_k} \otimes \hat{H}^{R_k} +
\sum_i \lambda_i  \left(\sum_{\alpha_k} \hat{h}^{L_k}_{\alpha_k} U_{\alpha_k i}\right)  \left(\sum_{\beta_k} V^\dag_{i \beta_k} \hat{h}^{R_k}_{\beta_k }\right),\label{eq:lrafter_decomp_compression}
  \end{align} 
and the corresponding transformation of the $\hat{W}[k]$ matrices appearing in Eq.(\ref{eq:renormrule}) is
\begin{align}
\hat{O}^k &\to \hat{O}^k \\
\hat{A} &\to V^\dag_{k-1} \hat{A} V_k \\
\hat{B} &\to V^\dag_{k-1} \hat{B} \\
\hat{C} &\to \hat{C} V_k
\end{align}
Note that the left-right decomposition has the same summation structure as in the standard DMRG representation, only the
number of indices summed over is smaller, since small singular values $\lambda_i$ are dropped. Consequently, standard strategies
for parallelization in DMRG which involve parallelizing over the left-right decomposition sum (see Sec. \ref{sec:parallel}) may be used without modification with the compressed representation.

\begin{figure}
  \includegraphics[width=4in]{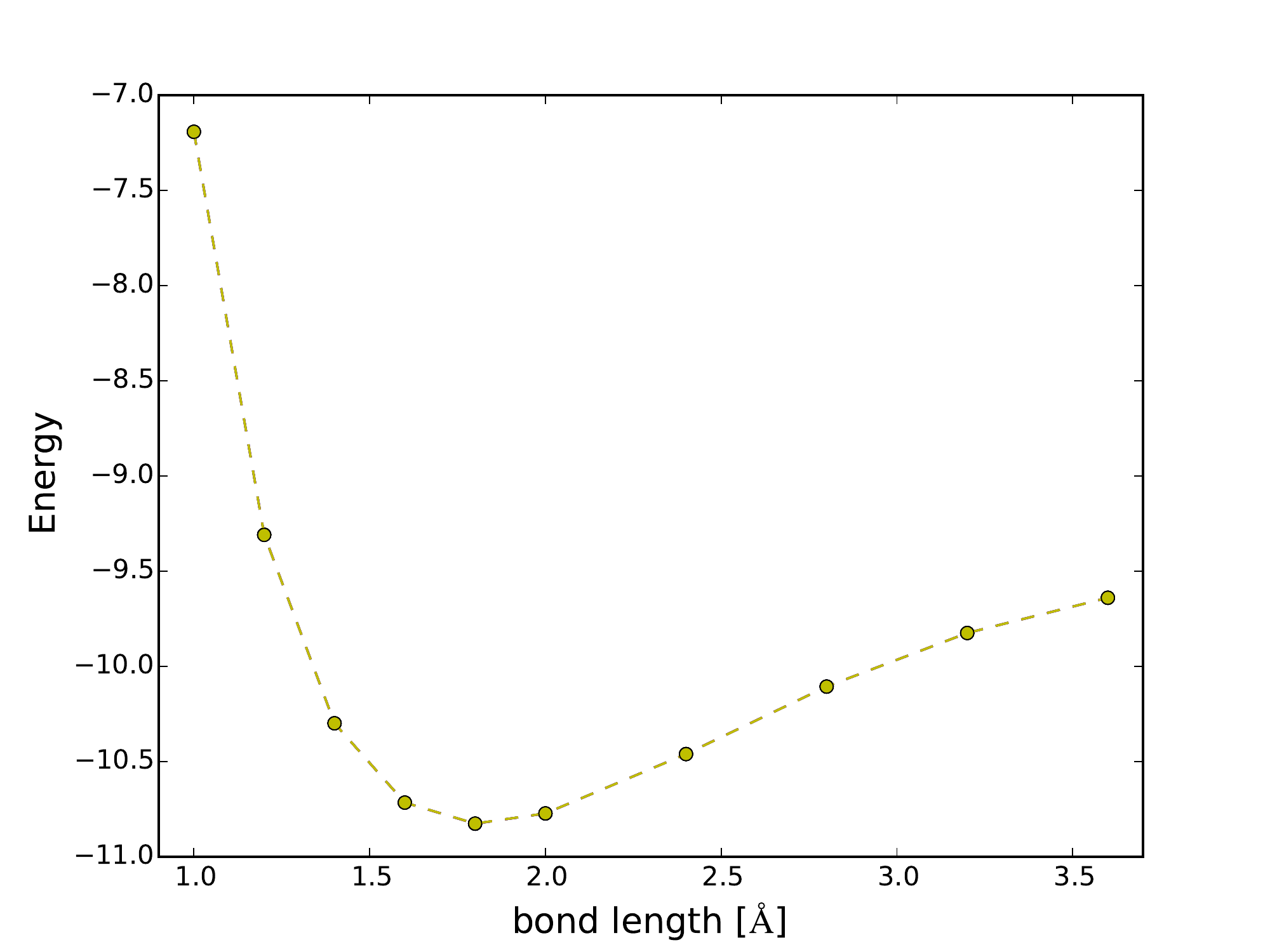}
  \caption{\label{fig:hydrogen1} Exact energy versus bond-length for the symmetric stretch of a chain of 20 hydrogen atoms.}
  \end{figure}

\begin{figure}
\centerline{  \includegraphics[width=4in]{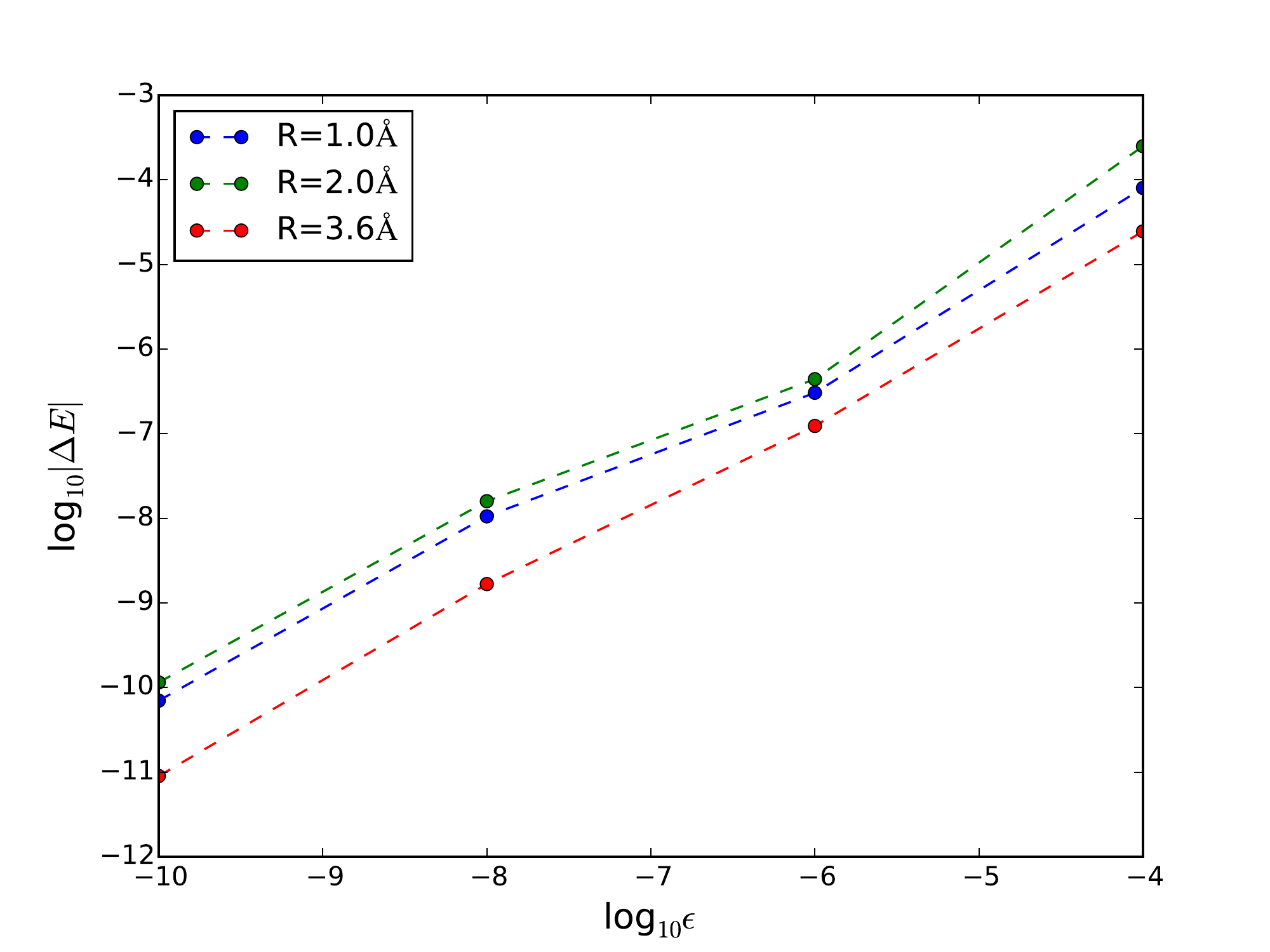}}
\caption{\label{fig:Hcompressed} Total energy error corresponding to a given singular
  value truncation threshold at bond-lengths 1.0\AA, 2.0\AA, and 3.6\AA.}
\end{figure}

\begin{figure}
\centerline{  \includegraphics[width=4in]{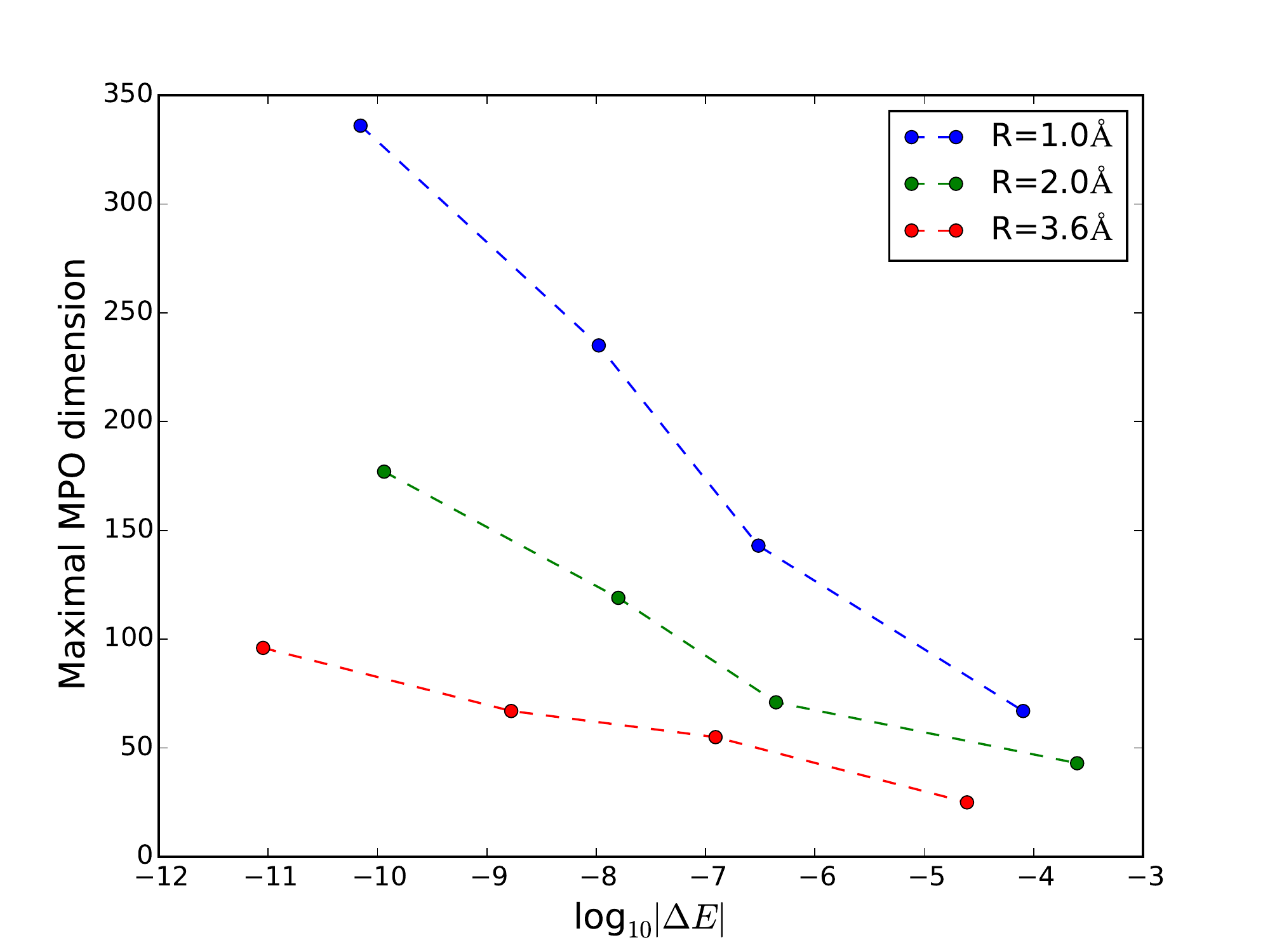}}
  \caption{\label{fig:Hcompressed2} Bond dimension corresponding to a given total energy error
at bond-lengths 1.0\AA,  2.0\AA, and 3.6\AA.}
\end{figure}

To illustrate this compression in an ab initio quantum chemistry context, we have implemented the above scheme
to compute the variational energy of
a linear chain of 20 equally spaced hydrogen atoms in the minimal STO-3G basis. Shown in Fig.~\ref{fig:hydrogen1} is the exact energy
versus bond-length curve computer using a DMRG calculation with $M=1000$. Also shown are the errors of
using an approximate compressed MPO, with the error shown versus the truncation threshold of the MPO (Fig.~\ref{fig:Hcompressed}), as well as the bond-dimension of the MPO (Fig.~\ref{fig:Hcompressed2}), for spacing $R= 1.0$\AA, 2.0\AA, 3.6\AA. We see that the
error in the energy is proportional to the truncation threshold, and exponentially decreases
with the bond-dimension of the MPO. Note that the full bond-dimension of the MPO in our choice
of gauge in this system varies between 43084 (bond-length of 1.0\AA) to 15096 (bond-length of 3.6\AA). However, to obtain an error of $10^{-6}$ $E_h$, it is sufficient to use an
MPO bond-dimension less than 200. Given that the cost of each step in the DMRG sweep
is proportional to the bond-dimension of the MPO, this is a factor of 100 in savings.

\subsection{Efficient sum of operators representation}

\label{sec:sum_over_ops}

In section~\ref{sec:efficient} we saw that a naive implementation
of DMRG using an MPO representation
with dense matrices leads to an incorrect scaling algorithm, and that the standard ab initio DMRG
algorithm corresponds to encoding the sparse matrix multiplications of the MPO to obtain
an optimal scaling.

There is, however, a different and quite simple way  to formulate an MPO representation
which, even when using naive dense matrix algebra, recovers the
the correct $O(K^4)$ scaling in a quantum chemistry algorithm. This
is achieved by abandoning a single MPO expression for
the Hamiltonian, and instead  rewriting the Hamiltonian as a sum of sub-Hamiltonians $\hat{H}_m$,
where each term is separately represented by an MPO.
Each sub-Hamiltonian $\hat{H}_m$ in Eq.~(\ref{eq:subH}) is defined as a Hamiltonian
where the integrals have a restriction on the first one- or two-electron integral index,
%% \begin{align}
%% \hat{H} &= \sum_m \hat{H}_m \\
%% \hat{H}_m &= \sum_{i} t_{im} a^\dag_i a_m + \frac{1}{2}\sum_{ijk} v_{ijkm} a^\dag_i a^\dag_j a_k a_m \label{eq:subH}
%% \end{align}
\begin{align}
\hat{H} &= \sum_m \hat{H}_m \\
\hat{H}_m &= \sum_{q} t_{mq} a^\dag_m a_q + \frac{1}{2}\sum_{qrs} v_{mqrs} a^\dag_m a^\dag_q a_r a_s \label{eq:subH}
\end{align}
The MPO representation of $\hat{H}_m$ has bond-dimension $O(K)$. We can see this by once again working with the left-right decomposition,
writing
\begin{align}
\hat{H}_m &= a^\dag_m \hat{T}_m \\
  \hat{T}_{m} &=  \sum_{q} t_{mq} a_q + \frac{1}{2}\sum_{qrs} v_{mqrs} a^\dag_q a_r a_s\\
&= \hat{T}_m^{L_k}\otimes \hat{{1}}^{R_k}+
\hat{{1}}^{L_k} \otimes \hat{T}^{R_k}_m  \notag\\
&+
\frac{1}{2}\sum_{q\in L_k}[a^\dag_q \hat{P}_{mq}^{R_k}- a_q\hat{Q}_{mq}^{R_k}]\notag\\
&
+\frac{1}{2}\sum_{q\in R_k}[\hat{P}_{mq}^{L_k} a^\dag_q- \hat{Q}_{mq}^{L_k} a_q]. \label{eq:hmrenorm}
\end{align}
$T_m$ is a sum over $O(K)$ terms and thus has bond-dimension $O(K)$.
Since $a^\dag_m$ is an MPO of bond-dimension 1, $\hat{H}_m$ (as a product
of $a^\dag_m$ and $T_m$) is also of bond-dimension $O(K)$.

Note that the above is not the only way to split $\hat{H}$ into sub-Hamiltonians.
For example, $\hat{H}_m$ could alternatively be defined as a collection of terms sharing the same rightmost operator
on the one-dimensional orbital lattice. The same scaling of the bond-dimension $O(K)$ is obtained, but
the bond dimensions to the
right of the site $m$ then become simply 1. This lowers the average bond-dimension of the MPO across
the lattice. (For details, see Appendix \ref{sec:app}.)

An immediate consequence of rewriting the Hamiltonian as a sum over
the $K$ MPO operators $\hat{H}_m$ of bond-dimension $O(K)$, is that the naive (dense matrix algebra) cost of working
with the MPO retains the {\it correct} $O(K^4)$ scaling of quantum chemistry algorithms.
For example, consider the reconstruction of $\hat{H}$ from its $\hat{W}$ matrix decomposition Eq.~(\ref{eq:wproduct}). For each $\hat{H}_m$ we
have
\begin{align}
\hat{H}_m = \hat{W}_m[1] \hat{W}_m[2] \cdots \hat{W}_m[K] \label{eq:wproduct_forhm}
\end{align}
where each $\hat{W}_m[k]$ matrix is an $O(K) \times O(K)$ matrix. Even if we manipulate
 each $\hat{W}_m[k]$ matrix as a dense matrix, the cost
of multiplying out the terms in Eq.~(\ref{eq:wproduct_forhm}) is $O(K^3)$ for each $\hat{H}_m$, and thus $O(K^4)$ cost when considering all $K$ $\hat{H}_m$ operators.
This is the correct physical scaling as contrasted with the $O(K^5)$ scaling with the naive MPO representation algorithm.
In a similar fashion, the cost to evaluate the energy expectation value in the sum of Hamiltonians representation
is $O(M^3 K^3) + O(M^2 K^4)$ i.e. the same scaling as the quantum chemistry DMRG algorithm.

The decomposition of the Hamiltonian into $\hat{H}_m$ can be seen as a way of using the inherent sparsity in the MPO representation of $\hat{H}$, to recover the correct scaling.
However,  although the correct scaling is achieved even when using dense matrix algebra in this representation,
the prefactor is significantly larger than the standard ab initio DMRG algorithm, if we do not use
additional sparsity in the $\hat{W}_m$ matrices. Consider, for example, the renormalization rule for $\hat{P}_{mq}^{R_k}$ in Eq.~(\ref{eq:hmrenorm}),
given by Eq.~(\ref{eq:pwform}).
Here, since there are only $O(K)$ $\hat{P}_{mq}^{R_k}$ operators,
 the $\hat{A}_1$ matrix is an $O(K) \times O(K)$ identity matrix. The dense multiplication of this matrix is only of $O(K^3)$ cost and
leads to a physically correct $O(K^4)$ scaling when all $\hat{W}_m$ matrices are multiplied over. However, it is clear that
without taking into account the zeros in the identity matrix, we will still perform too many operations.

\subsection{Perfect parallelism in the sum of operators formulation}

\label{sec:parallel}

Parallelization is a key component of practical DMRG calculations in quantum chemistry. There are
three principal sources of parallelism that have been so far been considered in DMRG calculations~\cite{chan2004algorithm,hager2004parallelization,kurashige2009high,rincon2010improved,stoudenmire2013real,nemes2014density}: (i) parallelism over the
left-right decomposition of the Hamiltonian~\cite{chan2004algorithm}, (ii) parallelism over ``quantum numbers'' in the DMRG
renormalized operator matrices~\cite{kurashige2009high}, and (iii) parallelism over sites in the sweep algorithm~\cite{stoudenmire2013real}.
Out of these three sources, only (i) and (ii) have been actually implemented in the context of quantum chemistry.
The sources of parallelism are largely independent and can be combined to give multiplicative
speed up in a parallel DMRG implementation and utilized in modern implementations.
%This is, for example, utilized in the implementations in the \textsc{Block} and \textsc{CheMPS2} codes

For typical systems, the largest source of parallelism is source (i), i.e. the left-right decomposition. In this case
parallelism is expressed over the loop over the normal and complementary operators appearing
in Eq.~(\ref{eq:lrdecomp}), i.e.
\begin{align}
\sum_{\alpha_k} \hat{h}^{L_k}_{\alpha_k} \hat{h}^{R_k}_{\alpha_k} \to \sum_{\text{proc}} \sum_{\alpha_k \in {\text{proc}}} \hat{h}^{L_k}_{\alpha_k} \hat{h}^{R_k}_{\alpha_k}
\end{align}
where different $\hat{h}^{L_k}_{\alpha_k}, \hat{h}^{R_k}_{\alpha_k}$ are stored and manipulated on different cores/processors.
This is an efficient source of parallelism because there are $O(K^2)$ terms in the sum, thus even for
a modest number of orbitals (e.g. $K=50$) it is possible to distribute operations over a large number cores. However, there still
remain important communication steps, as the renormalization rules (see e.g. Eq.~(\ref{eq:prenormrule}) and Eqs. (A1)-(A10) in Ref.~\cite{chan2004algorithm}) to build the different normal and complementary operators defined in Eq.~(\ref{eq:norm_complementary}), involve several different kinds of normal and complementary operators.
For example, in Eq.~(\ref{eq:prenormrule}), to
construct $P_{ij}^{R_k}$ we need not only $P_{ij}^{R_{k+1}}$, but also
the identity operator (which is trivial), as well as $a_k$ and $a_n$ operator matrices. If the $a_n$ operators are not stored
on the processor that also stores $P_{ij}^{R_k}$, then it must be communicated.

An important advantage of the sum over operators formulation in section \ref{sec:sum_over_ops} is that each sub-Hamiltonian term $\hat{H}_m$ can be manipulated completely independently of any other term. Thus the construction of $\hat{H}_m$, the associated renormalized operators, and renormalized operator
matrices for each $\hat{H}_m$, can be carried out independently of every other $\hat{H}_m$. This leads to a different organization of the
parallelization of the DMRG algorithm, which is highly scalable up to $O(K)$ processes. Compared to the
  most common parallelization strategy (i) there is no need to communicate renormalized operators between
  processes; only the renormalized wavefunction need be communicated, leading to a substantial decrease in
  communication cost, while the leading order memory and computation requirements remain unaffected. Further, for
  each sub-Hamiltonian, one may further parallelize its operations through  strategies (i), (ii), and (iii) above.
 The investigation of the scalability of the promising sum over operator parallelization is thus of interest in future work.

\section{Conclusions}
\label{sec:conclusions}

In this work we had three goals, namely to (i) explain
how to efficiently implement the ab initio DMRG in the language of matrix product operators
and matrix product states, in particular  highlighting the connection
to the original description of the ab initio DMRG sweep algorithm, (ii) to discuss
the implementation of  more general matrix product operator/matrix product state
algebra within the context of a DMRG sweep with renormalized operators, and (iii) to describe some ways
that thinking about matrix product operators can lead to new formulations
of the DMRG, using compression and parallelism as examples.

In recent years, many extensions of the ab initio DMRG have appeared which are motivated
by the very convenient matrix product operator/matrix product state
formalism. As these developments continue, the connections established in this work provide a bridge to
translate these conceptual advances into efficient implementations, using the long-standing technology
of the DMRG.

%{\color{blue}
\newpage
\section*{Appendix: Analysis of the distributions of bond dimensions
for different choices of intermediates}\label{sec:app}

As discussed in the main text, for two-body Hamiltonians
there is freedom to choose different normal
and complementary operator intermediates, all of which result in the same scaling of $O(K^2)$ for the bond dimension.
Here, we analyze in more detail how different
choices of intermediates lead to different distributions
of the leading bond dimensions $D(k)$ along the one-dimensional array of orbital partitions, $k\in\{1,\cdots,K-1\}$. Only the two body terms in Eq. \eqref{eq:longrangeh} will be examined, as the inclusion of the one-body term does not change
the leading bond dimensions $D(k)$.
To further simplify the discussion, we use the following form of
two-electron integrals, viz.,
$\hat{H}_2=\frac{1}{2}v_{pqrs} a_p^\dagger a_q^\dagger a_r a_s
=g_{pqrs}a_p^\dagger a_q^\dagger a_r a_s$,
where the Einstein summation convention for repeated indices has been assumed,
and the tensor $g_{pqrs}$ represents the unique two-electron integrals,
whose number is of $O(K^4/4)$,
\begin{eqnarray}
g_{pqrs}
=\left\{
\begin{array}{cc}
w_{pqrs},& p<q,\;r<s\\
0,& \text{otherwise}
\end{array}\right..
\end{eqnarray}
To examine $D(k)$ for $\hat{H}_2$, we consider the left-right bipartition of orbitals, in which case
$\hat{H}_2$ can be written as
\begin{eqnarray}
\hat{H}_2 &=&\hat{H}^L_2 + \hat{H}^R_2\nonumber\\
&-&
g_{p_Lq_Rr_Ls_R}(a_{p_L}^\dagger a_{r_L})(a_{q_R}^\dagger a_{s_R})+g_{p_Lq_Lr_Rs_R}(a_{p_L}^\dagger a_{q_L}^\dagger)(a_{r_R} a_{s_R})+g_{p_Rq_Rr_Ls_L}(a_{r_L} a_{s_L})(a_{p_R}^\dagger a_{q_R}^\dagger)\nonumber\\
&+&
g_{p_Lq_Rr_Rs_R}a_{p_L}^\dagger(a_{q_R}^\dagger a_{r_R} a_{s_R})
+g_{p_Lq_Rr_Ls_L}(a_{p_L}^\dagger a_{r_L} a_{s_L})a_{q_R}^\dagger \nonumber\\
&+&
g_{p_Rq_Rr_Ls_R}a_{r_L}(a_{p_R}^\dagger a_{q_R}^\dagger a_{s_R})
+
g_{p_Lq_Lr_Ls_R}(a_{p_L}^\dagger a_{q_L}^\dagger a_{r_L}) a_{s_R}.\label{eq:H2a}
\end{eqnarray}
To minimize the bond dimensions across the left and right blocks, the
unambiguous choice is
to define the following intermediates for the last four terms,
\begin{eqnarray}
(\hat{T}1)_{p_L}^R &\;\triangleq&\; g_{p_Lq_Rr_Rs_R} (a_{q_R}^\dagger a_{r_R} a_{s_R}),\\
(\hat{T}2)_{q_R}^L &\;\triangleq&\; g_{p_Lq_Rr_Ls_L}(a_{p_L}^\dagger a_{r_L} a_{s_L}), \\
(\hat{T}3)_{r_L}^R &\;\triangleq&\; g_{p_Rq_Rr_Ls_R} (a_{p_R}^\dagger a_{q_R}^\dagger a_{s_R}), \\
(\hat{T}4)_{s_R}^L &\;\triangleq&\; g_{p_Lq_Lr_Ls_R}(a_{p_L}^\dagger a_{q_L}^\dagger a_{r_L}),
\end{eqnarray}
such that Eq. \eqref{eq:H2a} becomes
\begin{eqnarray}
\hat{H}_2 &=&\hat{H}^L_2 + \hat{H}^R_2\nonumber\\
&-&
g_{p_Lq_Rr_Ls_R}(a_{p_L}^\dagger a_{r_L})(a_{q_R}^\dagger a_{s_R})
+
g_{p_Lq_Lr_Rs_R}(a_{p_L}^\dagger a_{q_L}^\dagger)(a_{r_R} a_{s_R})
+
g_{p_Rq_Rr_Ls_L}(a_{r_L} a_{s_L})(a_{p_R}^\dagger a_{q_R}^\dagger)\nonumber\\
&+&
a_{p_L}^\dagger(\hat{T}1)_{p_L}^R+(\hat{T}2)_{q_R}^L a_{q_R}^\dagger
+a_{r_L}(\hat{T}3)_{r_L}^R+(\hat{T}4)_{s_R}^L a_{s_R}\label{eq:H2b}
\end{eqnarray}
This reduces the bond dimensions for the last four terms in Eq. \eqref{eq:H2a} to $O(K)$. {If we
disregard the contributions from the one-body integrals, the last four terms involving the $\hat{T}1, \hat{T}2, \hat{T}3, \hat{T}4$ operators are equivalent
to the terms involving the $\hat{S}_p^{L_k/R_k}$ operators in Eq. \eqref{eq:Hdecomp}.}

For the remaining two-body terms that involve two left and two right integral
indices,
the integrals can be collected with either the left or the right operators.
However, regardless of the different choices, the bond dimension for the two-body Hamiltonian
clearly scales as $O(K^2)$, although the actual distributions $D(k)$ across
the partitions of the orbitals can be different as demonstrated below.

To analyze the different possibilities, we consider the recursion rules
that link the right operators in Eq. \eqref{eq:H2b} to the next site. Let $R=CR'$, where $C$ denotes the new  site that is being added
to the right block and $R'$ denotes the remaining sites, then the complementary operators $\hat{T}1$ and $\hat{T}3$ defined on the right block $R$ become
\begin{eqnarray}
(\hat{T}1)_{p_L}^{CR'} &=& (\hat{T}1)_{p_L}^{C} +(\hat{T}1)_{p_L}^{R'} \nonumber\\
&+&
(g_{p_Lq_{R'}r_C s_C} a_{r_C}a_{s_{C}})a_{q_{R'}}^\dagger
+(g_{p_Lq_Cr_Cs_{R'}} a_{q_C}^\dagger a_{r_C}) a_{s_{R'}} \nonumber\\
&+&
g_{p_Lq_{C}r_{R'}s_{R'}} a_{q_C}^\dagger (a_{r_{R'}} a_{s_{R'}})
+
g_{p_Lq_{R'}r_{C}s_{R'}}(-a_{r_C})(a_{q_{R'}}^\dagger a_{s_{R'}}),\label{T1}
\end{eqnarray}
and
\begin{eqnarray}
(\hat{T}3)_{r_L}^{CR'} &=& (\hat{T}3)_{r_L}^{C}+(\hat{T}3)_{r_L}^{R'}\nonumber\\
&+&
(-g_{p_{C}q_{R'}r_Ls_{C}}a_{p_{C}}^\dagger a_{s_{C}})a_{q_{R'}}^\dagger
+(g_{p_Cq_Cr_Ls_{R'}}a_{p_C}^\dagger a_{q_C}^\dagger) a_{s_R'}
\nonumber\\
&+&
g_{p_{R'}q_{R'}r_Ls_{C}} a_{s_{C}}(a_{p_{R'}}^\dagger a_{q_{R'}}^\dagger)
+g_{p_Cq_{R'}r_Ls_{R'}} a_{p_C}^\dagger(a_{q_{R'}}^\dagger a_{s_{R'}}),\label{T3}
\end{eqnarray}
respectively. Similar to Eq. \eqref{eq:H2b}, the integrals $g_{pqrs}$ in the last lines
of both Eqs. \eqref{T3} and \eqref{T1} can either be collected with the operators in $C$ or $R'$,
without changing the leading bond dimension of $O(K^2)$ for $\hat{H}_2$.
However, in order to maximally reuse common intermediates, the choice here for $\hat{T}1$ and $\hat{T}3$ also affects the assignment of integrals in Eq. \eqref{eq:H2b} for $\hat{H}_2$.

We first examine the case where the unassigned integrals in
Eqs. \eqref{eq:H2b},
\eqref{T1}, and \eqref{T3} are all combined with the right operators.
This is the choice of complementary operators as introduced in Eq.
\eqref{eq:norm_complementary}.
In this case, the following complementary operators can be defined,
\begin{eqnarray}
\hat{Q}_{p_L r_L}^R&=&g_{p_Lq_Rr_Ls_R}(a_{q_R}^\dagger a_{s_R}),\\
\hat{\underline{P}}^R_{p_L q_L} &=& g_{p_Lq_Lr_Rs_R}(a_{r_R} a_{s_R}),\\
\hat{\overline{P}}^R_{r_L s_L} &=& g_{p_Rq_Rr_Ls_L}(a_{p_R}^\dagger a_{q_R}^\dagger).
\end{eqnarray}
such that the related terms contributing to $\hat{H}_2$ \eqref{eq:H2b}, $\hat{T}1$ \eqref{T1}, and $\hat{T}3$ \eqref{T3} can be
rewritten as
\begin{eqnarray}
\hat{H}_2 &\Leftarrow&
-
(a_{p_L}^\dagger a_{r_L})\hat{Q}^{R}_{p_L r_L}
+
(a_{p_L}^\dagger a_{q_L}^\dagger)\hat{\underline{P}}^R_{p_L q_L}
+
(a_{r_L} a_{s_L})\hat{\overline{P}}^R_{r_L s_L}, \\
(\hat{T}1)_{p_L}^{CR'} &\Leftarrow &
a_{q_C}^\dagger \hat{\underline{P}}^{R'}_{p_Lq_{C}}+(-a_{r_C})\hat{Q}^{R'}_{p_L r_C},\label{t1a}\\
(\hat{T}3)_{r_L}^{CR'}  &\Leftarrow &
a_{s_{C}}\hat{\overline{P}}_{r_L s_C}^{R'}+ a_{p_C}^\dagger \hat{Q}^{R'}_{p_C r_L},
\end{eqnarray}
respectively. Meanwhile, since the expansions of the one-body terms $\hat{Q}$, $\hat{\underline{P}}$, and $\hat{\overline{P}}$
for $R=CR'$ do not require new intermediates, the recursion basis for the recursion to
the rightmost site is complete and given by
\begin{eqnarray}
\left(\hat{H}^R_2,a^{R\dagger},a^{R},(\hat{T}1)^R_{L},(\hat{T}3)^R_{L},\hat{Q}^R_{LL},\hat{\underline{P}}^R_{LL},
\hat{\overline{P}}^R_{LL},\hat{I}^R\right).\label{r1}
\end{eqnarray}
(This basis corresponds to the elements of the vector
in Eq.~(\ref{eq:pwform})).
The leading bond dimension determined by the triple
$(\hat{Q}^R_{LL},\hat{\underline{P}}^R_{LL},\hat{\overline{P}}^R_{LL})$ is $D_1(k)=O(k^2+k^2/2*2)=O(2k^2)$
along the one-dimensional array of orbitals. The averaged value along all the
sites is given by $\bar{D}_1=2/3K^2$.

Instead of using the complementary operators $(\hat{Q}^R_{LL},\hat{\underline{P}}^R_{LL},\hat{\overline{P}}^R_{LL})$, the integrals can also be collected with the left operators, viz.,
\begin{eqnarray}
H_2 &\Leftarrow&
(-g_{p_Lq_Rr_Ls_R}a_{p_L}^\dagger a_{r_L})(a_{q_R}^\dagger a_{s_R})+
(g_{p_Lq_Lr_Rs_R}a_{p_L}^\dagger a_{q_L}^\dagger)(a_{r_R} a_{s_R})+
(g_{p_Rq_Rr_Ls_L}a_{r_L} a_{s_L})(a_{p_R}^\dagger a_{q_R}^\dagger), \\
(\hat{T}1)_{p_L}^{CR'}&\Leftarrow&
(g_{p_Lq_{C}r_{R'}s_{R'}} a_{q_C}^\dagger)(a_{r_{R'}} a_{s_{R'}})+
(-g_{p_Lq_{R'}r_{C}s_{R'}}a_{r_C})(a_{q_{R'}}^\dagger a_{s_{R'}}),\label{t1b}\\
(\hat{T}3)_{r_L}^{CR'}  &\Leftarrow&
(g_{p_{R'}q_{R'}r_Ls_{C}} a_{s_{C}})(a_{p_{R'}}^\dagger a_{q_{R'}}^\dagger)
+(g_{p_Cq_{R'}r_Ls_{R'}} a_{p_C}^\dagger)(a_{q_{R'}}^\dagger a_{s_{R'}}).\label{t3b}
\end{eqnarray}
With this choice, the basis for recursion to the rightmost site becomes
\begin{eqnarray}
\left(\hat{H}^R_2,a^{R\dagger},a^{R},(\hat{T}1)^R_{L},(\hat{T}3)^R_{L},
\hat{B}^R,\hat{\underline{A}}^R,\hat{\overline{A}}^R,\hat{I}^R\right),\label{r2}
\end{eqnarray}
where the bare operators are defined as
\begin{eqnarray}
\hat{B}^R_{q_Rs_R} &=& a_{q_R}^\dagger a_{s_R},\\
\hat{\underline{A}}^R_{r_Rs_R} &=& a_{r_R}a_{s_R},\\
\hat{\overline{A}}^R_{p_Rq_R} &=& a_{p_R}^\dagger a_{q_R}^\dagger,
\end{eqnarray}
which determines the leading bond dimension as $D_2(k)=O\left((K-k)^2+(K-k)^2/2*2\right)=O\left(2(K-k)^2\right)$
and $\bar{D}_2=2/3K^2$.

Alternatively, mixed schemes that use different combinations
of the pairs ($\hat{B}^R$-$\hat{Q}^R_{LL}$, $\hat{\underline{A}}^R$-$\hat{\underline{P}}^R_{LL}$,
and $\hat{\overline{A}}^R$-$\hat{\overline{P}}^R_{LL}$) are also possible.
For instance, the recursion using $\hat{B}^R$, $\hat{\underline{P}}^R_{LL}$, and $\hat{\overline{P}}^R_{LL}$
gives the basis for recursion as
\begin{eqnarray}
\left(\hat{H}^R_2,a^{R\dagger},a^{R},(\hat{T}1)^R_{L},(\hat{T}3)^R_{L},
\hat{B}^R,\hat{\underline{P}}^R_{LL},\hat{\overline{P}}^R_{LL},\hat{I}^R\right)\label{r3}
\end{eqnarray}
which yields the leading bond dimension $D_3(k)=O\left((K-k)^2+k^2/2*2\right)=O\left((K-k)^2+k^2\right)$ and
$\bar{D}_3=2/3K^2$. Similarly,
the basis for the recursion using $\hat{Q}^R_{LL}$, $\hat{\underline{A}}^R$, and $\hat{\overline{A}}^R$ reads
\begin{eqnarray}
\left(\hat{H}^{R}_2,a^{R\dagger},a^R,(\hat{T}1)^{R}_L,(\hat{T}3)^{R}_L,
\hat{Q}^{R}_{LL},\hat{\underline{A}}^R,\hat{\overline{A}}^R,\hat{I}^{R}\right),\label{r4}
\end{eqnarray}
and the leading bond dimension is also $D_3(k)$, the same as that for Eq. \eqref{r3}.

\begin{figure}
\centering
\begin{tabular}{l}
{\resizebox{!}{0.2\textheight}{\includegraphics{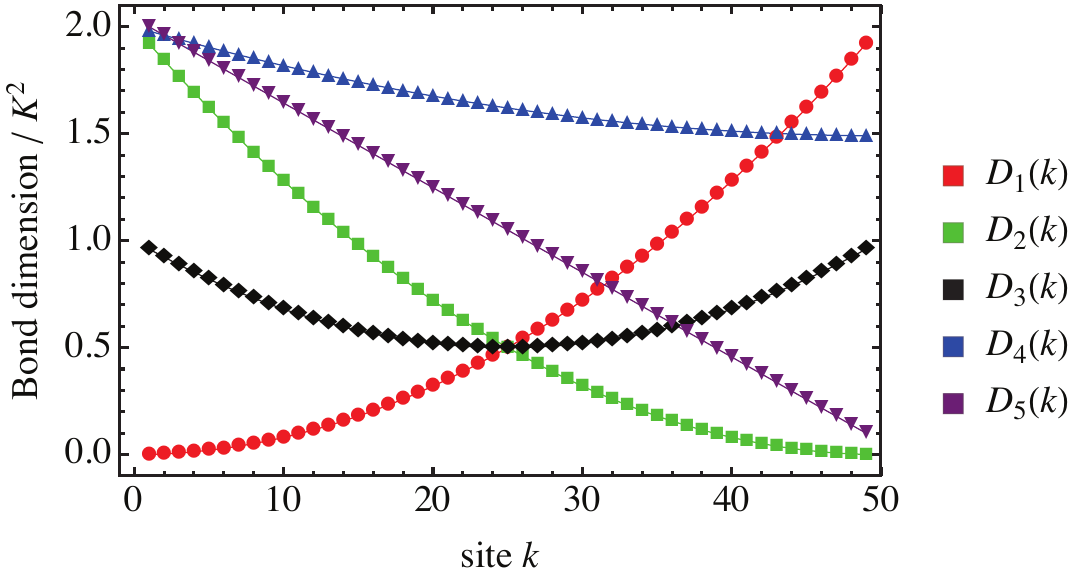}}}\\
\end{tabular}
\caption{Different distributions of the leading bond dimensions $D_n(k)$ ($n=1,2,3,4,5$) for $K=50$
derived based on the recursions in Eqs. \eqref{r1}, \eqref{r2}, \eqref{r3}, \eqref{d4}, and \eqref{r5}, respectively.
}\label{fig:distribution}
\end{figure}

The different distributions $D_{1,2,3}(k)$ discussed so far are compared in Figure \ref{fig:distribution} for $K=50$.
It is shown that the mixed schemes with $D_3(k)$ lead to a more balanced distribution of bond dimensions,
although $D_1(k)$, $D_2(k)$, and $D_3(k)$ all share the same averaged value $2/3K^2$.
Note also that the different recursions can also be changed at different sites, such as
the central site $k=K/2$, resulting in a
centrosymmetric distribution. The conventional DMRG algorithm using complementary operators
employs this fact and uses the biased distributions $D_1(k)$ and $D_2(k)$. Specifically,
DMRG follows $D_1(k)$ in the left part of sites, and changes to $D_2(k)$ after the middle site, $k > K/2$.
This gives the smallest computational cost in practice.

Next, we examine the MPO construction based on $\hat{H}_m=a_m^\dagger \hat{T}_m$ introduced in Eq. \eqref{eq:subH}.
The analysis for $\hat{T}_m$, with the first index fixed to be $m$, is very similar to that for $(\hat{T}1)^R_{p_L}$ in Eq. \eqref{t1a},
except that the index $p_L=m$ now runs through all sites. The recursion basis for $(\hat{T}1)^R_{m}$ can be
deduced from Eq. \eqref{t1a} as
\begin{eqnarray}
\left((\hat{T}1)^R_m,a^{R\dagger},a^{R},\hat{Q}^R_{mL},\hat{\underline{P}}^R_{mL},\hat{I}^R\right),\label{d4}
\end{eqnarray}
while Eq. \eqref{t1b} for $(\hat{T}1)^R_{m}$ leads to a larger bond dimension with scaling $O(K^2)$.
Thus, in such construction of MPO, there is no ambiguity for the choice of an optimal recursion basis.
For given $m$, the leading bond dimension for Eq. \eqref{d4} can be found to be $D(m,k)=2(K-k)+k+\left\{\begin{array}{cc} k-m, & k> m \\ 0, & k \le m \end{array}\right.$, where the last $m$-dependent term arises from
$\hat{\underline{P}}^R_{mL}$, which has nonzero contributions only for $m<k$.
Thus, for each $m$ the bond dimension for $(\hat{T}1)^R_{m}$ and hence $\hat{H}_m$ is of $O(K)$.

Although the separate sub-Hamiltonians $\hat{H}_m$ constitute separate MPO's which can be independently manipulated, to provide
a point of comparison with the earlier distributions $D_{1,2,3}(k)$ for the single $\hat{H}$ MPO, we can
compute the sum of the bond dimensions of all the sub-Hamiltonians, viz.,
%% To be compared with the preceding , the leading bond dimension denoted by $D_4(k)$ is computed by summing
%% over all $m$, viz.,
$D_4(k)=\sum_{m=1}^{K}D(m,k)=O(2K^2-kK+k^2/2)$.
The averaged value of $D_4(k)$ is $\bar{D}_4=5/3K^2$ and the
distribution for $K=50$ is also shown in Figure \ref{fig:distribution}.
$D_4(k)$ is significantly larger then $D_{1,2,3}(k)$. {This redundancy
is due to the repeated use of $a^{R\dagger}$ and $a^{R}$ in Eq. \eqref{d4} for all sub-Hamiltonians $\hat{H}_m$,
while in the former case only a single instance of these operators is required in
the recursion rules for $\hat{H}$ and thus they  do not contribute multiple times
to the leading bond dimension $D_{1,2,3}(k)$.} Indeed, the increase of $\bar{D}_4=5/3K^2$ by
$K^2$ as compared to $\bar{D}_{1,2,3}=2/3K^2$ is attributable to these two terms, whose contribution to
$\bar{D}_4$ is $1/K\sum_{k=1}(\sum_{m=1}^{K}2(K-k))=O(K^2)$. However, it is important to note that
$D_4(k)$ does not constitute a true computational bond dimension, as in practice, the different
$\hat{H}_m$ are manipulated separately and the combined bond dimension does not appears in an
actual calculation.

As discussed in the main text, there
is an alternative definition of the sub-Hamiltonian $\hat{H}_m$ as a collection of terms sharing the same rightmost operator.
In this definition, the redundancy in the recursion rules from the repeated use of $a^{R\dagger}$ and $a^R$ can be partially mitigated,
as the 'delocalization' of $a^{R\dagger}$ and $a^R$ onto sites $k>m$ is removed.
By this definition, the reduction of the (combined for all $\hat{H}_m$) bond dimension for each site index $k$ can be estimated as follows:
for $m<k$  the number of $a^{R\dagger}$ and $a^R$  eliminated is $2(K-k)$,
while for $m>k$ the number of necessary $a^{R\dagger}$ and $a^R$ to represent $\hat{H}_m$
is only $2(m-k)$, and hence the number of unnecessary $a^{R\dagger}$ and $a^R$ is $2(K-m)$.
Then, the averaged reduction is $1/K\sum_{k=1}^{K}\left(\sum_{m=1}^{k}2(K-k)+\sum_{m=k}^{K}2(K-m)\right)=O(2/3K^2)$,
and the averaged bond dimension for $\hat{H}_m$ becomes $5/3K^2-2/3K^2=K^2$.
We can explicitly demonstrate that the above estimates are correct.
Since, by definition, $\hat{H}_m$ contains terms having at least one index on the site $m$, and hence,
in terms of the bipartition, if the right block is the site $m$ and the left
block contains the sites from 1 to $m-1$, $\hat{H}_m$ collects all terms in
Eq. \eqref{eq:H2a} except for $\hat{H}^L_2$. Thus, to analyze the bond dimension for $\hat{H}_m$, we can simply
use the results for $\hat{H}_2$. Specifically, only the recursions for
$(\hat{T}2)_{q_R}^L$ and $(\hat{T}4)_{s_R}^L$ from the site $m-1$ to the leftmost site is relevant
to estimating the leading bond dimensions $D(k)$. Let $L=L'C$, the relevant recursions are found as
\begin{eqnarray}
(\hat{T}2)_{q_R}^{L'C} &\Leftarrow& \hat{\underline{P}}^{L'}_{p_Cq_R}a_{p_C}^\dagger+ \hat{Q}^{L'}_{q_Rs_C} a_{s_C}, \\
(\hat{T}4)_{s_R}^{L'C} &\Leftarrow& \hat{Q}_{q_{C}s_R}^{L'}(-a_{q_C}^\dagger ) +\hat{\overline{P}}_{r_Cs_R}^{L'} a_{r_C},
\end{eqnarray}
where $\hat{Q}^L_{qs}\triangleq\sum_{pr\in L}g_{pqrs}a^\dagger_p a_r$. Thus, similar to Eq. \eqref{r1},
the recursion basis is
\begin{eqnarray}
\left((\hat{T}2)^L_m,(\hat{T}4)^L_m,a^{L\dagger},a^{L},\hat{Q}^L_{mR},
\hat{Q}^L_{Rm},\hat{\underline{P}}^L_{Rm},\hat{\overline{P}}^L_{Rm},\hat{I}^R\right),\label{r5}
\end{eqnarray}
where $R$ denotes the sites between $k$ and $m$ in this expression. The leading
bond dimension becomes $D(m,k)=2k+2(m-k)+2(m-k)=2(2m-k)$ for $k<m$,
while the bond dimension for $k>m$ is simply 1. Thus it is seen that $\hat{H}_m$ defined
in this way also has a bond dimension of $O(K)$.
The leading bond dimension $D_5(k)$ obtained by summing over $m$
can be estimated as $D_5(k)=\sum_{m=1}^{K}D(m,k)=\sum_{m=k}^{K}D(m,k)=O(2K^2-2kK)$,
which decays linearly with the increase of $k$.
Its averaged value is $\bar{D}_5=K^2$, which agrees with our estimates from the
consideration of redundancies. This value is much smaller than $\bar{D}_4=5/3K^2$, but
larger than $\bar{D}_{1,2,3}=2/3K^2$. The distribution $D_5(k)$ is displayed
in Figure \ref{fig:distribution}. While $D_5(k)$ represents a combined bond dimension
and does not directly reflect the computational structure where the $\hat{H}_m$ are manipulated separately,
we nonetheless expect there will be a computational gain from using this second definition of $\hat{H}_m$
compared to the use of $\hat{H}_m$ in Eq.~\eqref{eq:subH} in practical computations.
%}

\acknowledgments
G. K.-L. Chan would like to acknowledge the US Department of Energy for funding primarily through  DE-SC0010530, with additional funding provided by DE-SC0008624. Z. Li was supported by the Simons Foundation through the Simons Collaboration on the Many-Electron problem. S. R. White would like to acknowledge funding from the Simons Foundation through the Simons Collaboration on the Many-Electron problem.

%\bibliography{abinitiodmrg}

\begin{thebibliography}{78}%
\makeatletter
\providecommand \@ifxundefined [1]{%
 \@ifx{#1\undefined}
}%
\providecommand \@ifnum [1]{%
 \ifnum #1\expandafter \@firstoftwo
 \else \expandafter \@secondoftwo
 \fi
}%
\providecommand \@ifx [1]{%
 \ifx #1\expandafter \@firstoftwo
 \else \expandafter \@secondoftwo
 \fi
}%
\providecommand \natexlab [1]{#1}%
\providecommand \enquote  [1]{``#1''}%
\providecommand \bibnamefont  [1]{#1}%
\providecommand \bibfnamefont [1]{#1}%
\providecommand \citenamefont [1]{#1}%
\providecommand \href@noop [0]{\@secondoftwo}%
\providecommand \href [0]{\begingroup \@sanitize@url \@href}%
\providecommand \@href[1]{\@@startlink{#1}\@@href}%
\providecommand \@@href[1]{\endgroup#1\@@endlink}%
\providecommand \@sanitize@url [0]{\catcode `\\12\catcode `\$12\catcode
  `\&12\catcode `\#12\catcode `\^12\catcode `\_12\catcode `\%12\relax}%
\providecommand \@@startlink[1]{}%
\providecommand \@@endlink[0]{}%
\providecommand \url  [0]{\begingroup\@sanitize@url \@url }%
\providecommand \@url [1]{\endgroup\@href {#1}{\urlprefix }}%
\providecommand \urlprefix  [0]{URL }%
\providecommand \Eprint [0]{\href }%
\providecommand \doibase [0]{http://dx.doi.org/}%
\providecommand \selectlanguage [0]{\@gobble}%
\providecommand \bibinfo  [0]{\@secondoftwo}%
\providecommand \bibfield  [0]{\@secondoftwo}%
\providecommand \translation [1]{[#1]}%
\providecommand \BibitemOpen [0]{}%
\providecommand \bibitemStop [0]{}%
\providecommand \bibitemNoStop [0]{.\EOS\space}%
\providecommand \EOS [0]{\spacefactor3000\relax}%
\providecommand \BibitemShut  [1]{\csname bibitem#1\endcsname}%
\let\auto@bib@innerbib\@empty
%</preamble>
\bibitem [{\citenamefont {White}(1992)}]{white1992density}%
  \BibitemOpen
  \bibfield  {author} {\bibinfo {author} {\bibfnamefont {S.~R.}\ \bibnamefont
  {White}},\ }\href@noop {} {\bibfield  {journal} {\bibinfo  {journal}
  {Physical Review Letters}\ }\textbf {\bibinfo {volume} {69}},\ \bibinfo
  {pages} {2863} (\bibinfo {year} {1992})}\BibitemShut {NoStop}%
\bibitem [{\citenamefont {White}(1993)}]{white1993density}%
  \BibitemOpen
  \bibfield  {author} {\bibinfo {author} {\bibfnamefont {S.~R.}\ \bibnamefont
  {White}},\ }\href@noop {} {\bibfield  {journal} {\bibinfo  {journal}
  {Physical Review B}\ }\textbf {\bibinfo {volume} {48}},\ \bibinfo {pages}
  {10345} (\bibinfo {year} {1993})}\BibitemShut {NoStop}%
\bibitem [{\citenamefont {Xiang}(1996)}]{xiang1996density}%
  \BibitemOpen
  \bibfield  {author} {\bibinfo {author} {\bibfnamefont {T.}~\bibnamefont
  {Xiang}},\ }\href@noop {} {\bibfield  {journal} {\bibinfo  {journal}
  {Physical Review B}\ }\textbf {\bibinfo {volume} {53}},\ \bibinfo {pages}
  {R10445} (\bibinfo {year} {1996})}\BibitemShut {NoStop}%
\bibitem [{\citenamefont {White}\ and\ \citenamefont
  {Martin}(1999)}]{white1999ab}%
  \BibitemOpen
  \bibfield  {author} {\bibinfo {author} {\bibfnamefont {S.~R.}\ \bibnamefont
  {White}}\ and\ \bibinfo {author} {\bibfnamefont {R.~L.}\ \bibnamefont
  {Martin}},\ }\href@noop {} {\bibfield  {journal} {\bibinfo  {journal} {The
  Journal of Chemical Physics}\ }\textbf {\bibinfo {volume} {110}},\ \bibinfo
  {pages} {4127} (\bibinfo {year} {1999})}\BibitemShut {NoStop}%
\bibitem [{\citenamefont {Daul}\ \emph {et~al.}(2000)\citenamefont {Daul},
  \citenamefont {Ciofini}, \citenamefont {Daul},\ and\ \citenamefont
  {White}}]{daul2000full}%
  \BibitemOpen
  \bibfield  {author} {\bibinfo {author} {\bibfnamefont {S.}~\bibnamefont
  {Daul}}, \bibinfo {author} {\bibfnamefont {I.}~\bibnamefont {Ciofini}},
  \bibinfo {author} {\bibfnamefont {C.}~\bibnamefont {Daul}}, \ and\ \bibinfo
  {author} {\bibfnamefont {S.~R.}\ \bibnamefont {White}},\ }\href@noop {}
  {\bibfield  {journal} {\bibinfo  {journal} {International Journal of Quantum
  Chemistry}\ }\textbf {\bibinfo {volume} {79}},\ \bibinfo {pages} {331}
  (\bibinfo {year} {2000})}\BibitemShut {NoStop}%
\bibitem [{\citenamefont {Chan}\ and\ \citenamefont
  {Head-Gordon}(2002)}]{chan2002highly}%
  \BibitemOpen
  \bibfield  {author} {\bibinfo {author} {\bibfnamefont {G.~K.-L.}\
  \bibnamefont {Chan}}\ and\ \bibinfo {author} {\bibfnamefont {M.}~\bibnamefont
  {Head-Gordon}},\ }\href@noop {} {\bibfield  {journal} {\bibinfo  {journal}
  {The Journal of Chemical Physics}\ }\textbf {\bibinfo {volume} {116}},\
  \bibinfo {pages} {4462} (\bibinfo {year} {2002})}\BibitemShut {NoStop}%
\bibitem [{\citenamefont {Chan}(2004)}]{chan2004algorithm}%
  \BibitemOpen
  \bibfield  {author} {\bibinfo {author} {\bibfnamefont {G.~K.-L.}\
  \bibnamefont {Chan}},\ }\href@noop {} {\bibfield  {journal} {\bibinfo
  {journal} {The Journal of Chemical Physics}\ }\textbf {\bibinfo {volume}
  {120}},\ \bibinfo {pages} {3172} (\bibinfo {year} {2004})}\BibitemShut
  {NoStop}%
\bibitem [{\citenamefont {Legeza}, \citenamefont {R{\"o}der},\ and\
  \citenamefont {Hess}(2003{\natexlab{a}})}]{legeza2003controlling}%
  \BibitemOpen
  \bibfield  {author} {\bibinfo {author} {\bibfnamefont {{\"O}.}~\bibnamefont
  {Legeza}}, \bibinfo {author} {\bibfnamefont {J.}~\bibnamefont {R{\"o}der}}, \
  and\ \bibinfo {author} {\bibfnamefont {B.}~\bibnamefont {Hess}},\ }\href@noop
  {} {\bibfield  {journal} {\bibinfo  {journal} {Physical Review B}\ }\textbf
  {\bibinfo {volume} {67}},\ \bibinfo {pages} {125114} (\bibinfo {year}
  {2003}{\natexlab{a}})}\BibitemShut {NoStop}%
\bibitem [{\citenamefont {Legeza}, \citenamefont {R{\"o}der},\ and\
  \citenamefont {Hess}(2003{\natexlab{b}})}]{legeza2003qc}%
  \BibitemOpen
  \bibfield  {author} {\bibinfo {author} {\bibfnamefont {{\"O}.}~\bibnamefont
  {Legeza}}, \bibinfo {author} {\bibfnamefont {J.}~\bibnamefont {R{\"o}der}}, \
  and\ \bibinfo {author} {\bibfnamefont {B.}~\bibnamefont {Hess}},\ }\href@noop
  {} {\bibfield  {journal} {\bibinfo  {journal} {Molecular Physics}\ }\textbf
  {\bibinfo {volume} {101}},\ \bibinfo {pages} {2019} (\bibinfo {year}
  {2003}{\natexlab{b}})}\BibitemShut {NoStop}%
\bibitem [{\citenamefont {Mitrushenkov}\ \emph {et~al.}(2001)\citenamefont
  {Mitrushenkov}, \citenamefont {Fano}, \citenamefont {Ortolani}, \citenamefont
  {Linguerri},\ and\ \citenamefont {Palmieri}}]{mitrushenkov2001quantum}%
  \BibitemOpen
  \bibfield  {author} {\bibinfo {author} {\bibfnamefont {A.~O.}\ \bibnamefont
  {Mitrushenkov}}, \bibinfo {author} {\bibfnamefont {G.}~\bibnamefont {Fano}},
  \bibinfo {author} {\bibfnamefont {F.}~\bibnamefont {Ortolani}}, \bibinfo
  {author} {\bibfnamefont {R.}~\bibnamefont {Linguerri}}, \ and\ \bibinfo
  {author} {\bibfnamefont {P.}~\bibnamefont {Palmieri}},\ }\href@noop {}
  {\bibfield  {journal} {\bibinfo  {journal} {Journal of Chemical Physics}\
  }\textbf {\bibinfo {volume} {115}},\ \bibinfo {pages} {6815} (\bibinfo {year}
  {2001})}\BibitemShut {NoStop}%
\bibitem [{\citenamefont {Chan}\ and\ \citenamefont
  {Head-Gordon}(2003)}]{chan2003exact}%
  \BibitemOpen
  \bibfield  {author} {\bibinfo {author} {\bibfnamefont {G.~K.-L.}\
  \bibnamefont {Chan}}\ and\ \bibinfo {author} {\bibfnamefont {M.}~\bibnamefont
  {Head-Gordon}},\ }\href@noop {} {\bibfield  {journal} {\bibinfo  {journal}
  {The Journal of Chemical Physics}\ }\textbf {\bibinfo {volume} {118}},\
  \bibinfo {pages} {8551} (\bibinfo {year} {2003})}\BibitemShut {NoStop}%
\bibitem [{\citenamefont {Legeza}\ and\ \citenamefont
  {S{\'o}lyom}(2003)}]{legeza2003optimizing}%
  \BibitemOpen
  \bibfield  {author} {\bibinfo {author} {\bibfnamefont {{\"O}.}~\bibnamefont
  {Legeza}}\ and\ \bibinfo {author} {\bibfnamefont {J.}~\bibnamefont
  {S{\'o}lyom}},\ }\href@noop {} {\bibfield  {journal} {\bibinfo  {journal}
  {Physical Review B}\ }\textbf {\bibinfo {volume} {68}},\ \bibinfo {pages}
  {195116} (\bibinfo {year} {2003})}\BibitemShut {NoStop}%
\bibitem [{\citenamefont {Legeza}\ and\ \citenamefont
  {S{\'o}lyom}(2004)}]{legeza2004quantum}%
  \BibitemOpen
  \bibfield  {author} {\bibinfo {author} {\bibfnamefont {{\"O}.}~\bibnamefont
  {Legeza}}\ and\ \bibinfo {author} {\bibfnamefont {J.}~\bibnamefont
  {S{\'o}lyom}},\ }\href@noop {} {\bibfield  {journal} {\bibinfo  {journal}
  {Physical Review B}\ }\textbf {\bibinfo {volume} {70}},\ \bibinfo {pages}
  {205118} (\bibinfo {year} {2004})}\BibitemShut {NoStop}%
\bibitem [{\citenamefont {Mitrushenkov}\ \emph {et~al.}(2003)\citenamefont
  {Mitrushenkov}, \citenamefont {Linguerri}, \citenamefont {Palmieri},\ and\
  \citenamefont {Fano}}]{mitrushenkov2003quantum}%
  \BibitemOpen
  \bibfield  {author} {\bibinfo {author} {\bibfnamefont {A.}~\bibnamefont
  {Mitrushenkov}}, \bibinfo {author} {\bibfnamefont {R.}~\bibnamefont
  {Linguerri}}, \bibinfo {author} {\bibfnamefont {P.}~\bibnamefont {Palmieri}},
  \ and\ \bibinfo {author} {\bibfnamefont {G.}~\bibnamefont {Fano}},\
  }\href@noop {} {\bibfield  {journal} {\bibinfo  {journal} {The Journal of
  Chemical Physics}\ }\textbf {\bibinfo {volume} {119}},\ \bibinfo {pages}
  {4148} (\bibinfo {year} {2003})}\BibitemShut {NoStop}%
\bibitem [{\citenamefont {Chan}, \citenamefont {K{\'a}llay},\ and\
  \citenamefont {Gauss}(2004)}]{chan2004state}%
  \BibitemOpen
  \bibfield  {author} {\bibinfo {author} {\bibfnamefont {G.~K.-L.}\
  \bibnamefont {Chan}}, \bibinfo {author} {\bibfnamefont {M.}~\bibnamefont
  {K{\'a}llay}}, \ and\ \bibinfo {author} {\bibfnamefont {J.}~\bibnamefont
  {Gauss}},\ }\href@noop {} {\bibfield  {journal} {\bibinfo  {journal} {The
  Journal of Chemical Physics}\ }\textbf {\bibinfo {volume} {121}},\ \bibinfo
  {pages} {6110} (\bibinfo {year} {2004})}\BibitemShut {NoStop}%
\bibitem [{\citenamefont {Chan}\ and\ \citenamefont
  {Van~Voorhis}(2005)}]{chan2005density}%
  \BibitemOpen
  \bibfield  {author} {\bibinfo {author} {\bibfnamefont {G.~K.-L.}\
  \bibnamefont {Chan}}\ and\ \bibinfo {author} {\bibfnamefont {T.}~\bibnamefont
  {Van~Voorhis}},\ }\href@noop {} {\bibfield  {journal} {\bibinfo  {journal}
  {The Journal of Chemical Physics}\ }\textbf {\bibinfo {volume} {122}},\
  \bibinfo {pages} {204101} (\bibinfo {year} {2005})}\BibitemShut {NoStop}%
\bibitem [{\citenamefont {Moritz}\ and\ \citenamefont
  {Reiher}(2006)}]{moritz2006construction}%
  \BibitemOpen
  \bibfield  {author} {\bibinfo {author} {\bibfnamefont {G.}~\bibnamefont
  {Moritz}}\ and\ \bibinfo {author} {\bibfnamefont {M.}~\bibnamefont
  {Reiher}},\ }\href@noop {} {\bibfield  {journal} {\bibinfo  {journal} {The
  Journal of Chemical Physics}\ }\textbf {\bibinfo {volume} {124}},\ \bibinfo
  {pages} {034103} (\bibinfo {year} {2006})}\BibitemShut {NoStop}%
\bibitem [{\citenamefont {Hachmann}, \citenamefont {Cardoen},\ and\
  \citenamefont {Chan}(2006)}]{hachmann2006multireference}%
  \BibitemOpen
  \bibfield  {author} {\bibinfo {author} {\bibfnamefont {J.}~\bibnamefont
  {Hachmann}}, \bibinfo {author} {\bibfnamefont {W.}~\bibnamefont {Cardoen}}, \
  and\ \bibinfo {author} {\bibfnamefont {G.~K.-L.}\ \bibnamefont {Chan}},\
  }\href@noop {} {\bibfield  {journal} {\bibinfo  {journal} {The Journal of
  Chemical Physics}\ }\textbf {\bibinfo {volume} {125}},\ \bibinfo {pages}
  {144101} (\bibinfo {year} {2006})}\BibitemShut {NoStop}%
\bibitem [{\citenamefont {Marti}\ \emph {et~al.}(2008)\citenamefont {Marti},
  \citenamefont {Ond{\'\i}k}, \citenamefont {Moritz},\ and\ \citenamefont
  {Reiher}}]{marti2008density}%
  \BibitemOpen
  \bibfield  {author} {\bibinfo {author} {\bibfnamefont {K.~H.}\ \bibnamefont
  {Marti}}, \bibinfo {author} {\bibfnamefont {I.~M.}\ \bibnamefont
  {Ond{\'\i}k}}, \bibinfo {author} {\bibfnamefont {G.}~\bibnamefont {Moritz}},
  \ and\ \bibinfo {author} {\bibfnamefont {M.}~\bibnamefont {Reiher}},\
  }\href@noop {} {\bibfield  {journal} {\bibinfo  {journal} {The Journal of
  Chemical Physics}\ }\textbf {\bibinfo {volume} {128}},\ \bibinfo {pages}
  {014104} (\bibinfo {year} {2008})}\BibitemShut {NoStop}%
\bibitem [{\citenamefont {Ghosh}\ \emph {et~al.}(2008)\citenamefont {Ghosh},
  \citenamefont {Hachmann}, \citenamefont {Yanai},\ and\ \citenamefont
  {Chan}}]{ghosh2008orbital}%
  \BibitemOpen
  \bibfield  {author} {\bibinfo {author} {\bibfnamefont {D.}~\bibnamefont
  {Ghosh}}, \bibinfo {author} {\bibfnamefont {J.}~\bibnamefont {Hachmann}},
  \bibinfo {author} {\bibfnamefont {T.}~\bibnamefont {Yanai}}, \ and\ \bibinfo
  {author} {\bibfnamefont {G.~K.-L.}\ \bibnamefont {Chan}},\ }\href@noop {}
  {\bibfield  {journal} {\bibinfo  {journal} {The Journal of Chemical Physics}\
  }\textbf {\bibinfo {volume} {128}},\ \bibinfo {pages} {144117} (\bibinfo
  {year} {2008})}\BibitemShut {NoStop}%
\bibitem [{\citenamefont {Chan}(2008)}]{chan2008density}%
  \BibitemOpen
  \bibfield  {author} {\bibinfo {author} {\bibfnamefont {G.~K.-L.}\
  \bibnamefont {Chan}},\ }\href@noop {} {\bibfield  {journal} {\bibinfo
  {journal} {Physical Chemistry Chemical Physics}\ }\textbf {\bibinfo {volume}
  {10}},\ \bibinfo {pages} {3454} (\bibinfo {year} {2008})}\BibitemShut
  {NoStop}%
\bibitem [{\citenamefont {Zgid}\ and\ \citenamefont
  {Nooijen}(2008)}]{zgid2008obtaining}%
  \BibitemOpen
  \bibfield  {author} {\bibinfo {author} {\bibfnamefont {D.}~\bibnamefont
  {Zgid}}\ and\ \bibinfo {author} {\bibfnamefont {M.}~\bibnamefont {Nooijen}},\
  }\href@noop {} {\bibfield  {journal} {\bibinfo  {journal} {The Journal of
  Chemical Physics}\ }\textbf {\bibinfo {volume} {128}},\ \bibinfo {pages}
  {144115} (\bibinfo {year} {2008})}\BibitemShut {NoStop}%
\bibitem [{\citenamefont {Marti}\ and\ \citenamefont
  {Reiher}(2010)}]{marti2010density}%
  \BibitemOpen
  \bibfield  {author} {\bibinfo {author} {\bibfnamefont {K.~H.}\ \bibnamefont
  {Marti}}\ and\ \bibinfo {author} {\bibfnamefont {M.}~\bibnamefont {Reiher}},\
  }\href@noop {} {\bibfield  {journal} {\bibinfo  {journal} {Zeitschrift
  f{\"u}r Physikalische Chemie International Journal of Research in Physical
  Chemistry and Chemical Physics}\ }\textbf {\bibinfo {volume} {224}},\
  \bibinfo {pages} {583} (\bibinfo {year} {2010})}\BibitemShut {NoStop}%
\bibitem [{\citenamefont {Luo}, \citenamefont {Qin},\ and\ \citenamefont
  {Xiang}(2010)}]{luo2010optimizing}%
  \BibitemOpen
  \bibfield  {author} {\bibinfo {author} {\bibfnamefont {H.-G.}\ \bibnamefont
  {Luo}}, \bibinfo {author} {\bibfnamefont {M.-P.}\ \bibnamefont {Qin}}, \ and\
  \bibinfo {author} {\bibfnamefont {T.}~\bibnamefont {Xiang}},\ }\href@noop {}
  {\bibfield  {journal} {\bibinfo  {journal} {Physical Review B}\ }\textbf
  {\bibinfo {volume} {81}},\ \bibinfo {pages} {235129} (\bibinfo {year}
  {2010})}\BibitemShut {NoStop}%
\bibitem [{\citenamefont {Marti}\ and\ \citenamefont
  {Reiher}(2011)}]{marti2011new}%
  \BibitemOpen
  \bibfield  {author} {\bibinfo {author} {\bibfnamefont {K.~H.}\ \bibnamefont
  {Marti}}\ and\ \bibinfo {author} {\bibfnamefont {M.}~\bibnamefont {Reiher}},\
  }\href@noop {} {\bibfield  {journal} {\bibinfo  {journal} {Physical Chemistry
  Chemical Physics}\ }\textbf {\bibinfo {volume} {13}},\ \bibinfo {pages}
  {6750} (\bibinfo {year} {2011})}\BibitemShut {NoStop}%
\bibitem [{\citenamefont {Kurashige}\ and\ \citenamefont
  {Yanai}(2011)}]{kurashige2011second}%
  \BibitemOpen
  \bibfield  {author} {\bibinfo {author} {\bibfnamefont {Y.}~\bibnamefont
  {Kurashige}}\ and\ \bibinfo {author} {\bibfnamefont {T.}~\bibnamefont
  {Yanai}},\ }\href@noop {} {\bibfield  {journal} {\bibinfo  {journal} {The
  Journal of Chemical Physics}\ }\textbf {\bibinfo {volume} {135}},\ \bibinfo
  {pages} {094104} (\bibinfo {year} {2011})}\BibitemShut {NoStop}%
\bibitem [{\citenamefont {Sharma}\ and\ \citenamefont
  {Chan}(2012)}]{sharma2012spin}%
  \BibitemOpen
  \bibfield  {author} {\bibinfo {author} {\bibfnamefont {S.}~\bibnamefont
  {Sharma}}\ and\ \bibinfo {author} {\bibfnamefont {G.~K.-L.}\ \bibnamefont
  {Chan}},\ }\href@noop {} {\bibfield  {journal} {\bibinfo  {journal} {The
  Journal of Chemical Physics}\ }\textbf {\bibinfo {volume} {136}},\ \bibinfo
  {pages} {124121} (\bibinfo {year} {2012})}\BibitemShut {NoStop}%
\bibitem [{\citenamefont {Chan}(2012)}]{chan2012low}%
  \BibitemOpen
  \bibfield  {author} {\bibinfo {author} {\bibfnamefont {G.~K.}\ \bibnamefont
  {Chan}},\ }\href@noop {} {\bibfield  {journal} {\bibinfo  {journal} {Wiley
  Interdisciplinary Reviews: Computational Molecular Science}\ }\textbf
  {\bibinfo {volume} {2}},\ \bibinfo {pages} {907} (\bibinfo {year}
  {2012})}\BibitemShut {NoStop}%
\bibitem [{\citenamefont {Wouters}\ \emph {et~al.}(2012)\citenamefont
  {Wouters}, \citenamefont {Limacher}, \citenamefont {Van~Neck},\ and\
  \citenamefont {Ayers}}]{wouters2012longitudinal}%
  \BibitemOpen
  \bibfield  {author} {\bibinfo {author} {\bibfnamefont {S.}~\bibnamefont
  {Wouters}}, \bibinfo {author} {\bibfnamefont {P.~A.}\ \bibnamefont
  {Limacher}}, \bibinfo {author} {\bibfnamefont {D.}~\bibnamefont {Van~Neck}},
  \ and\ \bibinfo {author} {\bibfnamefont {P.~W.}\ \bibnamefont {Ayers}},\
  }\href@noop {} {\bibfield  {journal} {\bibinfo  {journal} {The Journal of
  Chemical Physics}\ }\textbf {\bibinfo {volume} {136}},\ \bibinfo {pages}
  {134110} (\bibinfo {year} {2012})}\BibitemShut {NoStop}%
\bibitem [{\citenamefont {Mizukami}, \citenamefont {Kurashige},\ and\
  \citenamefont {Yanai}(2012)}]{mizukami2012more}%
  \BibitemOpen
  \bibfield  {author} {\bibinfo {author} {\bibfnamefont {W.}~\bibnamefont
  {Mizukami}}, \bibinfo {author} {\bibfnamefont {Y.}~\bibnamefont {Kurashige}},
  \ and\ \bibinfo {author} {\bibfnamefont {T.}~\bibnamefont {Yanai}},\
  }\href@noop {} {\bibfield  {journal} {\bibinfo  {journal} {Journal of
  Chemical Theory and Computation}\ }\textbf {\bibinfo {volume} {9}},\ \bibinfo
  {pages} {401} (\bibinfo {year} {2012})}\BibitemShut {NoStop}%
\bibitem [{\citenamefont {Kurashige}, \citenamefont {Chan},\ and\ \citenamefont
  {Yanai}(2013)}]{kurashige2013entangled}%
  \BibitemOpen
  \bibfield  {author} {\bibinfo {author} {\bibfnamefont {Y.}~\bibnamefont
  {Kurashige}}, \bibinfo {author} {\bibfnamefont {G.~K.-L.}\ \bibnamefont
  {Chan}}, \ and\ \bibinfo {author} {\bibfnamefont {T.}~\bibnamefont {Yanai}},\
  }\href@noop {} {\bibfield  {journal} {\bibinfo  {journal} {Nature Chemistry}\
  }\textbf {\bibinfo {volume} {5}},\ \bibinfo {pages} {660} (\bibinfo {year}
  {2013})}\BibitemShut {NoStop}%
\bibitem [{\citenamefont {Sharma}\ \emph {et~al.}(2014)\citenamefont {Sharma},
  \citenamefont {Sivalingam}, \citenamefont {Neese},\ and\ \citenamefont
  {Chan}}]{sharma2014low}%
  \BibitemOpen
  \bibfield  {author} {\bibinfo {author} {\bibfnamefont {S.}~\bibnamefont
  {Sharma}}, \bibinfo {author} {\bibfnamefont {K.}~\bibnamefont {Sivalingam}},
  \bibinfo {author} {\bibfnamefont {F.}~\bibnamefont {Neese}}, \ and\ \bibinfo
  {author} {\bibfnamefont {G.~K.-L.}\ \bibnamefont {Chan}},\ }\href@noop {}
  {\bibfield  {journal} {\bibinfo  {journal} {Nature Chemistry}\ }\textbf
  {\bibinfo {volume} {6}},\ \bibinfo {pages} {927} (\bibinfo {year}
  {2014})}\BibitemShut {NoStop}%
\bibitem [{\citenamefont {Wouters}\ and\ \citenamefont
  {Van~Neck}(2014)}]{wouters2014density}%
  \BibitemOpen
  \bibfield  {author} {\bibinfo {author} {\bibfnamefont {S.}~\bibnamefont
  {Wouters}}\ and\ \bibinfo {author} {\bibfnamefont {D.}~\bibnamefont
  {Van~Neck}},\ }\href@noop {} {\bibfield  {journal} {\bibinfo  {journal} {The
  European Physical Journal D}\ }\textbf {\bibinfo {volume} {68}},\ \bibinfo
  {pages} {1} (\bibinfo {year} {2014})}\BibitemShut {NoStop}%
\bibitem [{\citenamefont {Wouters}\ \emph {et~al.}(2014)\citenamefont
  {Wouters}, \citenamefont {Poelmans}, \citenamefont {Ayers},\ and\
  \citenamefont {Van~Neck}}]{wouters2014chemps2}%
  \BibitemOpen
  \bibfield  {author} {\bibinfo {author} {\bibfnamefont {S.}~\bibnamefont
  {Wouters}}, \bibinfo {author} {\bibfnamefont {W.}~\bibnamefont {Poelmans}},
  \bibinfo {author} {\bibfnamefont {P.~W.}\ \bibnamefont {Ayers}}, \ and\
  \bibinfo {author} {\bibfnamefont {D.}~\bibnamefont {Van~Neck}},\ }\href@noop
  {} {\bibfield  {journal} {\bibinfo  {journal} {Computer Physics
  Communications}\ }\textbf {\bibinfo {volume} {185}},\ \bibinfo {pages} {1501}
  (\bibinfo {year} {2014})}\BibitemShut {NoStop}%
\bibitem [{\citenamefont {Fertitta}\ \emph {et~al.}(2014)\citenamefont
  {Fertitta}, \citenamefont {Paulus}, \citenamefont {Barcza},\ and\
  \citenamefont {Legeza}}]{fertitta2014investigation}%
  \BibitemOpen
  \bibfield  {author} {\bibinfo {author} {\bibfnamefont {E.}~\bibnamefont
  {Fertitta}}, \bibinfo {author} {\bibfnamefont {B.}~\bibnamefont {Paulus}},
  \bibinfo {author} {\bibfnamefont {G.}~\bibnamefont {Barcza}}, \ and\ \bibinfo
  {author} {\bibfnamefont {{\"O}.}~\bibnamefont {Legeza}},\ }\href@noop {}
  {\bibfield  {journal} {\bibinfo  {journal} {Physical Review B}\ }\textbf
  {\bibinfo {volume} {90}},\ \bibinfo {pages} {245129} (\bibinfo {year}
  {2014})}\BibitemShut {NoStop}%
\bibitem [{\citenamefont {Knecht}, \citenamefont {Legeza},\ and\ \citenamefont
  {Reiher}(2014)}]{knecht2014communication}%
  \BibitemOpen
  \bibfield  {author} {\bibinfo {author} {\bibfnamefont {S.}~\bibnamefont
  {Knecht}}, \bibinfo {author} {\bibfnamefont {{\"O}.}~\bibnamefont {Legeza}},
  \ and\ \bibinfo {author} {\bibfnamefont {M.}~\bibnamefont {Reiher}},\
  }\href@noop {} {\bibfield  {journal} {\bibinfo  {journal} {The Journal of
  Chemical Physics}\ }\textbf {\bibinfo {volume} {140}},\ \bibinfo {pages}
  {041101} (\bibinfo {year} {2014})}\BibitemShut {NoStop}%
\bibitem [{\citenamefont {Szalay}\ \emph {et~al.}(2015)\citenamefont {Szalay},
  \citenamefont {Pfeffer}, \citenamefont {Murg}, \citenamefont {Barcza},
  \citenamefont {Verstraete}, \citenamefont {Schneider},\ and\ \citenamefont
  {Legeza}}]{szalay2015tensor}%
  \BibitemOpen
  \bibfield  {author} {\bibinfo {author} {\bibfnamefont {S.}~\bibnamefont
  {Szalay}}, \bibinfo {author} {\bibfnamefont {M.}~\bibnamefont {Pfeffer}},
  \bibinfo {author} {\bibfnamefont {V.}~\bibnamefont {Murg}}, \bibinfo {author}
  {\bibfnamefont {G.}~\bibnamefont {Barcza}}, \bibinfo {author} {\bibfnamefont
  {F.}~\bibnamefont {Verstraete}}, \bibinfo {author} {\bibfnamefont
  {R.}~\bibnamefont {Schneider}}, \ and\ \bibinfo {author} {\bibfnamefont
  {{\"O}.}~\bibnamefont {Legeza}},\ }\href@noop {} {\bibfield  {journal}
  {\bibinfo  {journal} {International Journal of Quantum Chemistry}\ }\textbf
  {\bibinfo {volume} {115}},\ \bibinfo {pages} {1342} (\bibinfo {year}
  {2015})}\BibitemShut {NoStop}%
\bibitem [{\citenamefont {Yanai}\ \emph {et~al.}(2015)\citenamefont {Yanai},
  \citenamefont {Kurashige}, \citenamefont {Mizukami}, \citenamefont
  {Chalupsk{\`y}}, \citenamefont {Lan},\ and\ \citenamefont
  {Saitow}}]{yanai2015density}%
  \BibitemOpen
  \bibfield  {author} {\bibinfo {author} {\bibfnamefont {T.}~\bibnamefont
  {Yanai}}, \bibinfo {author} {\bibfnamefont {Y.}~\bibnamefont {Kurashige}},
  \bibinfo {author} {\bibfnamefont {W.}~\bibnamefont {Mizukami}}, \bibinfo
  {author} {\bibfnamefont {J.}~\bibnamefont {Chalupsk{\`y}}}, \bibinfo {author}
  {\bibfnamefont {T.~N.}\ \bibnamefont {Lan}}, \ and\ \bibinfo {author}
  {\bibfnamefont {M.}~\bibnamefont {Saitow}},\ }\href@noop {} {\bibfield
  {journal} {\bibinfo  {journal} {International Journal of Quantum Chemistry}\
  }\textbf {\bibinfo {volume} {115}},\ \bibinfo {pages} {283} (\bibinfo {year}
  {2015})}\BibitemShut {NoStop}%
\bibitem [{\citenamefont {Olivares-Amaya}\ \emph {et~al.}(2015)\citenamefont
  {Olivares-Amaya}, \citenamefont {Hu}, \citenamefont {Nakatani}, \citenamefont
  {Sharma}, \citenamefont {Yang},\ and\ \citenamefont {Chan}}]{olivares2015ab}%
  \BibitemOpen
  \bibfield  {author} {\bibinfo {author} {\bibfnamefont {R.}~\bibnamefont
  {Olivares-Amaya}}, \bibinfo {author} {\bibfnamefont {W.}~\bibnamefont {Hu}},
  \bibinfo {author} {\bibfnamefont {N.}~\bibnamefont {Nakatani}}, \bibinfo
  {author} {\bibfnamefont {S.}~\bibnamefont {Sharma}}, \bibinfo {author}
  {\bibfnamefont {J.}~\bibnamefont {Yang}}, \ and\ \bibinfo {author}
  {\bibfnamefont {G.~K.-L.}\ \bibnamefont {Chan}},\ }\href@noop {} {\bibfield
  {journal} {\bibinfo  {journal} {The Journal of Chemical Physics}\ }\textbf
  {\bibinfo {volume} {142}},\ \bibinfo {pages} {034102} (\bibinfo {year}
  {2015})}\BibitemShut {NoStop}%
\bibitem [{\citenamefont {Li}\ and\ \citenamefont
  {Chan}(2016)}]{li2016hilbert}%
  \BibitemOpen
  \bibfield  {author} {\bibinfo {author} {\bibfnamefont {Z.}~\bibnamefont
  {Li}}\ and\ \bibinfo {author} {\bibfnamefont {G.~K.-L.}\ \bibnamefont
  {Chan}},\ }\href {\doibase http://dx.doi.org/10.1063/1.4942174} {\bibfield
  {journal} {\bibinfo  {journal} {The Journal of Chemical Physics}\ }\textbf
  {\bibinfo {volume} {144}},\ \bibinfo {pages} {084103} (\bibinfo {year}
  {2016})}\BibitemShut {NoStop}%
\bibitem [{\citenamefont {Guo}\ \emph {et~al.}(2016)\citenamefont {Guo},
  \citenamefont {Watson}, \citenamefont {Hu}, \citenamefont {Sun},\ and\
  \citenamefont {Chan}}]{guo2016nelectron}%
  \BibitemOpen
  \bibfield  {author} {\bibinfo {author} {\bibfnamefont {S.}~\bibnamefont
  {Guo}}, \bibinfo {author} {\bibfnamefont {M.~A.}\ \bibnamefont {Watson}},
  \bibinfo {author} {\bibfnamefont {W.}~\bibnamefont {Hu}}, \bibinfo {author}
  {\bibfnamefont {Q.}~\bibnamefont {Sun}}, \ and\ \bibinfo {author}
  {\bibfnamefont {G.~K.-L.}\ \bibnamefont {Chan}},\ }\href@noop {} {\bibfield
  {journal} {\bibinfo  {journal} {Journal of Chemical Theory and Computation}\
  }\textbf {\bibinfo {volume} {12}},\ \bibinfo {pages} {1583} (\bibinfo {year}
  {2016})}\BibitemShut {NoStop}%
\bibitem [{\citenamefont {{\"O}stlund}\ and\ \citenamefont
  {Rommer}(1995)}]{ostlund1995thermodynamic}%
  \BibitemOpen
  \bibfield  {author} {\bibinfo {author} {\bibfnamefont {S.}~\bibnamefont
  {{\"O}stlund}}\ and\ \bibinfo {author} {\bibfnamefont {S.}~\bibnamefont
  {Rommer}},\ }\href@noop {} {\bibfield  {journal} {\bibinfo  {journal}
  {Physical Review Letters}\ }\textbf {\bibinfo {volume} {75}},\ \bibinfo
  {pages} {3537} (\bibinfo {year} {1995})}\BibitemShut {NoStop}%
\bibitem [{\citenamefont {Rommer}\ and\ \citenamefont
  {{\"O}stlund}(1997)}]{rommer1997class}%
  \BibitemOpen
  \bibfield  {author} {\bibinfo {author} {\bibfnamefont {S.}~\bibnamefont
  {Rommer}}\ and\ \bibinfo {author} {\bibfnamefont {S.}~\bibnamefont
  {{\"O}stlund}},\ }\href@noop {} {\bibfield  {journal} {\bibinfo  {journal}
  {Physical Review B}\ }\textbf {\bibinfo {volume} {55}},\ \bibinfo {pages}
  {2164} (\bibinfo {year} {1997})}\BibitemShut {NoStop}%
\bibitem [{\citenamefont {Verstraete}, \citenamefont {Garcia-Ripoll},\ and\
  \citenamefont {Cirac}(2004)}]{verstraete2004matrix}%
  \BibitemOpen
  \bibfield  {author} {\bibinfo {author} {\bibfnamefont {F.}~\bibnamefont
  {Verstraete}}, \bibinfo {author} {\bibfnamefont {J.~J.}\ \bibnamefont
  {Garcia-Ripoll}}, \ and\ \bibinfo {author} {\bibfnamefont {J.~I.}\
  \bibnamefont {Cirac}},\ }\href@noop {} {\bibfield  {journal} {\bibinfo
  {journal} {Physical Review Letters}\ }\textbf {\bibinfo {volume} {93}},\
  \bibinfo {pages} {207204} (\bibinfo {year} {2004})}\BibitemShut {NoStop}%
\bibitem [{\citenamefont {McCulloch}(2007)}]{mcculloch2007density}%
  \BibitemOpen
  \bibfield  {author} {\bibinfo {author} {\bibfnamefont {I.~P.}\ \bibnamefont
  {McCulloch}},\ }\href@noop {} {\bibfield  {journal} {\bibinfo  {journal}
  {Journal of Statistical Mechanics: Theory and Experiment}\ }\textbf {\bibinfo
  {volume} {2007}},\ \bibinfo {pages} {P10014} (\bibinfo {year}
  {2007})}\BibitemShut {NoStop}%
\bibitem [{\citenamefont {Verstraete}, \citenamefont {Murg},\ and\
  \citenamefont {Cirac}(2008)}]{verstraete2008matrix}%
  \BibitemOpen
  \bibfield  {author} {\bibinfo {author} {\bibfnamefont {F.}~\bibnamefont
  {Verstraete}}, \bibinfo {author} {\bibfnamefont {V.}~\bibnamefont {Murg}}, \
  and\ \bibinfo {author} {\bibfnamefont {J.~I.}\ \bibnamefont {Cirac}},\
  }\href@noop {} {\bibfield  {journal} {\bibinfo  {journal} {Advances in
  Physics}\ }\textbf {\bibinfo {volume} {57}},\ \bibinfo {pages} {143}
  (\bibinfo {year} {2008})}\BibitemShut {NoStop}%
\bibitem [{\citenamefont {Pirvu}\ \emph {et~al.}(2010)\citenamefont {Pirvu},
  \citenamefont {Murg}, \citenamefont {Cirac},\ and\ \citenamefont
  {Verstraete}}]{pirvu2010matrix}%
  \BibitemOpen
  \bibfield  {author} {\bibinfo {author} {\bibfnamefont {B.}~\bibnamefont
  {Pirvu}}, \bibinfo {author} {\bibfnamefont {V.}~\bibnamefont {Murg}},
  \bibinfo {author} {\bibfnamefont {J.~I.}\ \bibnamefont {Cirac}}, \ and\
  \bibinfo {author} {\bibfnamefont {F.}~\bibnamefont {Verstraete}},\
  }\href@noop {} {\bibfield  {journal} {\bibinfo  {journal} {New Journal of
  Physics}\ }\textbf {\bibinfo {volume} {12}},\ \bibinfo {pages} {025012}
  (\bibinfo {year} {2010})}\BibitemShut {NoStop}%
\bibitem [{\citenamefont {Vidal}(2004)}]{vidal2004efficient}%
  \BibitemOpen
  \bibfield  {author} {\bibinfo {author} {\bibfnamefont {G.}~\bibnamefont
  {Vidal}},\ }\href@noop {} {\bibfield  {journal} {\bibinfo  {journal}
  {Physical Review Letters}\ }\textbf {\bibinfo {volume} {93}},\ \bibinfo
  {pages} {040502} (\bibinfo {year} {2004})}\BibitemShut {NoStop}%
\bibitem [{\citenamefont {Daley}\ \emph {et~al.}(2004)\citenamefont {Daley},
  \citenamefont {Kollath}, \citenamefont {Schollw{\"o}ck},\ and\ \citenamefont
  {Vidal}}]{daley2004time}%
  \BibitemOpen
  \bibfield  {author} {\bibinfo {author} {\bibfnamefont {A.~J.}\ \bibnamefont
  {Daley}}, \bibinfo {author} {\bibfnamefont {C.}~\bibnamefont {Kollath}},
  \bibinfo {author} {\bibfnamefont {U.}~\bibnamefont {Schollw{\"o}ck}}, \ and\
  \bibinfo {author} {\bibfnamefont {G.}~\bibnamefont {Vidal}},\ }\href@noop {}
  {\bibfield  {journal} {\bibinfo  {journal} {Journal of Statistical Mechanics:
  Theory and Experiment}\ }\textbf {\bibinfo {volume} {2004}},\ \bibinfo
  {pages} {P04005} (\bibinfo {year} {2004})}\BibitemShut {NoStop}%
\bibitem [{\citenamefont {White}\ and\ \citenamefont
  {Feiguin}(2004)}]{white2004real}%
  \BibitemOpen
  \bibfield  {author} {\bibinfo {author} {\bibfnamefont {S.~R.}\ \bibnamefont
  {White}}\ and\ \bibinfo {author} {\bibfnamefont {A.~E.}\ \bibnamefont
  {Feiguin}},\ }\href@noop {} {\bibfield  {journal} {\bibinfo  {journal}
  {Physical Review Letters}\ }\textbf {\bibinfo {volume} {93}},\ \bibinfo
  {pages} {076401} (\bibinfo {year} {2004})}\BibitemShut {NoStop}%
\bibitem [{\citenamefont {Vidal}(2007)}]{vidal2007classical}%
  \BibitemOpen
  \bibfield  {author} {\bibinfo {author} {\bibfnamefont {G.}~\bibnamefont
  {Vidal}},\ }\href@noop {} {\bibfield  {journal} {\bibinfo  {journal}
  {Physical Review Letters}\ }\textbf {\bibinfo {volume} {98}},\ \bibinfo
  {pages} {070201} (\bibinfo {year} {2007})}\BibitemShut {NoStop}%
\bibitem [{\citenamefont {Orus}\ and\ \citenamefont
  {Vidal}(2008)}]{orus2008infinite}%
  \BibitemOpen
  \bibfield  {author} {\bibinfo {author} {\bibfnamefont {R.}~\bibnamefont
  {Orus}}\ and\ \bibinfo {author} {\bibfnamefont {G.}~\bibnamefont {Vidal}},\
  }\href@noop {} {\bibfield  {journal} {\bibinfo  {journal} {Physical Review
  B}\ }\textbf {\bibinfo {volume} {78}},\ \bibinfo {pages} {155117} (\bibinfo
  {year} {2008})}\BibitemShut {NoStop}%
\bibitem [{\citenamefont {McCulloch}(2008)}]{mcculloch2008infinite}%
  \BibitemOpen
  \bibfield  {author} {\bibinfo {author} {\bibfnamefont {I.~P.}\ \bibnamefont
  {McCulloch}},\ }\href@noop {} {\bibfield  {journal} {\bibinfo  {journal}
  {arXiv preprint arXiv:0804.2509}\ } (\bibinfo {year} {2008})}\BibitemShut
  {NoStop}%
\bibitem [{\citenamefont {Feiguin}\ and\ \citenamefont
  {White}(2005)}]{feiguin2005finite}%
  \BibitemOpen
  \bibfield  {author} {\bibinfo {author} {\bibfnamefont {A.~E.}\ \bibnamefont
  {Feiguin}}\ and\ \bibinfo {author} {\bibfnamefont {S.~R.}\ \bibnamefont
  {White}},\ }\href@noop {} {\bibfield  {journal} {\bibinfo  {journal}
  {Physical Review B}\ }\textbf {\bibinfo {volume} {72}},\ \bibinfo {pages}
  {220401} (\bibinfo {year} {2005})}\BibitemShut {NoStop}%
\bibitem [{\citenamefont {Nishino}\ \emph {et~al.}(2001)\citenamefont
  {Nishino}, \citenamefont {Hieida}, \citenamefont {Okunishi}, \citenamefont
  {Maeshima}, \citenamefont {Akutsu},\ and\ \citenamefont
  {Gendiar}}]{nishino2001two}%
  \BibitemOpen
  \bibfield  {author} {\bibinfo {author} {\bibfnamefont {T.}~\bibnamefont
  {Nishino}}, \bibinfo {author} {\bibfnamefont {Y.}~\bibnamefont {Hieida}},
  \bibinfo {author} {\bibfnamefont {K.}~\bibnamefont {Okunishi}}, \bibinfo
  {author} {\bibfnamefont {N.}~\bibnamefont {Maeshima}}, \bibinfo {author}
  {\bibfnamefont {Y.}~\bibnamefont {Akutsu}}, \ and\ \bibinfo {author}
  {\bibfnamefont {A.}~\bibnamefont {Gendiar}},\ }\href@noop {} {\bibfield
  {journal} {\bibinfo  {journal} {Progress of Theoretical Physics}\ }\textbf
  {\bibinfo {volume} {105}},\ \bibinfo {pages} {409} (\bibinfo {year}
  {2001})}\BibitemShut {NoStop}%
\bibitem [{\citenamefont {Verstraete}\ and\ \citenamefont
  {Cirac}(2004)}]{verstraete2004renormalization}%
  \BibitemOpen
  \bibfield  {author} {\bibinfo {author} {\bibfnamefont {F.}~\bibnamefont
  {Verstraete}}\ and\ \bibinfo {author} {\bibfnamefont {J.~I.}\ \bibnamefont
  {Cirac}},\ }\href@noop {} {\bibfield  {journal} {\bibinfo  {journal} {arXiv
  preprint cond-mat/0407066}\ } (\bibinfo {year} {2004})}\BibitemShut {NoStop}%
\bibitem [{\citenamefont {Or{\'u}s}(2014)}]{orus2014practical}%
  \BibitemOpen
  \bibfield  {author} {\bibinfo {author} {\bibfnamefont {R.}~\bibnamefont
  {Or{\'u}s}},\ }\href@noop {} {\bibfield  {journal} {\bibinfo  {journal}
  {Annals of Physics}\ }\textbf {\bibinfo {volume} {349}},\ \bibinfo {pages}
  {117} (\bibinfo {year} {2014})}\BibitemShut {NoStop}%
\bibitem [{\citenamefont {Murg}\ \emph {et~al.}(2010)\citenamefont {Murg},
  \citenamefont {Verstraete}, \citenamefont {Legeza},\ and\ \citenamefont
  {Noack}}]{murg2010simulating}%
  \BibitemOpen
  \bibfield  {author} {\bibinfo {author} {\bibfnamefont {V.}~\bibnamefont
  {Murg}}, \bibinfo {author} {\bibfnamefont {F.}~\bibnamefont {Verstraete}},
  \bibinfo {author} {\bibfnamefont {{\"O}.}~\bibnamefont {Legeza}}, \ and\
  \bibinfo {author} {\bibfnamefont {R.}~\bibnamefont {Noack}},\ }\href@noop {}
  {\bibfield  {journal} {\bibinfo  {journal} {Physical Review B}\ }\textbf
  {\bibinfo {volume} {82}},\ \bibinfo {pages} {205105} (\bibinfo {year}
  {2010})}\BibitemShut {NoStop}%
\bibitem [{\citenamefont {Nakatani}\ and\ \citenamefont
  {Chan}(2013)}]{nakatani2013efficient}%
  \BibitemOpen
  \bibfield  {author} {\bibinfo {author} {\bibfnamefont {N.}~\bibnamefont
  {Nakatani}}\ and\ \bibinfo {author} {\bibfnamefont {G.~K.-L.}\ \bibnamefont
  {Chan}},\ }\href@noop {} {\bibfield  {journal} {\bibinfo  {journal} {The
  Journal of Chemical Physics}\ }\textbf {\bibinfo {volume} {138}},\ \bibinfo
  {pages} {134113} (\bibinfo {year} {2013})}\BibitemShut {NoStop}%
\bibitem [{\citenamefont {Murg}\ \emph {et~al.}(2015)\citenamefont {Murg},
  \citenamefont {Verstraete}, \citenamefont {Schneider}, \citenamefont {Nagy},\
  and\ \citenamefont {Legeza}}]{murg2015tree}%
  \BibitemOpen
  \bibfield  {author} {\bibinfo {author} {\bibfnamefont {V.}~\bibnamefont
  {Murg}}, \bibinfo {author} {\bibfnamefont {F.}~\bibnamefont {Verstraete}},
  \bibinfo {author} {\bibfnamefont {R.}~\bibnamefont {Schneider}}, \bibinfo
  {author} {\bibfnamefont {P.}~\bibnamefont {Nagy}}, \ and\ \bibinfo {author}
  {\bibfnamefont {O.}~\bibnamefont {Legeza}},\ }\href@noop {} {\bibfield
  {journal} {\bibinfo  {journal} {Journal of Chemical Theory and Computation}\
  }\textbf {\bibinfo {volume} {11}},\ \bibinfo {pages} {1027} (\bibinfo {year}
  {2015})}\BibitemShut {NoStop}%
\bibitem [{\citenamefont {Schollw{\"o}ck}(2011)}]{schollwock2011density}%
  \BibitemOpen
  \bibfield  {author} {\bibinfo {author} {\bibfnamefont {U.}~\bibnamefont
  {Schollw{\"o}ck}},\ }\href@noop {} {\bibfield  {journal} {\bibinfo  {journal}
  {Annals of Physics}\ }\textbf {\bibinfo {volume} {326}},\ \bibinfo {pages}
  {96} (\bibinfo {year} {2011})}\BibitemShut {NoStop}%
\bibitem [{\citenamefont {Nakatani}()}]{Naoki}%
  \BibitemOpen
  \bibfield  {author} {\bibinfo {author} {\bibfnamefont {N.}~\bibnamefont
  {Nakatani}},\ }\href@noop {} {\enquote {\bibinfo {title} {{MPSXX}: Mps/mpo
  multi-linear library implemented in c++11.}}\ }\bibinfo {howpublished}
  {\url{https://github.com/naokin/mpsxx}}\BibitemShut {NoStop}%
\bibitem{keller2014determining} S. Keller and M. Reiher, CHIMIA International Journal for Chemistry
{\bf 68}, 200 (2014).
\bibitem [{\citenamefont {Keller}\ \emph {et~al.}(2015)\citenamefont {Keller},
  \citenamefont {Dolfi}, \citenamefont {Troyer},\ and\ \citenamefont
  {Reiher}}]{keller2015efficient}%
  \BibitemOpen
  \bibfield  {author} {\bibinfo {author} {\bibfnamefont {S.}~\bibnamefont
  {Keller}}, \bibinfo {author} {\bibfnamefont {M.}~\bibnamefont {Dolfi}},
  \bibinfo {author} {\bibfnamefont {M.}~\bibnamefont {Troyer}}, \ and\ \bibinfo
  {author} {\bibfnamefont {M.}~\bibnamefont {Reiher}},\ }\href@noop {}
  {\bibfield  {journal} {\bibinfo  {journal} {The Journal of Chemical Physics}\
  }\textbf {\bibinfo {volume} {143}},\ \bibinfo {pages} {244118} (\bibinfo
  {year} {2015})}\BibitemShut {NoStop}%
\bibitem [{\citenamefont {Keller}\ and\ \citenamefont
  {Reiher}(2016)}]{keller2016spin}%
  \BibitemOpen
  \bibfield  {author} {\bibinfo {author} {\bibfnamefont {S.}~\bibnamefont
  {Keller}}\ and\ \bibinfo {author} {\bibfnamefont {M.}~\bibnamefont
  {Reiher}},\ }\href@noop {} {\bibfield  {journal} {\bibinfo  {journal} {The
  Journal of Chemical Physics}\ }\textbf {\bibinfo {volume} {144}},\ \bibinfo
  {pages} {134101} (\bibinfo {year} {2016})}\BibitemShut {NoStop}%
\bibitem [{\citenamefont {Stoudenmire}\ and\ \citenamefont
  {White}()}]{iTensor}%
  \BibitemOpen
  \bibfield  {author} {\bibinfo {author} {\bibfnamefont {E.}~\bibnamefont
  {Stoudenmire}}\ and\ \bibinfo {author} {\bibfnamefont {S.~R.}\ \bibnamefont
  {White}},\ }\href@noop {} {\enquote {\bibinfo {title} {{ITensor}: A c++
  library for creating efficient and flexible physics simulations based on
  tensor product wavefunctions.}}\ }\bibinfo {howpublished}
  {\url{http://itensor.org/}}\BibitemShut {NoStop}%
\bibitem [{\citenamefont {Dolfi}\ \emph {et~al.}(2014)\citenamefont {Dolfi},
  \citenamefont {Bauer}, \citenamefont {Keller}, \citenamefont {Kosenkov},
  \citenamefont {Ewart}, \citenamefont {Kantian}, \citenamefont {Giamarchi},\
  and\ \citenamefont {Troyer}}]{dolfi2014matrix}%
  \BibitemOpen
  \bibfield  {author} {\bibinfo {author} {\bibfnamefont {M.}~\bibnamefont
  {Dolfi}}, \bibinfo {author} {\bibfnamefont {B.}~\bibnamefont {Bauer}},
  \bibinfo {author} {\bibfnamefont {S.}~\bibnamefont {Keller}}, \bibinfo
  {author} {\bibfnamefont {A.}~\bibnamefont {Kosenkov}}, \bibinfo {author}
  {\bibfnamefont {T.}~\bibnamefont {Ewart}}, \bibinfo {author} {\bibfnamefont
  {A.}~\bibnamefont {Kantian}}, \bibinfo {author} {\bibfnamefont
  {T.}~\bibnamefont {Giamarchi}}, \ and\ \bibinfo {author} {\bibfnamefont
  {M.}~\bibnamefont {Troyer}},\ }\href@noop {} {\bibfield  {journal} {\bibinfo
  {journal} {Computer Physics Communications}\ }\textbf {\bibinfo {volume}
  {185}},\ \bibinfo {pages} {3430} (\bibinfo {year} {2014})}\BibitemShut
  {NoStop}%
\bibitem [{\citenamefont {Sharma}\ and\ \citenamefont
  {Chan}(2014)}]{sharma2014communication}%
  \BibitemOpen
  \bibfield  {author} {\bibinfo {author} {\bibfnamefont {S.}~\bibnamefont
  {Sharma}}\ and\ \bibinfo {author} {\bibfnamefont {G.~K.-L.}\ \bibnamefont
  {Chan}},\ }\href@noop {} {\bibfield  {journal} {\bibinfo  {journal} {The
  Journal of Chemical Physics}\ }\textbf {\bibinfo {volume} {141}},\ \bibinfo
  {pages} {111101} (\bibinfo {year} {2014})}\BibitemShut {NoStop}%
\bibitem [{\citenamefont {Sharma}\ and\ \citenamefont
  {Alavi}(2015)}]{sharma2015multireference}%
  \BibitemOpen
  \bibfield  {author} {\bibinfo {author} {\bibfnamefont {S.}~\bibnamefont
  {Sharma}}\ and\ \bibinfo {author} {\bibfnamefont {A.}~\bibnamefont {Alavi}},\
  }\href@noop {} {\bibfield  {journal} {\bibinfo  {journal} {The Journal of
  Chemical Physics}\ }\textbf {\bibinfo {volume} {143}},\ \bibinfo {pages}
  {102815} (\bibinfo {year} {2015})}\BibitemShut {NoStop}%
\bibitem [{\citenamefont {Sharma}, \citenamefont {Jeanmairet},\ and\
  \citenamefont {Alavi}(2016)}]{sharma2016quasi}%
  \BibitemOpen
  \bibfield  {author} {\bibinfo {author} {\bibfnamefont {S.}~\bibnamefont
  {Sharma}}, \bibinfo {author} {\bibfnamefont {G.}~\bibnamefont {Jeanmairet}},
  \ and\ \bibinfo {author} {\bibfnamefont {A.}~\bibnamefont {Alavi}},\
  }\href@noop {} {\bibfield  {journal} {\bibinfo  {journal} {The Journal of
  Chemical Physics}\ }\textbf {\bibinfo {volume} {144}},\ \bibinfo {pages}
  {034103} (\bibinfo {year} {2016})}\BibitemShut {NoStop}%
\bibitem [{\citenamefont {Dorando}, \citenamefont {Hachmann},\ and\
  \citenamefont {Chan}(2009)}]{dorando2009analytic}%
  \BibitemOpen
  \bibfield  {author} {\bibinfo {author} {\bibfnamefont {J.~J.}\ \bibnamefont
  {Dorando}}, \bibinfo {author} {\bibfnamefont {J.}~\bibnamefont {Hachmann}}, \
  and\ \bibinfo {author} {\bibfnamefont {G.~K.-L.}\ \bibnamefont {Chan}},\
  }\href@noop {} {\bibfield  {journal} {\bibinfo  {journal} {The Journal of
  Chemical Physics}\ }\textbf {\bibinfo {volume} {130}},\ \bibinfo {pages}
  {184111} (\bibinfo {year} {2009})}\BibitemShut {NoStop}%
\bibitem [{\citenamefont {Nakatani}\ \emph {et~al.}(2014)\citenamefont
  {Nakatani}, \citenamefont {Wouters}, \citenamefont {Van~Neck},\ and\
  \citenamefont {Chan}}]{nakatani2014linear}%
  \BibitemOpen
  \bibfield  {author} {\bibinfo {author} {\bibfnamefont {N.}~\bibnamefont
  {Nakatani}}, \bibinfo {author} {\bibfnamefont {S.}~\bibnamefont {Wouters}},
  \bibinfo {author} {\bibfnamefont {D.}~\bibnamefont {Van~Neck}}, \ and\
  \bibinfo {author} {\bibfnamefont {G.~K.-L.}\ \bibnamefont {Chan}},\
  }\href@noop {} {\bibfield  {journal} {\bibinfo  {journal} {The Journal of
  Chemical Physics}\ }\textbf {\bibinfo {volume} {140}},\ \bibinfo {pages}
  {024108} (\bibinfo {year} {2014})}\BibitemShut {NoStop}%
\bibitem [{\citenamefont {Chan}\ \emph {et~al.}(2008)\citenamefont {Chan},
  \citenamefont {Dorando}, \citenamefont {Ghosh}, \citenamefont {Hachmann},
  \citenamefont {Neuscamman}, \citenamefont {Wang},\ and\ \citenamefont
  {Yanai}}]{chan2008introduction}%
  \BibitemOpen
  \bibfield  {author} {\bibinfo {author} {\bibfnamefont {G.~K.-L.}\
  \bibnamefont {Chan}}, \bibinfo {author} {\bibfnamefont {J.~J.}\ \bibnamefont
  {Dorando}}, \bibinfo {author} {\bibfnamefont {D.}~\bibnamefont {Ghosh}},
  \bibinfo {author} {\bibfnamefont {J.}~\bibnamefont {Hachmann}}, \bibinfo
  {author} {\bibfnamefont {E.}~\bibnamefont {Neuscamman}}, \bibinfo {author}
  {\bibfnamefont {H.}~\bibnamefont {Wang}}, \ and\ \bibinfo {author}
  {\bibfnamefont {T.}~\bibnamefont {Yanai}},\ }in\ \href@noop {} {\emph
  {\bibinfo {booktitle} {Frontiers in Quantum Systems in Chemistry and
  Physics}}}\ (\bibinfo  {publisher} {Springer Netherlands},\ \bibinfo {year}
  {2008})\ pp.\ \bibinfo {pages} {49--65}\BibitemShut {NoStop}%
\bibitem [{\citenamefont {Chan}\ and\ \citenamefont
  {Sharma}(2011)}]{chan2011density}%
  \BibitemOpen
  \bibfield  {author} {\bibinfo {author} {\bibfnamefont {G.~K.-L.}\
  \bibnamefont {Chan}}\ and\ \bibinfo {author} {\bibfnamefont {S.}~\bibnamefont
  {Sharma}},\ }\href@noop {} {\bibfield  {journal} {\bibinfo  {journal} {Annual
  Review of Physical Chemistry}\ }\textbf {\bibinfo {volume} {62}},\ \bibinfo
  {pages} {465} (\bibinfo {year} {2011})}\BibitemShut {NoStop}%
\bibitem [{\citenamefont {Hager}\ \emph {et~al.}(2004)\citenamefont {Hager},
  \citenamefont {Jeckelmann}, \citenamefont {Fehske},\ and\ \citenamefont
  {Wellein}}]{hager2004parallelization}%
  \BibitemOpen
  \bibfield  {author} {\bibinfo {author} {\bibfnamefont {G.}~\bibnamefont
  {Hager}}, \bibinfo {author} {\bibfnamefont {E.}~\bibnamefont {Jeckelmann}},
  \bibinfo {author} {\bibfnamefont {H.}~\bibnamefont {Fehske}}, \ and\ \bibinfo
  {author} {\bibfnamefont {G.}~\bibnamefont {Wellein}},\ }\href@noop {}
  {\bibfield  {journal} {\bibinfo  {journal} {Journal of Computational
  Physics}\ }\textbf {\bibinfo {volume} {194}},\ \bibinfo {pages} {795}
  (\bibinfo {year} {2004})}\BibitemShut {NoStop}%
\bibitem [{\citenamefont {Kurashige}\ and\ \citenamefont
  {Yanai}(2009)}]{kurashige2009high}%
  \BibitemOpen
  \bibfield  {author} {\bibinfo {author} {\bibfnamefont {Y.}~\bibnamefont
  {Kurashige}}\ and\ \bibinfo {author} {\bibfnamefont {T.}~\bibnamefont
  {Yanai}},\ }\href@noop {} {\bibfield  {journal} {\bibinfo  {journal} {The
  Journal of Chemical Physics}\ }\textbf {\bibinfo {volume} {130}},\ \bibinfo
  {pages} {234114} (\bibinfo {year} {2009})}\BibitemShut {NoStop}%
\bibitem [{\citenamefont {Rinc{\'o}n}, \citenamefont {Garc{\'\i}a},\ and\
  \citenamefont {Hallberg}(2010)}]{rincon2010improved}%
  \BibitemOpen
  \bibfield  {author} {\bibinfo {author} {\bibfnamefont {J.}~\bibnamefont
  {Rinc{\'o}n}}, \bibinfo {author} {\bibfnamefont {D.}~\bibnamefont
  {Garc{\'\i}a}}, \ and\ \bibinfo {author} {\bibfnamefont {K.}~\bibnamefont
  {Hallberg}},\ }\href@noop {} {\bibfield  {journal} {\bibinfo  {journal}
  {Computer Physics Communications}\ }\textbf {\bibinfo {volume} {181}},\
  \bibinfo {pages} {1346} (\bibinfo {year} {2010})}\BibitemShut {NoStop}%
\bibitem [{\citenamefont {Stoudenmire}\ and\ \citenamefont
  {White}(2013)}]{stoudenmire2013real}%
  \BibitemOpen
  \bibfield  {author} {\bibinfo {author} {\bibfnamefont {E.}~\bibnamefont
  {Stoudenmire}}\ and\ \bibinfo {author} {\bibfnamefont {S.~R.}\ \bibnamefont
  {White}},\ }\href@noop {} {\bibfield  {journal} {\bibinfo  {journal}
  {Physical Review B}\ }\textbf {\bibinfo {volume} {87}},\ \bibinfo {pages}
  {155137} (\bibinfo {year} {2013})}\BibitemShut {NoStop}%
\bibitem [{\citenamefont {Nemes}\ \emph {et~al.}(2014)\citenamefont {Nemes},
  \citenamefont {Barcza}, \citenamefont {Nagy}, \citenamefont {Legeza},\ and\
  \citenamefont {Szolgay}}]{nemes2014density}%
  \BibitemOpen
  \bibfield  {author} {\bibinfo {author} {\bibfnamefont {C.}~\bibnamefont
  {Nemes}}, \bibinfo {author} {\bibfnamefont {G.}~\bibnamefont {Barcza}},
  \bibinfo {author} {\bibfnamefont {Z.}~\bibnamefont {Nagy}}, \bibinfo {author}
  {\bibfnamefont {{\"O}.}~\bibnamefont {Legeza}}, \ and\ \bibinfo {author}
  {\bibfnamefont {P.}~\bibnamefont {Szolgay}},\ }\href@noop {} {\bibfield
  {journal} {\bibinfo  {journal} {Computer Physics Communications}\ }\textbf
  {\bibinfo {volume} {185}},\ \bibinfo {pages} {1570} (\bibinfo {year}
  {2014})}\BibitemShut {NoStop}%
\end{thebibliography}

\end{document}